\documentclass[11pt, aps,showpacs,nofootinbib,prd,aps,epsf,floats,
               amsmath,amssymb,amsfonts,axodraw,floatfix,graphicx]{revtex4}
\usepackage{amsmath, amssymb}
\newcommand{\mathsym}[1]{{}}

\usepackage{graphicx}
\usepackage{amsmath}
\usepackage{amssymb}
\usepackage{bm}
\newcommand{\be}{\begin{equation}}
\newcommand{\ee}{\end{equation}}
\newcommand{\bea}{\begin{eqnarray}}
\newcommand{\eea}{\end{eqnarray}}

\newcommand{\rem}[1]{}
\newsavebox{\PSLASH}
 \sbox{\PSLASH}{$p$\hspace{-1.8mm}/}
 
\renewcommand{\theequation}{\thesection.\arabic{equation}}
\newcounter{saveeqn}
\newcommand{\add}{\addtocounter{equation}{1}}
\newcommand{\alpheqn}{\setcounter{saveeqn}{\value{equation}}%
\setcounter{equation}{0}%
\renewcommand{\theequation}{\mbox{\thesection.\arabic{saveeqn}{\alph{equation}}}}}
\newcommand{\reseteqn}{\setcounter{equation}{\value{saveeqn}}%
\renewcommand{\theequation}{\thesection.\arabic{equation}}}

 \newsavebox{\notrightarrow}
 \sbox{\notrightarrow}{$\to$\hspace{-4mm}/}
 
 \newsavebox{\PARTIALSLASH}
 \sbox{\PARTIALSLASH}{$\partial$\hspace{-1.6mm}/}
 
 \newsavebox{\ASLASH}
 \sbox{\ASLASH}{$A$\hspace{-2.1mm}/}
 
 \newsavebox{\KSLASH}
 \sbox{\KSLASH}{$k$\hspace{-1.8mm}/}
 
 \newsavebox{\LSLASH}
 \sbox{\LSLASH}{$\ell$\hspace{-1.8mm}/}
 
 \newsavebox{\QSLASH}
 \sbox{\QSLASH}{$q$\hspace{-1.8mm}/}
 
 \newsavebox{\DSLASH}
 \sbox{\DSLASH}{$D$\hspace{-2.2mm}/}
 
 \newsavebox{\DbfSLASH}
 \sbox{\DbfSLASH}{${\mathbf D}$\hspace{-2.8mm}/}
 
 \newsavebox{\DELVECRIGHT}
 \sbox{\DELVECRIGHT}{$\stackrel{\rightarrow}{\partial}$}
 
 \newcommand{\blue}{\IfColor{\textCadetBlue}{}}
\newcommand{\black}{\IfColor{\textBlack}{}}
\newcommand{\red}{\IfColor{\textRed}{}}
\newcommand{\green}{\IfColor{\textOliveGreen}{}}
\newcommand{\lila}{\IfColor{\textRedViolet}{}}

\newcommand{\nc}{\newcommand}
\nc{\Gz}{\fullg}
\nc{\Gvc}{\boldsymbol{G}^c}
\nc{\Gam}{\boldsymbol{\Delta}}
\nc{\Sig}{\boldsymbol{\Sigma}}
\nc{\TS}{\tilde{\Sig}}
\nc{\TG}{\tilde{\mbox{\boldmath $G$}}}
\nc{\gam}{\boldsymbol{\gamma}}
\nc{\alp}{\boldsymbol{\alpha}}
\nc{\hs}{\hspace*{1mm}}
\nc{\ti}{\tilde}
\nc{\p}{\mathcal{P}}
\nc{\si}{\Sigma}
\nc{\beq}{\begin{eqnarray}}
\nc{\eeq}{\end{eqnarray}}
\nc{\ba}{\begin{array}}
\nc{\ea}{\end{array}}
\nc{\la}{\label}
\nc{\no}{\nonumber}
\nc{\btab}{\begin{tabular}}
\nc{\etab}{\end{tabular}}
\nc{\ci}{\cite}
\nc{\ps}{\slash\hspace{-0.20cm}p}
\nc{\ells}{\slash\hspace{-0.23cm}\ell}
\nc{\ks}{\slash\hspace{-0.23cm}k}
\nc{\Ss}{\slash\hspace{-0.23cm}\Sigma}
\nc{\ms}{\slash\hspace{-0.25cm}\mathfrak{P}}
\nc{\trace}{{\rm Tr\,}}
\nc{\se}{\section}
\nc{\ptildes}{\slash\hspace{-0.20cm}\widetilde{p}}
\nc{\ktildes}{\slash\hspace{-0.18cm}\widetilde{k}}
\newcommand{\qpar}{{\mathbf{q}}_{\|}^{2}}
\begin{document}
\title{
Dilepton production rate in a hot and magnetized quark-gluon plasma}
\author{N. Sadooghi}\email{sadooghi@physics.sharif.ir (Corresponding-Author)}
\author{F. Taghinavaz}\email{taghinavaz@physics.sharif.ir}
\affiliation{Department of Physics, Sharif University of Technology,
P.O. Box 11155-9161, Tehran, Iran}
\begin{abstract}
The differential multiplicity of dileptons in a hot and magnetized quark-gluon plasma, $\Delta_{B}\equiv dN_{B}/d^{4}xd^{4}q$, is derived from first principles. The constant magnetic field $B$ is assumed to be aligned in a fixed spatial direction. It is shown that the anisotropy induced by the $B$ field is mainly reflected in the general structure of photon spectral density function. This is related to the imaginary part of the vacuum polarization tensor, $\mbox{Im}[\Pi^{\mu\nu}]$, which is derived in a first order  perturbative approximation. َAs expected, the final analytical expression for $\Delta_{B}$ includes a trace over the product of a photonic part, $\mbox{Im}[\Pi^{\mu\nu}]$, and a leptonic part, ${\cal{L}}_{\mu\nu}$. It is shown that $\Delta_{B}$ consists of two parts, $\Delta_{B}^{\|}$ and $\Delta_{B}^{\perp}$, arising from the components $(\mu,\nu)=(\|,\|)$ and $(\mu,\nu)=(\perp,\perp)$ of $\mbox{Im}[\Pi^{\mu\nu}]$ and ${\cal{L}}_{\mu\nu}$. Here, the transverse and longitudinal directions are defined with respect to the direction of the $B$ field. Combining $\Delta_{B}^{\|}$ and $\Delta_{B}^{\perp}$, a novel anisotropy factor $\nu_{B}$ is introduced. Using the final analytical expression of $\Delta_{B}$, the possible interplay between the temperature $T$ and the magnetic field strength $eB$ on the ratio $\Delta_{B}/\Delta_{0}$ and  $\nu_{B}$ is numerically studied. Here, $\Delta_{0}$ is the Born approximated dilepton multiplicity in the absence of external magnetic fields. It is, in particular, shown that for each fixed $T$ and $B$, in the vicinity of certain threshold energies of virtual photons, $\Delta_{B}\gg \Delta_{0}$ and $\Delta_{B}^{\perp}\gg \Delta_{B}^{\|}$.  The latter anisotropy may be interpreted as one of the microscopic sources of the macroscopic anisotropies, reflecting themselves, e.g., in the elliptic asymmetry factor $v_{2}$ of dileptons.
\end{abstract}
  \maketitle

\section{Introduction}\label{introduction}\label{sec1}
\par\noindent
The ultimate goal of modern experiments of Heavy Ion Collision (HIC) is to create a (macroscopic) state of deconfined quarks and gluons in local thermal equilibrium. The nuclei that are accelerated to (ultra-) relativistic energies are directed towards each other, and create, after the collision, a fireball of hot and dense nuclear matter. The fireball is believed to consist of a plasma of quarks and gluons, and goes through several stages once it cools through rapid expansion under its own pressure.  Electromagnetic probes, i.e. real and virtual photons (dileptons), are used to convey information about the entire fireball evolution. A major advantage of these probes over the majority of hadronic observables is that they are emitted during all stages of the reaction, and, once produced, only participate in electromagnetic and weak interactions for which the mean free paths are much larger than the size and the lifetime of the fireball. They can thus leave the zone of hot and dense matter without suffering from final-state interactions \cite{shen2015, rapp2015, endres2015, rapp2013}. The measurable dilepton invariant mass spectra and photon transverse mass spectra show characteristic features, which represent the main links between experimental observables and the microscopic structure of strongly interacting Quark-Gluon Plasma (QGP) created in early stages of HICs.
\par
Historically, photon and dilepton production rates are calculated by a number of authors \cite{mclerran1985, weldon1990}.
In the present paper, we will follow the method  introduced by H. A. Weldon in \cite{weldon1990}, and will derive the differential multiplicity of dileptons in a hot QGP in the presence of strong magnetic fields. Strong magnetic fields are believed to be created in early stages of noncentral HICs \cite{skokov2012}. Depending on the collision energies and impact parameters of the
collisions, the strength of these fields are estimated
to be of the order $eB \simeq 0.03$ GeV$^{2}$ at RHIC and
$eB \simeq 0.3$ GeV$^{2}$ at LHC \cite{skokov2009, kharzeev2007}.\footnote{$eB= 1$ GeV$^{2}$ corresponds to $B \simeq  1.7 \times 10^{20}$ Gau\ss.} These magnetic fields are, in principle, time-dependent and rapidly decay after $\tau\sim 1$-$2$ fm.
Most theoretical studies deal nevertheless with the idealized limit of
constant and homogeneous magnetic fields. This turns out to be a good approximation because, as
it is argued in \cite{tuchin2013}, due to the electrical conductivity of the QGP medium, the external magnetic field is essentially frozen, and its decay is thus substantially delayed \cite{tuchin2013,  rajagopal2014}.
Uniform magnetic fields affect, in particular, the phase diagram of Quantum Chromodynamics (QCD),\footnote{See, e.g. \cite{fayazbakhsh2011, fayazbakhsh2010} for a complete analysis of the effect of strong magnetic fields on various phases of quark
matters, including chiral and color superconducting phases. See also \cite{andersen2014} for the most recent review of the
effects induced by magnetic catalysis \cite{klimenko1992, miransky1995} and inverse magnetic catalysis \cite{bali2011} on QCD phase diagram.}
and play a significant role in the physics of
relativistic fermions at zero and nonzero temperatures (see \cite{shovkovy2015} for a recent review). 
\par
Another important effect of spatially fixed magnetic fields is the appearance of certain anisotropies in the dynamics of magnetized fermions in the longitudinal and
transverse directions with respect to the direction of these
external fields. They include anisotropies
arising in group velocities, refraction indices and decay
constants of mesons in a hot and magnetized quark matter
\cite{fayazbakhsh2013, fayazbakhsh2012}. Pressure anisotropies \cite{fayazbakhsh2014} and paramagnetic effects, which are also induced by external magnetic fields, are
supposed to have significant effects on the elliptic flow $v_2$ in
HICs  \cite{bali2014}. In \cite{kharzeev2012, basar2014}, novel photon production mechanisms arising from the conformal anomaly of QCD$\times$QED and magneto-sono-luminescence are introduced. In \cite{kharzeev2012}, e.g., it is shown that the presence of a strong magnetic field provides a positive contribution to the azimuthal anisotropy of photons in noncentral HICs.
\par
Finally, external magnetic fields modify the energy dispersion relations of relativistic particles in hot QED and QCD plasmas. In \cite{taghinavaz2015}, we have systematically explored the complete quasiparticle spectrum of a magnetized electromagnetic plasma at finite temperature. We have shown that in addition to the expected normal modes, nontrivial collective excitations arise as new poles of the one-loop corrected fermion propagator at finite temperatures and in the presence of constant magnetic fields. We refer to these new excitations as \textit{hot magnetized plasminos}. Hot plasminos are familiar from the literature \cite{klimov1982, weldon1982}. They have, in particular, important effects on the production rate of dileptons in a hot QGP \cite{pisarski1990}. Here, it is shown that unexpected sharp peaks due to van Hove singularities, appear in the partial annihilation and decay rate of soft quarks and antiquarks (plasminos). These sharp structures are believed to provide a unique signature for the presence of deconfined collective quarks in a QCD plasma \cite{thoma1999}.
\par
In the present paper, motivated, on the one hand, by our recent studies on the role played by constant magnetic fields in the creation of ``hot magnetized plasminos'' \cite{taghinavaz2015}, and, on the other hand, by the effect of ``hot plasminos'' on modifying the dilepton production rate (DPR) in a hot QGP \cite{pisarski1990}, we will compute the differential (local) production rate of dileptons in a hot and magnetized QGP, $\Delta_{B}$, in a first order perturbative approximation.\footnote{Studying the effect of hot magnetized plasminos on DPR is rather involved, and will be postponed to our future works.} After presenting the analytical expression for $\Delta_{B}$ up to a summation over all Landau levels, we will numerically compare $\Delta_{B}$ of dielectrons, with the one-loop (Born) approximated dielectron production rate in the absence of external magnetic fields, $\Delta_{0}$. We will show that near certain threshold energies of (virtual) photons, $\Delta_{B}$ is larger than $\Delta_{0}$. In the limit of large photon energies, $q_{0}\gg T$, however, $\Delta_{B}$ turns out to be even smaller than $\Delta_{0}$. Having in mind that a certain Fourier transformation of DPR leads to flow coefficients $v_{n}$, this specific behavior is thus expected to be reflected in the dependence of these coefficients on the (virtual) photon energy. Surprisingly, this expectation, arising from our present computation, is similar to the recently suggested behavior of the elliptic flow of heavy quarks as a function of their transverse momentum in the presence of large magnetic fields \cite{fukushima2015}. Here, the diffusion coefficient (drag force) of heavy quarks is computed in a hot magnetized QGP in a LLL approximation by making use of a weak coupling expansion. The magnetic field is shown to induce a certain anisotropy in the diffusion coefficients in the transverse and longitudinal directions with respect to this background field. Based on these results, the authors argue that whereas in the quarks' low transverse momentum limit the elliptic flow in the presence of the external $B$ field, $v_{2}^{B}$, is larger than the same quantity in the absence of the $B$ field, $v_{2}^{0}$, for large transverse momenta, $v_{2}^{B}\leq v_{2}^{0}$. Although our results from the DPR in a hot magnetized QGP can be interpreted in the same manner, more profound studies are to be performed to find a rigorous link between our results and the flow coefficients in the presence of (approximately) constant magnetic fields.
Let us only notice at this stage that although the relation between $v_{2}^{B}$ and $\Delta_{B}$ seems to be
given by a Fourier transformation, as aforementioned, but the rigorous computation of $v^{B}_{2}$ from $\Delta_{B}$ requires many ingredients, which are either unknown or not yet well established. One of these ingredients is the space-time evolution of the magnetic field within an expanding QGP. This may arise from the solution of the corresponding relativistic magnetohydrodynamical equations \cite{rischke2016}. Another important ingredient is the exact $(T, eB)$ dependence of the sound velocity in a hot and magnetized QGP, which requires the knowledge about its equation of state. Although there are many attempts in lattice gauge theory to determine this specific quantity (see e.g. \cite{bruckmann2014}), but these kinds of discussions are far from the scope of this paper. It seems therefore to be difficult to present an exact analytical, even numerical result for $v_{2}^{B}$ defined from $\Delta_{B}$. We will therefore postpone this specific computation to our future works, and will present in this paper only the corresponding results for $\Delta_{B}$.
\par
This paper is organized as follows: In section \ref{sec2}, we will briefly review the properties of fermions in the presence of constant magnetic fields by presenting a summary of the Ritus eigenfunction method \cite{ritus1972}. The latter leads to the exact solution of the relativistic Dirac equation in a uniform magnetic field.
The general structure of the production rates of dileptons in a hot magnetized QGP, $\Delta_{B}$, will be derived in section \ref{sec3}. We will show, that similar to $\Delta_{0}$
in the absence of magnetic fields, it consists of a trace over the product of a leptonic and a photonic part. The leptonic part, ${\cal{L}}_{\mu\nu}$  includes certain basis tensors,
which can be separated into four groups, depending on whether the indices $\mu$ and $\nu$ are parallel or perpendicular to the direction of the background magnetic field. Similarly,
the photonic part, consisting of the imaginary part of the vacuum polarization tensor $\Pi_{\mu\nu}$, is characterized by the same grouping of $\mu$ and $\nu$ indices.
The analytical result for the one-loop approximated expression for $\Delta_{B}$ will be presented in section \ref{sec4}. We will show that $\Delta_{B}$ can be separated
into two parts, $\Delta_{B}^{\|}$ and $\Delta_{B}^{\perp}$, which arise from the components $(\mu,\nu)=(\|,\|)$ and $(\mu,\nu)=(\perp,\perp)$ of the leptonic part ${\cal{L}}_{\mu\nu}$
and the photonic part $\mbox{Im}[\Pi^{\mu\nu}]$. Combining these two contributions, a novel anisotropy factor, 
\begin{eqnarray}\label{O9}
\nu_{B}\equiv \frac{\Delta_{B}^{\perp}-\Delta_{B}^{\|}}{\Delta_{B}^{\perp}+\Delta_{B}^{\|}},
\end{eqnarray}
between $\Delta_{B}^{\|}$ and $\Delta_{B}^{\perp}$ will be introduced. The dependence of $\nu_{B}$ on photon energy may be brought into relation with the dependence of flow coefficients $v_{n}$ of dileptons on their invariant masses. In section \ref{sec5}, we will
numerically evaluate the ratio of $\Delta_{B}/\Delta_{0}$ for dielectrons and dimuons and the anisotropy factor $\nu_{B}$ for dielectrons as functions of rescaled photon energy $q_{0}/T$. We will, in particular, show that, near certain threshold energies of (virtual) photons,\footnote{This threshold is different from the ``minimum'' production threshold of dilepton production as it will be discussed in section \ref{sec5}.} $\Delta_{B}^{\perp}\gg \Delta_{B}^{\|}$. Having in mind that dilepton production rates are directly related to flow coefficients $v_{n}$ of QGP, this specific anisotropy may be interpreted as one of the microscopic sources of the macroscopic anisotropies observed in dilepton flow coefficients at RHIC and LHC.
A summary of our results, together with a number of concluding remarks will be presented in section \ref{sec6}.
\section{Magnetized fermions in a QED plasma at zero temperature}\label{sec2}
\setcounter{equation}{0}
\par\noindent
The Ritus eigenfunction method \cite{ritus1972} is along with the Schwinger proper-time formalism \cite{schwinger1951}, a commonly used method to solve the Dirac equation of charged fermions in the presence of a constant magnetic field. In this section, we will review this Ritus' method, and present the eigenvalues as well as eigenfunctions of the Dirac equation. We will also present the propagator of charged fermions in a multi-flavor model.
\par
To start, let us consider the Dirac equation
\begin{eqnarray}\label{S1}
(\gamma\cdot \Pi^{(q)}-m_{q})\psi^{(q)}=0,
\end{eqnarray}
for fermions of mass $m_{q}$ in a constant magnetic field. Here, $\Pi_{\mu}^{(q)}\equiv i\partial_{\mu}+eq_{f} A_{\mu}^{\mbox{\scriptsize{ext.}}}$, with $e>0$ and $q_{f}$ the charge of the fermions. To describe a magnetic field ${\mathbf{B}}$ aligned in the third direction, ${\mathbf{B}}=B{\mathbf{e}}_{3}$, the gauge field $A_{\mu}^{\mbox{\scriptsize{ext.}}}$ is chosen to be
$A_{\mu}^{\mbox{\scriptsize{ext.}}}=(0,0,Bx_{1},0)$, with $B>0$. As it is shown in \cite{taghinavaz2012, warringa2009} for a one-flavor system and in \cite{fayazbakhsh2013, fayazbakhsh2012, taghinavaz2015} for a multi-flavor system, \eqref{S1} is solved by making use of the ansatz $\psi^{(q)}=\mathbb{E}_{p}^{(q)}u(\tilde{p})$ for a Dirac fermion with charge $q_{f}$. Here, $\mathbb{E}_{p}^{(q)}$ is the Ritus eigenfunction satisfying
\begin{eqnarray}\label{S2}
(\gamma\cdot \Pi^{(q)})\mathbb{E}_{p}^{(q)}=\mathbb{E}_{p}^{(q)}\left(\gamma\cdot \tilde{p}_{q}\right),
\end{eqnarray}
and $u(\tilde{p})$ is the free Dirac spinor satisfying $(\tilde{\ps}_{q}-m_{q})u(\tilde{p})=0$. In \eqref{S2}, the Ritus momentum turns out to be given by
\begin{eqnarray}\label{S3}
\tilde{p}_{q}^{\mu}\equiv \left(p_{0}, 0,-s_{q}\sqrt{2p|q_{f}eB|}, p^{3}\right),
\end{eqnarray}
with $p$ labeling the Landau levels in the external magnetic field ${\mathbf{B}}$, and $s_{q}\equiv \mbox{sign}(q_{f}eB)$. The Ritus eigenfunction $\mathbb{E}_{p}^{(q)}$ is then derived from \eqref{S2} and \eqref{S3}. It reads
\begin{eqnarray}\label{S4}
\mathbb{E}_{p}^{(q)}(\xi^{s_{q}}_{x,p})=e^{-i\bar{p}\cdot \bar{x}}P_{p}^{(q)}(\xi^{s_{q}}_{x,p}),
\end{eqnarray}
with $\bar{p}_{\mu}\equiv (p_{0},0,p_{2},p_{3})$, $\bar{x}_{\mu}\equiv (x_{0},0,x_{2},x_{3})$ and
\begin{eqnarray}\label{S5}
\xi_{x,p}^{s_{q}}\equiv \frac{x_{1}-s_{q}\ell_{B_q}^{2}p_{2}}{\ell_{B_q}},
\end{eqnarray}
where $\ell_{B_q}\equiv |q_{f}eB|^{-1/2}$. Moreover,
\begin{eqnarray}\label{S6}
\hspace{-0.5cm}P_{p}^{(q)}(\xi_{x,p}^{s_{q}})\equiv {\cal{P}}_{+}^{(q)}f_{p}^{+s_{q}}(\xi_{x,p}^{s_{q}})+\Pi_{p}{\cal{P}}_{-}^{(q)}f_{p}^{-s_{q}}(\xi_{x,p}^{s_{q}}),
\end{eqnarray}
with $\Pi_{p}\equiv 1-\delta_{p,0}$. For $eB>0$, the projectors ${\cal{P}}_{\pm}^{(q)}$ are defined by
\begin{eqnarray}\label{S7}
{\cal{P}}_{\pm}^{(q)}\equiv\frac{1\pm is_{q} \gamma_{1}\gamma_{2}}{2}.
\end{eqnarray}
According to this definition, ${\cal{P}}_{\pm}^{(+)}={\cal{P}}_{\pm}$ and ${\cal{P}}_{\pm}^{(-)}={\cal{P}}_{\mp}$. The projectors ${\cal{P}}_{\pm}=\frac{1}{2}(1\pm i\gamma_{1}\gamma_{2})$ are previously introduced in \cite{taghinavaz2015}.
The functions $f_{p}^{\pm s_{q}}(\xi_{x,p}^{s_{q}})$, appearing in \eqref{S6} are given by
\begin{eqnarray}\label{S8}
\begin{array}{rclcrcl}
f_{p}^{+s_{q}}(\xi_{x,p}^{s_{q}})&\equiv& \Phi_{p}(\xi_{x,p}^{s_{q}}),& \qquad& p&=&0,1,2,\cdots,\nonumber\\
f_{p}^{-s_{q}}(\xi_{x,p}^{s_{q}})&\equiv&\Phi_{p-1}(\xi_{x,p}^{s_{q}}),&\qquad& p&=&1,2,\cdots,
\end{array}\\
\end{eqnarray}
with $\Phi_{p}$ given in terms of Hermite polynomials
\begin{eqnarray}\label{S9}
\Phi_{p}(\xi_{x,p}^{s_{q}})\equiv a_{p}\exp\left(-\frac{(\xi_{x,p}^{s_{q}})^{2}}{2}\right)H_{p}(\xi_{x,p}^{s_{q}}),
\end{eqnarray}
with $a_{p}\equiv (2^{p}p!\sqrt{\pi}\ell_{B_q})^{-1/2}$.
The above results lead to the free fermion propagator for a multi-flavor system \cite{fayazbakhsh2013, fayazbakhsh2012, taghinavaz2015}
\begin{eqnarray}\label{S10}
\hspace{-1cm}S^{(q)}(x,y)&=&\sum_{p=0}^{\infty}\int \frac{d^{3}\bar{p}}{(2\pi)^{3}}e^{-i\bar{p}\cdot (\bar{x}-\bar{y})}P_{p}^{(q)}(\xi_{x,p}^{s_{q}})\frac{i}{\gamma\cdot \tilde{p}_{q}-m_{q}}P_{p}^{(q)}(\xi_{y,p}^{s_{q}}),
\end{eqnarray}
where $\bar{p}\equiv (p_{0},\bar{\mathbf{p}})$ with $\bar{\mathbf{p}}\equiv (0,p_{2},p_{3})$.
To show this, let us, for simplicity, assume $q_{f}=+1$ for a fermion with mass $m$, and introduce the following quantized fermionic fields in the presence of a constant magnetic field:
\begin{eqnarray}\label{S11}
\psi_{\beta}(x)&=&\frac{1}{V^{1/2}}\sum_{n,s}\int \frac{dp_{2}dp_{3}}{(2\pi)^{2}}\frac{1}{\sqrt{2\tilde{p}^{(0)}_{+}}}\bigg\{[P_{n}^{(+)}(\xi_{x}^{p})]_{\beta\sigma}u_{s,\sigma}(\tilde{p})a_{\bar{\mathbf{p}}}^{n,s}e^{-i\bar{p}\cdot \bar{x}}+\big[P_{n}^{(+)}(\bar{\xi}_{x}^{p})\big]_{\beta\sigma}v_{s,\sigma}(\tilde{p})b_{\bar{\mathbf{p}}}^{\dagger n,s}e^{+i\bar{p}\cdot \bar{x}}\bigg\},\nonumber\\
\bar{\psi}_{\alpha}(x)&=&\frac{1}{V^{1/2}}\sum\limits_{n,s}\int\frac{dp_{2}dp_{3}}{(2\pi)^{2}}\frac{1}{\sqrt{2\tilde{p}^{(0)}_{+}}}\bigg\{a_{\bar{\mathbf{p}}}^{\dagger n,s}\bar{u}_{s,\rho}(\tilde{p})\big[P_{n}^{(+)}(\xi_{x}^{p})\big]_{\rho\alpha}e^{+i\bar{p}\cdot \bar{x}}+b_{\bar{\mathbf{p}}}^{n,s}\bar{v}_{s,\rho}(\tilde{p})\big[P_{n}^{(+)}(\bar{\xi}_{x}^{p})\big]_{\rho\alpha}e^{-i\bar{p}\cdot \bar{x}}\bigg\},\nonumber\\
\end{eqnarray}
where the simplified notations $\xi_{x}^{p}\equiv \xi_{x,p}^{+}$ and $\bar{\xi}_{x}^{p}\equiv \xi_{x,p}^{-}$ are used. Here, $u(\tilde{k})$ and $v(\tilde{p})$ as well as $\bar{u}(\tilde{p})$ and $\bar{v}(\tilde{p})$ are free spinors, and $\tilde{p}^{(0)}_{+}=(p_{3}^{2}+2neB+m_{q}^{2})^{1/2}$ arises from the definition of the Ritus momentum in (\ref{S3}). Moreover, $a_{\bar{\mathbf{p}}}^{\dagger n,s}$ and $a_{\bar{\mathbf{p}}}^{n,s}$ as well as $b_{\bar{\mathbf{p}}}^{\dagger n,s}$ and $b_{\bar{\mathbf{p}}}^{n,s}$ are the creation and annihilation operators for particles as well as antiparticles with charge $q_{f}=+1$ and $q_{f}=-1$ in the $n$-th Landau level with spin $s$. To evaluate the Feynman propagator
\begin{eqnarray}\label{S12}\hspace{-0.5cm}
[S_{0}^{(+)}(x-y)]_{\alpha\beta}&\equiv &\theta(x_{0}-y_{0})\langle\psi_{\alpha}(x)\bar{\psi}_{\beta}(y)\rangle-\theta(y_{0}-x_{0})\langle \bar{\psi}_{\beta}(y)\psi_{\alpha}(x)\rangle,
\end{eqnarray}
we use the modified equal-time commutation relations
\begin{eqnarray}\label{S13}
\{a_{\bar{\mathbf{p}}}^{n,r}, a_{\bar{\mathbf{k}}}^{\dagger m,s}\}&=&(2\pi)^{2}V\delta^{2}(\bar{\mathbf{p}}-\bar{\mathbf{k}})\delta_{r,s}\delta_{m,n},\nonumber\\
\{b_{\bar{\mathbf{p}}}^{n,r}, b_{\bar{\mathbf{k}}}^{\dagger m,s}\}&=&(2\pi)^{2}V\delta^{2}(\bar{\mathbf{p}}-\bar{\mathbf{k}})\delta_{r,s}\delta_{m,n},
\end{eqnarray}
and arrive at
\begin{eqnarray}\label{S14}
\langle \psi_{\alpha}(x)\bar{\psi}_{\beta}(y)\rangle&=&\sum_{n}\int\frac{dp_{2}dp_{3}}{(2\pi)^{2}}\frac{1}{2\tilde{p}^{(0)}_{+}}e^{-i\bar{p}\cdot (\bar{x}-\bar{y})}[P_{n}^{(+)}(\xi_{x}^{p})]_{\alpha\sigma} (\gamma\cdot\tilde{p}_{+}+m)_{\sigma\rho}[P_{n}^{(+)}(\xi_{y}^{p})]_{\rho\beta},\nonumber\\
\langle \bar{\psi}_{\beta}(y)\psi_{\alpha}(x)\rangle&=&\sum_{n}\int \frac{dp_{2}dp_{3}}{(2\pi)^{2}}\frac{1}{2\tilde{p}^{(0)}_{+}}e^{+i\bar{p}\cdot(\bar{x}-\bar{y})}[P_{n}^{(+)}(\bar{\xi}_{x}^{p})]_{\alpha\sigma} (\gamma\cdot\tilde{p}_{-}-m)_{\sigma\rho}[P_{n}^{(+)}(\bar{\xi}_{x}^{p})]_{\rho\beta}.
\end{eqnarray}
In \eqref{S14}, the factors $(\gamma\cdot \tilde{p}_{\pm}\pm m)$ with
\begin{eqnarray}\label{S15}
\tilde{p}_{\pm}^{\mu}=(p_{0},0,\mp \sqrt{2neB},-p_{3}),
\end{eqnarray}
for positive and negative charges, arise from
\begin{eqnarray}\label{S16}
\sum_{s}{u}_{\alpha}^{s}(\tilde{p})\bar{u}_{\beta}^{s}(\tilde{p})&=&\left(\gamma\cdot \tilde{p}_{+}+m\right)_{\alpha\beta},\nonumber\\
\sum_{r}v_{\alpha}^{r}(\tilde{p})\bar{v}_{\beta}^{r}(\tilde{p})&=&\left(\gamma\cdot \tilde{p}_{-}-m\right)_{\alpha\beta}.
\end{eqnarray}
Plugging the expressions arising in \eqref{S14} into \eqref{S12}, and following the standard procedure to rewrite the two-dimensional integration over $p_{2}$ and $p_{3}$, appearing in \eqref{S14} as a three-dimensional integration over $p_{0},p_{2}$ and $p_{3}$, we arrive after some computations at
\begin{eqnarray}\label{S17}
S_{0}^{(+)}(x,y)&=&\sum_{n=0}^{\infty}\int \frac{d^{3}\bar{p}}{(2\pi)^{3}}e^{-i\bar{p}\cdot (\bar{x}-\bar{y})} P_{n}^{(+)}(\xi_{x}^{p})\frac{i}{\gamma\cdot \tilde{p}_{+}-m}P_{n}^{(+)}(\xi_{y}^{p}).
\end{eqnarray}
We can generalize the above arguments for an arbitrary charge $q_{f}$, and arrive at the fermion propagator \eqref{S10} for a nonzero magnetic field at zero temperature.
\par
In the imaginary-time formalism of thermal field theory, the free fermion propagator for nonvanishing magnetic fields is thus given by
\begin{eqnarray}\label{S18}
S_{T}^{(q)}(x,y)=iT\sum\limits_{p=0}^{\infty}\sum\limits_{\ell=-\infty}^{\infty}
\int\frac{dp_{2}dp_{3}}{(2\pi)^{2}}e^{-i\omega_{\ell}(\tau_{x}-\tau_{y})}e^{i\bar{\mathbf{p}}\cdot (\bar{\mathbf{x}}-\bar{\mathbf{y}})}
P_{p}^{(q)}(\xi_{x,p}^{s_{q}})\left(\gamma\cdot \tilde{p}_{q}+m_{q}\right)\Delta_{f}(i\omega_{\ell},E_{p})P_{p}^{(q)}(\xi_{y,p}^{s_{q}}), \nonumber\\
\end{eqnarray}
where $\omega_{\ell}=(2\ell+1)\pi T$ is the fermionic Matsubara frequency, and
\begin{eqnarray}\label{S19}
\Delta_{f}(i\omega_{\ell},E_{p})\equiv \frac{1}{\omega_{\ell}^{2}+E_{p}^{2}},
\end{eqnarray}
with $E_{p}\equiv\left(p_{3}^{2}+2p|q_{f}eB|+m_{q}^{2}\right)^{1/2}$.
\section{Dilepton production rate in a magnetized quark-gluon plasma: General considerations}\label{sec3}
\setcounter{equation}{0}
\par\noindent
In \cite{weldon1990}, the production rate of dilepton pairs in a hot relativistic plasma is derived by making use of general field theoretical arguments at finite temperature. The final result is then expressed in terms of the imaginary part of the full vacuum polarization tensor. In this section, we will follow the same method, and, after presenting a brief review of the main steps of the method presented in \cite{weldon1990}, will derive the exact expression for differential dilepton multiplicity in a hot and magnetized QCD plasma.
\par
\begin{figure}[hbt]
\centering
\includegraphics[width=10cm,height=15cm]{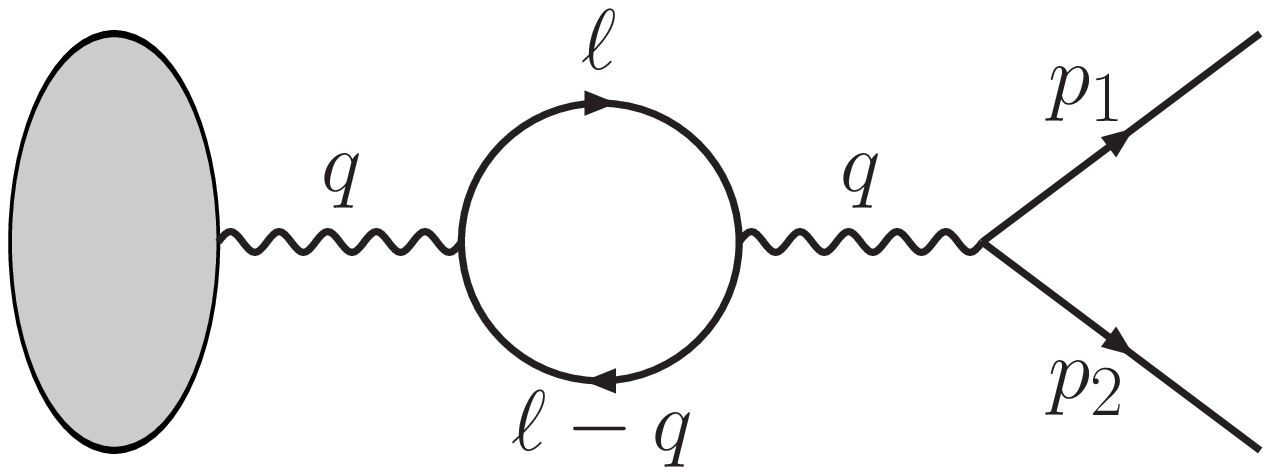}
\vspace{-11cm}
\caption{One-loop contribution to dilepton production in the QCD plasma.}\label{fig1}
\end{figure}

We start with the definition of the thermally averaged multiplicity in the local rest frame of the plasma\footnote{According to \cite{weldon1990}, if the plasma has four-velocity $u^{\mu}=\gamma(1,\mathbf{v})$ in the lab, then $E_{I}$ is to be replaced by 
$P_{I}\cdot u$ and $Z$ by $\mbox{Tr}[e^{-\beta P_{I}\cdot u}]$.}
\begin{eqnarray}\label{A1}
N\equiv \sum_{I,F}|\langle F\ell(p_{2})\bar{\ell}(p_{1})|S|I\rangle|^{2}\frac{e^{-\beta E_{I}}}{Z}\frac{V d^{3}p_{1}}{(2\pi)^{3}}\frac{V d^{3}p_{2}}{(2\pi)^{3}}.
\end{eqnarray}
In the lowest order of perturbative expansion, $S$ is given by
\begin{eqnarray}\label{Ax2}
S=e\int d^{4}x \bar{\psi}\gamma_{\mu}A^{\mu}(x)\psi(x),
\end{eqnarray}
where $\bar{\psi}$ and $\psi$ are solutions of the ordinary Dirac equation with zero external magnetic field. Moreover,
$Z\equiv \mbox{tr}[e^{-\beta H}]$ is the canonical partition function with $\beta\equiv T^{-1}$. 
Following the standard steps to evaluate (\ref{A1}) \cite{weldon1990}, we arrive after a straightforward computation at
\begin{eqnarray}\label{A2}
N=e^{2}L_{\mu\nu}M^{\mu\nu}\frac{d^{3}p_{1}}{(2\pi)^{3}E_{1}}\frac{d^{3}p_{2}}{(2\pi)^{3}E_{2}},
\end{eqnarray}
with the lepton tensor
\begin{eqnarray}\label{A3}
L_{\mu\nu}=\frac{1}{4}\sum_{\mbox{\tiny{spins}}}\mbox{tr}\big[\bar{u}(p_{2})\gamma_{\mu}v(p_{1})\bar{v}(p_{1})\gamma_{\nu}u(p_{2})\big]=
p_{1\mu}p_{2\nu}+p_{2\mu}p_{1\nu}-(p_{1}\cdot p_{2}+m_{\ell}^{2})g_{\mu\nu},
\end{eqnarray}
and the photon tensor
\begin{eqnarray}\label{A4}
M^{\mu\nu}=e^{-\beta q_{0}}\sum_{F}\int d^{4}x d^{4}y e^{iq\cdot (x-y)}
\langle F|A^{\mu}(x)A^{\nu}(y)|F\rangle\frac{e^{-\beta E_{F}}}{Z},
\end{eqnarray}
where $\langle F|A^{\mu}(x)A^{\nu}(y)|F\rangle$ is a Wightman function. According to figure \ref{fig1}, the energy of the virtual photon, $q_{0}=E_{I}-E_{F}$, is given in terms of the energies of an on the mass-shell lepton pair $p_{i}^{0}=({\mathbf{p}}_{i}^{2}+m_{\ell}^{2}), i=1,2$ with four-momenta $p_{1}$ and $p_{2}$ as well as lepton mass $m_{\ell}$.
\vspace{-1cm}
\par
Using at this stage the definition of the photon spectral density function at finite temperature
\begin{eqnarray}\label{A5}
\rho^{\mu\nu}(q)\equiv \int \frac{d^{4}x}{2\pi}e^{iq\cdot x}\sum_{F}\langle F|A^{\mu}(x)A^{\nu}(0)|F\rangle \frac{e^{-\beta E_{F}}}{Z},
\end{eqnarray}
the multiplicity per unit space-time volume is given by
\begin{eqnarray}\label{A6}
\frac{dN}{d^{4}x}=2\pi e^{2}e^{-\beta q_{0}}L_{\mu\nu}\rho^{\mu\nu}\frac{d^{3}p_{1}}{(2\pi)^{3}E_{1}}\frac{d^{3}p_{2}}{(2\pi)^{3}E_{2}}.
\end{eqnarray}
To obtain the DPR in terms of exact photon self-energy, the crucial point is to relate $\rho^{\mu\nu}(q)$ to the imaginary part of the retarded photon propagator $D_{R}^{\mu\nu}$
\begin{eqnarray}\label{A7}
\rho^{\mu\nu}(q_{0},{\mathbf{q}})=-\frac{1}{\pi}\frac{e^{\beta q_{0}}}{e^{\beta q_{0}}-1}\mbox{Im}[D_{R}^{\mu\nu}(q_{0},{\mathbf{q}})].
\end{eqnarray}
Using now the Schwinger-Dyson equation to arrive at the relation between $D_{R}^{\mu\nu}$  and the photon self-energy $\Pi^{\mu\nu}$, and replacing $D_{R}^{\mu\nu}$ appearing in \eqref{A7} with the following expression in terms of transverse ($T$) and longitudinal ($L$) parts of $\Pi^{\mu\nu}$, $\Pi_{T}$ and $\Pi_{L}$,\footnote{For the definition of transverse and longitudinal projectors, $P_{T}^{\mu\nu}$ and $P_{L}^{\mu\nu}$, see \cite{weldon1990}.}
\begin{eqnarray}\label{A8}
D_{R}^{\mu\nu}(q)=-\frac{P_{T}^{\mu\nu}}{q^{2}-\Pi_{T}(q^{2})}-\frac{P_{L}^{\mu\nu}}{q^{2}-\Pi_{L}(q^{2})}+q^{\mu}q^{\nu}~\mbox{terms},
\end{eqnarray}
one arrives after some works at the standard formula for the differential multiplicity in a hot relativistic plasma \cite{weldon1990}
\begin{eqnarray}\label{A9}
\Delta_{0}\equiv\frac{dN}{d^{4}x d^{4}q}=\frac{\alpha}{12\pi^{4}}\frac{\left(2{\cal{R}}_{T}+{\cal{R}}_{L}\right)}{(e^{\beta q_{0}}-1)}{\cal{F}}\left(m_{\ell}^{2}/q^{2}\right),
\end{eqnarray}
where $\alpha=1/137$ is the QED fine structure constant, and
\begin{eqnarray}\label{A10}
{\cal{R}}_{i}\equiv -\frac{q^{2}\mbox{Im}[\Pi_{i}]}{\left(q^{2}-\mbox{Re}[\Pi_{i}]\right)^{2}+\left(\mbox{Im}[\Pi_{i}]\right)^{2}},
\end{eqnarray}
with $i=T,L$. The function ${\cal{F}}(x)$ in (\ref{A9}) is defined by
\begin{eqnarray}\label{A11}
{\cal{F}}(x)\equiv\left(1+2x\right)\left(1-4 x\right)^{1/2}\Theta(1-4x).
\end{eqnarray}
From (\ref{A9}), it is clear that the first nonvanishing contribution to $\Delta_{0}$ arises from the one-loop vacuum polarization tensor $\Pi^
{\mu\nu}$, whose general form in terms of $\Pi_{T}$ and $\Pi_{L}$ is given by
$\Pi^{\mu\nu}=\Pi_{T}P_{T}^{\mu\nu}+\Pi_{L}P_{L}^{\mu\nu}$. Since, by definition, the first contribution to $\Pi_{T/L}$ arises from the photon self-energy including a fermion loop, the direct photon-to-dilepton process without the fermion loop does not contribute to $\Delta_{0}$. To simplify $\Delta_{0}$ from (\ref{A9}), we use at this stage the limit $\mbox{Re}[\Pi_{i}]\ll q^{2}$. Doing this,  ${\cal{R}}_{i}$ from \eqref{A10} can be approximately given by
\begin{eqnarray}\label{A12}
{\cal{R}}_{i}\approx -\frac{\mbox{Im}[\Pi_{i}]}{q^{2}}.
\end{eqnarray}
Plugging this expression into \eqref{A9}, the differential multiplicity is then given by \cite{weldon1990}
\begin{eqnarray}\label{A13}
\Delta_{0}=-\frac{\alpha}{12\pi^{4}q^{2}}\frac{\mbox{Im}[\Pi^{\mu}_{~\mu}]}{(e^{\beta q_{0}}-1)}{\cal{F}}\left(m_{\ell}^{2}/q^{2}\right),
\end{eqnarray}
where $\Pi^{\mu}_{~\mu}=2\Pi_{T}+\Pi_{L}$ is used. Later, we will use
\begin{eqnarray}\label{A14}
\Delta_{0}=\sum\limits_{q_{f}=\{u,d\}}\frac{\alpha^{2}q_{f}^{2} N_{c}T}{6\pi^{4}|{\mathbf{q}}|}\frac{(1+2\lambda_{\ell})(1+2\varsigma_{q}){Q_{0}}}{(e^{\beta q_{0}}-1)}
\ln\left(\frac{\cosh\frac{q_{0}+|{\mathbf{q}}|R_{0}}{4T}}{\cosh\frac{q_{0}-|{\mathbf{q}}|R_{0}}{4T}}\right)\Theta(q^{2}-4\tilde{m}^{2}),
\end{eqnarray}
which arises from a first order perturbative expansion of $\Pi^{\mu\nu}$ in (\ref{A13}), and will compare it with the differential multiplicity of dileptons in the presence of a constant magnetic field $B$. In (\ref{A14}), $\lambda_{\ell}\equiv m_{\ell}^{2}/q^{2}$, $\varsigma_{q}\equiv m_{q}^{2}/q^{2}$, with $m_{q}$ the bare quark mass, and $Q_{0}\equiv \left(1-4\lambda_{\ell}\right)^{1/2}$ as well as $R_{0}\equiv \left(1-4\varsigma_{q}\right)^{1/2}$. Moreover, $\tilde{m}\equiv\mbox{max}(m_{q},m_{\ell})$. The Heaviside $\Theta$-function arising on the r.h.s. of (\ref{A14}) is inserted to discard the kinematically forbidden regime $q^{2}<4\tilde{m}^{2}$ for the production of dileptons (see footnote 18).  Let us notice that, assuming isospin symmetry $m_{q}\equiv m_{u}=m_{d}$ and taking the limit of vanishing lepton and quark masses in (\ref{A14}), we arrive at the standard Born approximated result of local DPR for vanishing magnetic fields \cite{pisarski1990, greiner2010}.
At this stage, it is worth to remind that in order to arrive at the above
differential multiplicity $\Delta_{0}$, we have started with the thermally averaged multiplicity $N$ from (\ref{A1}), which is defined in the local rest frame of the plasma, and used, as is demonstrated in \cite{weldon1990}, the standard definition of 
$P_{T}^{\mu\nu}$ and $P_{L}^{\mu\nu}$ in the rest frame of the fluid. In a Lorentz
frame, where the plasma is not at rest, $q_0$ in (\ref{A14}) is to be replaced by $q\cdot u$, where $u_{\mu}$ is the four-velocity of the plasma \cite{weldon1990} (see also \cite{hatsuda2005} for more details on the comparison of this result
with those of HICs).
\par
For nonzero magnetic fields, because of the well-known dimensional reduction, which is also reflected in the definition of quantized $\psi$ and $\bar{\psi}$ from \eqref{S11}, we have to begin with a new definition for the multiplicity
\begin{eqnarray}\label{A15}
N_{B}\equiv\frac{eB}{2\pi}\sum\limits_{I,F}|\langle F\ell(p_{2})\bar{\ell}(p_{1})|S|I\rangle|^{2}\frac{e^{-\beta E_{I}}}{Z}\frac{Vdp_{1}^{(2)}dp_{1}^{(3)}}{(2\pi)^{2}}\frac{Vdp_{2}^{(2)}dp_{2}^{(3)}}{(2\pi)^{2}},
\end{eqnarray}
where the factor $\frac{eB}{2\pi}$ counts the number of Landau-quantized states.\footnote{Here, we are working with $eB>0$. In general $eB$ in front of (\ref{A15}) is to be replaced with $|eB|$.} From dimensional point of view, it replaces  the integration over $dp_{1}^{(1)} dp_{2}^{(1)}$ that does not appear in \eqref{A15} in comparison with \eqref{A1}. In the lowest order of perturbative expansion, $S$ from (\ref{A15}) is given by (\ref{Ax2}), with $\bar{\psi}$ and
 $\psi$ the solution of the Dirac equation with a nonzero external magnetic field (see Sec. II for a specific solution of this equation).
To determine the differential multiplicity $\frac{dN_{B}}{d^{4}x d^{4}q}$, let us first compute $|\langle F\ell(p_{2})\bar{\ell}(p_{1})|S|I)|^{2}$ in \eqref{A15}. To do this, it is necessary to define the quantum states $|\ell\rangle$ and $|\bar{\ell}\rangle$ in the presence of constant magnetic fields. Assuming that these states correspond to positively and negatively charged leptons with momentum $k$ and spin $s$, they are given by
\begin{eqnarray}\label{A16}
|\ell(\bar{\mathbf{k}};n,s)\rangle&\equiv& \frac{1}{V^{\frac{5}{6}}}a_{\bar{\mathbf{k}}}^{\dagger n,s}|0\rangle,\nonumber\\
|\bar{\ell}(\bar{\mathbf{k}};n,s)\rangle&\equiv&\frac{1}{ V^{\frac{5}{6}}}b_{\bar{\mathbf{k}}}^{\dagger n,s}|0\rangle,
\end{eqnarray}
where $n$ labels the Landau levels. The normalization factors are chosen so that the states remain dimensionless. Using now the relations
\begin{eqnarray}\label{A17}
\langle \ell(\bar{\mathbf{p}}_{2};n,s)|\bar{\psi}_{\alpha}(x)&=&\frac{1}{V^{\frac{1}{3}}\sqrt{2\tilde{p}_{2}^{(0)}}}\bar{u}_{s,\rho}(\tilde{p}_{2})e^{i\bar{p}_{2}\cdot \bar{x}}[P_{n}^{(+)}(\xi_{x}^{p_{2}})]_{\rho\alpha},\nonumber\\
\langle \bar{\ell}(\bar{\mathbf{p}}_{1};k,r)|\psi_{\beta}(x)&=&\frac{1}{V^{\frac{1}{3}}\sqrt{2\tilde{p}_{1}^{(0)}}}[P_{k}^{(+)}(\bar{\xi}_{x}^{p_{1}})]_{\beta\sigma}v_{r,\sigma}(\tilde{p}_{1})e^{i\bar{p}_{1}\cdot \bar{x}},
\end{eqnarray}
and (\ref{S16}) to sum over spins, we arrive first at
\begin{eqnarray}\label{A18}
\hspace{-0.3cm}\sum\limits_{I}|\langle F\ell(p_{2})\bar{\ell}(p_{1})|S|I \rangle|^{2}\frac{e^{-\beta E_{I}}}{Z}=\frac{e^{2}}{V^{\frac{4}{3}}}\frac{e^{-\beta q_{0}}}{\tilde{p}_{1}^{(0)}\tilde{p}_{2}^{(0)}}
\int d^{4}x
d^{4}y\frac{e^{-\beta E_{F}}}{Z}e^{i\bar{q}\cdot (\bar{x}-\bar{y})} \langle F|A_{\mu}(x)A_{\nu}(y)|F\rangle {\cal{L}}^{\mu\nu},
\end{eqnarray}
with $\bar{q}=\bar{p}_{1}+\bar{p}_{2}$, and the lepton tensor
\begin{eqnarray}\label{A19}
{\cal{L}}^{\mu\nu}=\frac{1}{4}\mbox{tr}\left((\gamma\cdot \tilde{p}_{2}+m_{\ell})P_{n}^{(+)}(\xi_{x}^{p_{2}})\gamma^{\mu}P_{k}^{(+)}(\bar{\xi}_{x}^{p_{1}})(\gamma\cdot \tilde{p}_{1}-m_{\ell})P_{k}^{(+)}(\bar{\xi}_{y}^{p_{1}})\gamma^{\nu}P_{n}^{(+)}(\xi_{y}^{p_{2}})\right).
\end{eqnarray}
In \eqref{A18} $\tilde{p}_{i}^{(0)}$ with $i=1,2$ are defined by
\begin{eqnarray}\label{A20}
\tilde{p}_{1}^{(0)}=[p_{1}^{(3)2}+M_{k}^{2}]^{1/2},\qquad\mbox{and}\qquad
\tilde{p}_{2}^{(0)}=[(q_{3}-p_{1}^{(3)})^{2}+M_{n}^{2}]^{1/2},
\end{eqnarray}
where the lepton magnetic mass $M_{j}^{2}\equiv m_{\ell}^{2}+2j|q_{f}eB|$.
Plugging then $P_{n}^{(+)}$ from \eqref{S6} into \eqref{A19}, and eventually performing the trace over Dirac matrices, we arrive after a lengthy computation at
\begin{eqnarray}\label{A21}
{\cal{L}}^{\mu\nu}_{}=\frac{1}{2}\sum\limits_{n,k=0}^{\infty}\sum\limits_{i,j=1}^{4}C_{ij}^{nk}
\Psi^{\mu\nu}_{ij,nk},
\end{eqnarray}
where the coefficients $C_{ij}^{nk}$ are given in (\ref{A22}) in appendix \ref{app1},
and the basis tensors $\Psi^{\mu\nu}_{ij,nk}$ by
\begin{eqnarray}\label{A24}
\Psi^{\mu\nu}_{11,nk}&\equiv&\tilde{p}_{1}^{\mu_{\|}}\tilde{p}_{2}^{\nu_{\|}}+
\tilde{p}_{1}^{\nu_{\|}}\tilde{p}_{2}^{\mu_{\|}}-g^{\mu_{\|}\nu_{\|}}(\tilde{\mathbf{p}}_{1}^{\|}\cdot\tilde{\mathbf{p}}_{2}^{\|}+m^{2}),\nonumber\\
\Psi^{\mu\nu}_{12,nk}&\equiv&\tilde{p}_{1}^{(2)}\tilde{p}_{2}^{(2)}
g^{\mu_{\|}\nu_{\|}},\nonumber\\
\Psi^{\mu\nu}_{21,nk}&\equiv&\tilde{p}_{1}^{(2)}p_{2}^{\mu_{\|}}(g^{2\nu_{\perp}}-ig^{1\nu_{\perp}}),\nonumber\\
\Psi^{\mu\nu}_{22,nk}&\equiv&\tilde{p}_{1}^{(2)}p_{2}^{\mu_{\|}}(g^{2\nu_{\perp}}+ig^{1\nu_{\perp}}),\nonumber\\
\Psi^{\mu\nu}_{23,nk}&\equiv&\tilde{p}_{2}^{(2)}p_{1}^{\mu_{\|}}(g^{2\nu_{\perp}}+ig^{1\nu_{\perp}}),\nonumber\\
\Psi^{\mu\nu}_{24,nk}&\equiv&\tilde{p}_{2}^{(2)}p_{1}^{\mu_{\|}}(g^{2\nu_{\perp}}-ig^{1\nu_{\perp}}),\nonumber\\
\Psi^{\mu\nu}_{31,nk}&\equiv&\tilde{p}_{2}^{(2)}p_{1}^{\nu_{\|}}(g^{2\mu_{\perp}}-ig^{1\mu_{\perp}}),\nonumber\\
\Psi^{\mu\nu}_{32,nk}&\equiv&\tilde{p}_{2}^{(2)}p_{1}^{\nu_{\|}}(g^{2\mu_{\perp}}+ig^{1\mu_{\perp}}),\nonumber\\
\Psi^{\mu\nu}_{33,nk}&\equiv&\tilde{p}_{1}^{(2)}p_{2}^{\nu_{\|}}(g^{2\mu_{\perp}}-ig^{1\mu_{\perp}}),\nonumber\\
\Psi^{\mu\nu}_{34,nk}&\equiv&\tilde{p}_{1}^{(2)}p_{2}^{\nu_{\|}}(g^{2\mu_{\perp}}+ig^{1\mu_{\perp}}),\nonumber\\
\Psi^{\mu\nu}_{41,nk}&\equiv&\tilde{p}_{1}^{(2)}\tilde{p}_{2}^{(2)}
\big[2g^{2\mu_{\perp}}g^{2\nu_{\perp}}+g^{\mu_{\perp}\nu_{\perp}}
-i(g^{1\mu_{\perp}}g^{2\nu_{\perp}}+g^{1\nu_{\perp}}g^{2\mu_{\perp}})\big],
\nonumber\\
\Psi^{\mu\nu}_{42,nk}&\equiv&
\tilde{p}_{1}^{(2)}\tilde{p}_{2}^{(2)}
\big[2g^{2\mu_{\perp}}g^{2\nu_{\perp}}+g^{\mu_{\perp}\nu_{\perp}}+i(g^{1\mu_{\perp}}g^{2\nu_{\perp}}+g^{1\nu_{\perp}}g^{2\mu_{\perp}})\big],
\nonumber\\
\Psi^{\mu\nu}_{43,nk}&\equiv&-(\tilde{\mathbf{p}}_{1}^{\|}\cdot \tilde{\mathbf{p}}_{2}^{\|}+m^{2})\big[g^{\mu_{\perp}\nu_{\perp}}+i(g^{1\mu_{\perp}}g^{2\nu_{\perp}}-g^{1\nu_{\perp}}g^{2\mu_{\perp}})\big],\nonumber\\
\Psi^{\mu\nu}_{44,nk}&\equiv&-(\tilde{\mathbf{p}}_{1}^{\|}\cdot \tilde{\mathbf{p}}_{2}^{\|}+m^{2})\big[g^{\mu_{\perp}\nu_{\perp}}-i(g^{1\mu_{\perp}}g^{2\nu_{\perp}}-g^{1\nu_{\perp}}g^{2\mu_{\perp}})\big].
\end{eqnarray}
Here, $\tilde{\mathbf{p}}^{\|}\equiv (p_{0},0,0,p_{3})$, $\tilde{p}_{1}^{(2)}=-\sqrt{2k|eB|}$ and $\tilde{p}_{2}^{(2)}=+\sqrt{2n|eB|}$. Let us notice that the basis tensors $\Psi^{\mu\nu}_{ij,nk}$ can be separated into four groups, depending on whether their $\mu$ and $\nu$ indices are parallel or perpendicular to the direction of the external $B$ field: According to \eqref{A24}, the $(\mu,\nu)$ indices in $\Psi_{1j,nk}^{\mu\nu}, j=1,2$ are both parallel to the $B$ field, whereas the $\mu$ index in $\Psi_{2j,nk}^{\mu\nu}, j=1,\cdots,4$ is parallel to the $B$ field, their $\nu$ indices are perpendicular to the magnetic field, etc. Since, according to \eqref{A18} these indices are to be contracted with the indices of the photon tensor $\langle F|A_{\mu}(x)A_{\nu}(y)|F\rangle$, it seems to be appropriate to define four Green's functions
\begin{eqnarray}\label{A25}
\begin{array}{rclcrcl}
G_{\mu\nu}^{(1)}(x,y)&\equiv&\langle F|A_{\mu_{\|}}(x)A_{\nu_{\|}}(y)|F\rangle,&\qquad&
G_{\mu\nu}^{(2)}(x,y)&\equiv&\langle F|A_{\mu_{\|}}(x)A_{\nu_{\perp}}(y)|F\rangle,\\
G_{\mu\nu}^{(3)}(x,y)&\equiv&\langle F|A_{\mu_{\perp}}(x)A_{\nu_{\|}}(y)|F\rangle,&\qquad&
G_{\mu\nu}^{(4)}(x,y)&\equiv&\langle F|A_{\mu_{\perp}}(x)A_{\nu_{\perp}}(y)|F\rangle,\\
\end{array}
\end{eqnarray}
in order to arrive at
\begin{eqnarray}\label{A26}
\lefteqn{
\sum\limits_{I}|\langle F\ell(p_{2})\bar{\ell}(p_{1})|S|I \rangle|^{2}\frac{e^{-\beta E_{I}}}{Z}}\nonumber\\
&&=\frac{e^{2}}{2V^{\frac{4}{3}}\tilde{p}_{1}^{(0)}\tilde{p}_{2}^{(0)}}e^{-\beta q_{0}}
\sum_{n,k=0}^{\infty}\sum_{i,j=1}^{4}\int d^{4}x
d^{4}y\frac{e^{-\beta E_{F}}}{Z}e^{i\bar{q}\cdot (\bar{x}-\bar{y})}G_{\mu\nu}^{(i)}(x,y)C_{ij}^{nk} \Psi_{ij,nk}^{\mu\nu}.
\end{eqnarray}
Plugging at this stage \eqref{A26} into \eqref{A15}, and using translational invariance, we obtain
\begin{eqnarray}\label{A27}
\frac{dN_{B}}{d^{4}x}=\frac{e^{-\beta q_{0}}e^{2}(eB)^{3}V^{2/3}}{2(2\pi)^{2}}
\sum\limits_{n,k=0}^{\infty}\sum\limits_{i,j=1}^{4}\int \frac{dp_{1}^{(3)}dp_{2}^{(3)}}{\tilde{p}_{1}^{(0)}\tilde{p}_{2}^{(0)}}\Psi_{ij,nk}^{\mu\nu}\rho_{\mu\nu}^{ij,nk}(q),
\end{eqnarray}
where the ``directional'' photon spectral density function at finite temperature and nonzero magnetic field is defined by
\begin{eqnarray}\label{A28}
\rho_{\mu\nu}^{ij,nk}(q)\equiv
(eB)^{-2}\sum\limits_{F}\int \frac{d^{4}x}{2\pi}\frac{e^{-\beta E_{F}}}{Z}
e^{i{\mathbf{q}}_{\|}\cdot {\mathbf{x}}_{\|}}\widetilde{C}_{ij}^{nk}(x,0;{\mathbf{q}}_{\perp})G^{(i)}_{\mu\nu}(x,0),
\end{eqnarray}
with
\begin{eqnarray}\label{A29}
\hspace{-0.8cm}\widetilde{C}_{ij}^{nk}(x,y;{\mathbf{q}}_{\perp})&\equiv &
\int \frac{dp_{1}^{(2)}dp_{2}^{(2)}}{(2\pi)^{2}}e^{-iq_{2}(x_{2}-y_{2})}C_{ij}^{nk}.
\end{eqnarray}
Similar to the method used in \cite{weldon1990}, we insert at this stage
\begin{eqnarray*}
1=\int d^{2}{\mathbf{q}}_{\|}\delta^{2}({\mathbf{q}}_{\|}-{\mathbf{p}}_{1}^{\|}-{\mathbf{p}}_{2}^{\|}),
\end{eqnarray*}
into the right hand side of \eqref{A27}, and arrive at the differential multiplicity
\begin{eqnarray}\label{A30}
\Delta_{B}\equiv \frac{dN_{B}}{d^{4}x~d^{4}q}=2\pi\alpha eB e^{-\beta q_{0}}
\sum\limits_{n,k=0}^{\infty}\sum\limits_{i,j=1}^{4}\int \frac{dp_{1}^{(3)}dp_{2}^{(3)}}{\tilde{p}_{1}^{(0)}\tilde{p}_{2}^{(0)}}\delta^{2}({\mathbf{q}}_{\|}-{\mathbf{p}}_{1}^{\|}-{\mathbf{p}}_{2}^{\|})\Psi_{ij,nk}^{\mu\nu}
\rho_{\mu\nu}^{ij,nk}(q).
\end{eqnarray}
Here, the factor $V^{2/3}\left(\frac{eB}{2\pi}\right)^{2}$ appearing in \eqref{A27} is replaced by $d^{2}q_{\perp}$ in $d^{4}q=d^{2}q_{\perp}d^{2}q_{\|}$. It counts the number of Landau-quantized states in the transverse directions with respect to the direction of the magnetic field. Using at this stage the relation
\begin{eqnarray}\label{A31}
\rho_{ij,nk}^{\mu\nu}(q)\equiv  -\frac{1}{\pi}\frac{e^{\beta q_{0}}}{e^{\beta q_{0}}-1}\mbox{Im}[D_{R}^{\mu\nu}(q)]_{ij}^{nk},
\end{eqnarray}
in order to express the directional photon spectral density function in terms of the retarded full photon propagator $D_{R}^{\mu\nu}(q)$, we arrive at the exact expression for the differential DPR in a relativistic hot and magnetized plasma
\begin{eqnarray}\label{A32}
\Delta_{B}=-\frac{2\alpha eB}{e^{\beta q_{0}}-1}\sum\limits_{n,k=0}^{\infty}\sum\limits_{i,j=1}^{4}\int \frac{dp_{1}^{(3)}dp_{2}^{(3)}}{\tilde{p}_{1}^{(0)}\tilde{p}_{2}^{(0)}}
\delta^{2}({\mathbf{q}}_{\|}-{\mathbf{p}}_{1}^{\|}-{\mathbf{p}}_{2}^{\|})\Psi_{ij,nk}^{\mu\nu}
\mbox{Im}[D_{\mu\nu}^{R}(q)]_{ij}^{nk}.
\end{eqnarray}
It is the aim of this paper to determine the above production rate in the first order of weak coupling expansion. To do this, we will determine $D_{\mu\nu}^{R}$ in terms of one-loop vacuum polarization tensor $\Pi_{\mu\nu}$ at finite $T$ and $eB$. Using then the approximation
\begin{eqnarray}\label{A33}
\mbox{Im}[D_{\mu\nu}^{R}(q)]\simeq -\frac{\mbox{Im}[\Pi_{\mu\nu}(q)]_{ij}^{nk}}{(q^{2})^{2}},
\end{eqnarray}
arising from a truncated Schwinger-Dyson series, the differential DPR in a hot and magnetized QGP, expressed in terms of the imaginary part of the one-loop polarization tensor is then given by
\begin{eqnarray}\label{A34}
\Delta_{B}=\frac{2\alpha eB}{(e^{\beta q_{0}}-1)}\frac{1}{(q^{2})^{2}}\sum\limits_{n,k=0}^{\infty}\sum\limits_{i,j=1}^{4}
\mbox{Im}[\Pi_{\mu\nu}(q)]_{ij}^{nk}{\cal{L}}_{ij,nk}^{\mu\nu},
\end{eqnarray}
 with
\begin{eqnarray}\label{A35}
{\cal{L}}_{ij,nk}^{\mu\nu}\equiv \int \frac{dp_{1}^{(3)}dp_{2}^{(3)}}{\tilde{p}_{1}^{(0)}\tilde{p}_{2}^{(0)}}\delta^{2}({\mathbf{q}}_{\|}-{\mathbf{p}}_{1}^{\|}-{\mathbf{p}}_{2}^{\|})
\Psi_{ij,nk}^{\mu\nu}.
\end{eqnarray}
In the next section, we will first present the analytical expression for the dilepton production
rate $\Delta_{B}$ from (\ref{A34}) in a one-loop perturbative expansion.\footnote{Let us notice that here, similar to the $B=0$ case, by definition, the first nonvanishing contribution to $\Pi_{\mu\nu}$ in (\ref{A34}) arises from the one-loop photon self-energy including the fermion loop demonstrated in figure \ref{fig1}. The direct free photon to magnetized leptons production channel is studied in \cite{harding1983}.} The production rates for dielectrons as well as dimuons will be then numerically determined in section \ref{sec5}.
We will, in particular, focus on the $T$ and $B$ dependence of the production rates, and will compare the results for $eB=0.02, 0.03$ GeV$^{2}$ and for $eB=0.2,0.3$ GeV$^{2}$ at fixed $T=200, 400$ MeV.
\section{Dilepton production rate in a hot and magnetized QCD plasma: Analytical results  at one-loop level}\label{sec4}
\setcounter{equation}{0}
\par\noindent
In what follows, we will analytically derive the differential multiplicity of dileptons $\Delta_{B}$  in a first order perturbative approximation.
According to (\ref{A34}), $\Delta_{B}$ consists of two parts, a photonic part,  $\mbox{Im}[\Pi_{\mu\nu}(q)]_{ij}^{nk}$, and
a leptonic part, ${\cal{L}}_{ij,nk}^{\mu\nu}$. We will first evaluate these two parts separately. We will then combine them, and present the analytical
expressions for $\Delta_{B}$ at one-loop level by building the trace over two matrices $\mbox{Im}[\Pi_{\mu\nu}(q)]$ and ${\cal{L}}^{\mu\nu}$ in the $ij$-space. Let us remind that, according
to our descriptions in the previous section, the indices $i$ and $j$ denote the orientation of $\mu$ and $\nu$ indices with respect to the direction of the external magnetic field. A lengthy but straightforward computation shows that the combinations $(\mu,\nu)=(\|,\perp)$ as well as $(\mu,\nu)=(\perp,\|)$ do not contribute to $\Delta_{B}$. Here, $\|$ and $\perp$ denote $(0,3)$ and $(1,2)$ directions, respectively. To simplify the presentation, we will therefore focus only on the relevant combinations, $(\mu,\nu)=(\|,\|)$ as well as $(\mu,\nu)=(\perp,\perp)$.
\subsection{Photonic part of ${\Delta_{B}}$}\label{sec4A}
\par\noindent
The one-loop contribution of the photon self-energy diagram of a two-flavor QCD in the presence of an external magnetic field is given by (see figure \ref{fig1})
\begin{eqnarray}\label{D1}
i\Pi_{\mu\nu}(i\omega_{q},{\mathbf{q}};i\omega_{k},{\mathbf{k}})&=&\sum\limits_{q_{f}=\{u,d\}}e^{2}q_{f}^{2} \int d^{3}x d^{3}y d\tau_{x} d\tau_{y} e^{i\omega_{q}\tau_{x}}e^{-i\omega_{k}\tau_{y}}
 e^{-iq_{1}x_{1}+ik_{1}y_{1}}e^{-iq_{2}x_{2}+ik_{2}y_{2}}e^{-iq_{3}x_{3}+ik_{3}y_{3}}
\nonumber\\
&&\times\mbox{tr}_{c,s}\left(\gamma_{\mu}S_{T}^{(q)}(x,y)\gamma_{\nu}S_{T}^{(q)}(y,x)\right).
\end{eqnarray}
Here, the symbol $q_{f}=\{u,d\}$ is used for a summation over the charges of up and down quarks, $q_{u}=2/3$ and $q_{d}=-1/3$.
Plugging $S_{T}^{(q)}(x,y)$ from (\ref{S18}) into (\ref{D1}), we arrive after some computations at\footnote{The energy $i\omega_{q}$ will be replaced by $q_{0}$, once the imaginary part of $\Pi_{\mu\nu}$ is computed. }
\begin{eqnarray}\label{D2}
\Pi_{\mu\nu}(i\omega_{q},{\mathbf{q}};i\omega_{k},{\mathbf{k}})=(2\pi)^{2}\delta(q_{2}-k_{2})\delta(q_{3}-k_{3})\beta\delta_{kq}\Pi_{\mu\nu}(i\omega_{q},{\mathbf{q}}),
\end{eqnarray}
with
\begin{eqnarray}\label{D3}
\lefteqn{\Pi_{\mu\nu}(i\omega_{q},{\mathbf{q}})=-\sum_{q_{f}=\{u,d\}}e^{2}q_{f}^{2}\int dx_{1}e^{-iq_{1}x_{1}}\int dy_{1}e^{ik_{1}y_{1}}\int \frac{d\ell_{2}}{2\pi}
}\nonumber\\
&&\times
T\sum\limits_{r=-\infty}^{\infty} \sum\limits_{\ell,p=0}^{\infty}\int \frac{d\ell_{3}}{2\pi}\Delta_{f}(i(\omega_{q}-\omega_{r});E_{p})\Delta_{f}(i\omega_{r};E_{\ell}){\cal{T}}_{\mu\nu}^{(q)}\bigg|_{p_{0}=i(\omega_{q}-\omega_{r}),p_{i}=\ell_{i}-q_{i}, i=2,3}^{\tilde{p}_{2}=-s_{q}\sqrt{2p|q_{f}eB|},\tilde{\ell}_{2}=-s_{q}\sqrt{2\ell|q_{f}eB|}}.
\end{eqnarray}
Here, $\Delta_{f}$'s are defined in \eqref{S19} with
\begin{eqnarray}\label{D4}
E_{\ell}=\sqrt{\ell_{3}^{2}+m_{q}^{2}+2\ell|q_{f}eB|},
\qquad\mbox{and}\qquad E_{p}=\sqrt{(\ell_{3}-q_{3})^{2}+m_{q}^{2}+2p|q_{f}eB|}.
\end{eqnarray}
Moreover, we have
\begin{eqnarray}\label{D5}
{\cal{T}}_{\mu\nu}^{(q)}\equiv
\mbox{tr}_{cs}\left(
(\gamma\cdot\tilde{p}_{q}+m_{q})
P_{p}^{(q)}(x_{1},p_{2})
\gamma_{\mu}P_{\ell}^{(q)}(x_{1},\ell_{2})(\gamma\cdot \tilde{\ell}_{q}+m_{q})P_{\ell}^{(q)}(y_{1},\ell_{2})\gamma_{\nu}P_{p}^{(q)}(y_{1},p_{2})
\right).
\end{eqnarray}
Plugging $P_{\ell}^{(q)}$ from (\ref{S6}) into \eqref{D5}, we arrive after some lengthy but straightforward computations at
\begin{eqnarray}\label{D6}
{\cal{T}}_{\mu\nu}^{(q)}=\sum_{i,j=1}^{4}{\cal{A}}_{ij}^{(q)}~\Xi_{\mu\nu}^{(q)ij},
\end{eqnarray}
with ${\cal{A}}_{ij}^{(q)}$ given in (\ref{D6}) from appendix \ref{app2a}.
The bases $\Xi_{\mu\nu}^{(q)ij}$, appearing in (\ref{D6}) read
\begin{eqnarray}\label{D10}
\Xi^{(q)\mu\nu}_{11}&\equiv&\left(\tilde{p}_{q}^{\mu_{\|}}
\tilde{\ell}_{q}^{\nu_{\|}}+
\tilde{p}_{q}^{\nu_{\|}}\tilde{\ell}_{q}^{\mu_{\|}}-g^{\mu_{\|}\nu_{\|}}(\tilde{\mathbf{p}}_{q}^{\|}\cdot\tilde{\boldsymbol{\ell}}_{q}^{\|}-m_{q}^{2})\right),\nonumber\\
\Xi_{12}^{(q)\mu\nu}&\equiv&\tilde{p}_{q}^{(2)}\tilde{\ell}_{q}^{(2)}
g^{\mu_{\|}\nu_{\|}},\nonumber\\
\Xi_{21}^{(q)\mu\nu}&\equiv&\tilde{\ell}_{q}^{(2)}p_{q}^{\mu_{\|}}(g^{2\nu_{\perp}}-is_{q}g^{1\nu_{\perp}}),\nonumber\\
\Xi_{22}^{(q)\mu\nu}&\equiv&\tilde{\ell}_{q}^{(2)}p_{q}^{\mu_{\|}}(g^{2\nu_{\perp}}+is_{q}g^{1\nu_{\perp}}),\nonumber\\
\Xi_{23}^{(q)\mu\nu}&\equiv&\tilde{p}_{q}^{(2)}\ell_{q}^{\mu_{\|}}(g^{2\nu_{\perp}}+is_{q}g^{1\nu_{\perp}}),\nonumber\\
\Xi_{24}^{(q)\mu\nu}&\equiv&\tilde{p}_{q}^{(2)}\ell_{q}^{\mu_{\|}}(g^{2\nu_{\perp}}-is_{q}g^{1\nu_{\perp}}),\nonumber\\
\Xi_{31}^{(q)\mu\nu}&\equiv&\tilde{p}_{q}^{(2)}\ell_{q}^{\nu_{\|}}(g^{2\mu_{\perp}}-is_{q}g^{1\mu_{\perp}}),\nonumber\\
\Xi_{32}^{(q)\mu\nu}&\equiv&\tilde{p}_{q}^{(2)}\ell_{q}^{\nu_{\|}}(g^{2\mu_{\perp}}+is_{q}g^{1\mu_{\perp}}),\nonumber\\
\Xi_{33}^{(q)\mu\nu}&\equiv&\tilde{\ell}_{q}^{(2)}p_{q}^{\nu_{\|}}(g^{2\mu_{\perp}}-is_{q}g^{1\mu_{\perp}}),\nonumber\\
\Xi_{34}^{(q)\mu\nu}&\equiv&\tilde{\ell}_{q}^{(2)}p_{q}^{\nu_{\|}}(g^{2\mu_{\perp}}+is_{q}g^{1\mu_{\perp}}),\nonumber\\
\Xi_{41}^{(q)\mu\nu}&\equiv&\tilde{\ell}_{q}^{(2)}\tilde{p}_{q}^{(2)}
\big[2g^{2\mu_{\perp}}g^{2\nu_{\perp}}+g^{\mu_{\perp}\nu_{\perp}}-is_{q}(g^{1\mu_{\perp}}g^{2\nu_{\perp}}+g^{1\nu_{\perp}}g^{2\mu_{\perp}})\big],
\nonumber\\
\Xi_{42}^{(q)\mu\nu}&\equiv&
\tilde{\ell}_{q}^{(2)}\tilde{p}_{q}^{(2)}
\big[2g^{2\mu_{\perp}}g^{2\nu_{\perp}}+g^{\mu_{\perp}\nu_{\perp}}+is_{q}(g^{1\mu_{\perp}}g^{2\nu_{\perp}}+g^{1\nu_{\perp}}g^{2\mu_{\perp}})\big],
\nonumber\\
\Xi_{43}^{(q)\mu\nu}&\equiv&-(\tilde{\mathbf{p}}_{q}^{\|}\cdot\tilde{\boldsymbol{\ell}}_{q}^{\|}-m_{q}^{2})\big[g^{\mu_{\perp}\nu_{\perp}}+is_{q}(g^{1\mu_{\perp}}g^{2\nu_{\perp}}-g^{1\nu_{\perp}}g^{2\mu_{\perp}})\big],\nonumber\\
\Xi_{44}^{(q)\mu\nu}&\equiv&-(\tilde{\mathbf{p}}_{q}^{\|}\cdot\tilde{\boldsymbol{\ell}}_{q}^{\|}-m_{q}^{2})\big[g^{\mu_{\perp}\nu_{\perp}}-is_{q}(g^{1\mu_{\perp}}g^{2\nu_{\perp}}-g^{1\nu_{\perp}}g^{2\mu_{\perp}})\big].
\end{eqnarray}
Here, $\tilde{p}_{q}^{(2)}=-s_{q}\sqrt{2p|q_{f}eB|}$ and $\tilde{\ell}_{q}^{(2)}=-s_{q}\sqrt{2\ell|q_{f}eB|}$ with $s_{q}=\mbox{sign}(q_{f}eB)$.
\par
To determine $\Pi_{\mu\nu}$ from (\ref{D3}), we will first perform the integration over $\ell_{2}$, $x_{1}$ and $y_{1}$, by making use of
\begin{eqnarray}\label{D11}
\int dx_{1}e^{-ik_{1}x_{1}}f_{p}(\xi_{x}^{p})f_{\ell}(\xi_{x}^{\ell})
&=&\frac{1}{\sqrt{2^{p+\ell}p!\ell!}}(\ell_{B_{+}})^{p+\ell}e^{-ik_{1}(p_{2}+\ell_{2})\frac{\ell_{B_{+}}^{2}}{2}} (\ell_{2}-p_{2}-ik_{1})^{p}
(p_{2}-\ell_{2}-ik_{1})^{\ell}\nonumber\\
&&\times e^{-\frac{\kappa_{+}}{2}}\kappa_{+}^{-m}U^{-m}_{M-m+1}(\kappa_{+}),
\end{eqnarray}
for positive charges and
\begin{eqnarray}\label{D12}
\int dx_{1}e^{-ik_{1}x_{1}}f_{p}(\bar{\xi}_{x}^{p})f_{\ell}(\bar{\xi}_{x}^{\ell})
&=&\frac{(-1)^{p+\ell}}{\sqrt{2^{p+\ell}p!\ell!}}(\ell_{B_{-}})^{p+\ell}e^{+ik_{1}(p_{2}+\ell_{2})\frac{\ell_{B_{-}}^{2}}{2}} (\ell_{2}-p_{2}+ik_{1})^{p}(p_{2}-\ell_{2}+ik_{1})^{\ell}\nonumber\\
&&\times e^{-\frac{\kappa_{-}}{2}}\kappa_{-}^{-m}U^{-m}_{M-m+1}(\kappa_{-}),
\end{eqnarray}
for negative charges. Here, $U_{a}^{b}(z)$ is the confluent hypergeometric function of second kind \cite{wolfram}, and
$m=\mbox{min}(p,\ell)$ as well as $M=\mbox{max}(p,\ell)$.\footnote{See \cite{taghinavaz2015} for a rigorous proof of \eqref{D11} and \eqref{D12}. } Moreover, $\kappa_{q}\equiv\frac{\ell_{B_{q}}^{2}{\mathbf{q}}_{\perp}^{2}}{2}$ with $\ell_{B_{q}}= |q_{f}eB|^{-1/2}$ and ${\mathbf{q}}_{\perp}^{2}\equiv q_{1}^{2}+q_{2}^{2}$. A lengthy but straightforward computation results in
\begin{eqnarray}\label{D13}
\Pi_{\mu\nu}(i\omega_{q},{\mathbf{q}};i\omega_{k},{\mathbf{k}})=(2\pi)^{3}\delta^{3}({\mathbf{q}}-{\mathbf{k}})\beta\delta_{kq}\Pi_{\mu\nu}(i\omega_{q},{\mathbf{q}}),
\end{eqnarray}
with
\begin{eqnarray}\label{D14}
[\Pi_{\mu\nu}(i\omega_{q},{\mathbf{q}})]^{p\ell}
=N_{c}\sum_{q_{f}=\{u,d\}}\sum\limits_{p,\ell=0}^{\infty}\sum\limits_{i,j=1}^{4}{\cal{K}}_{ij}^{(q)}[\tilde{\Xi}_{\mu\nu}^{(q)}]_{ij}^{p\ell},
\end{eqnarray}
and
\begin{eqnarray}\label{D15}
[\tilde{\Xi}_{\mu\nu}^{(q)}]_{ij}^{p\ell}\equiv -e^{2}q_{f}^{2}~
T\sum\limits_{r=-\infty}^{\infty}
\int \frac{d\ell_{3}}{(2\pi)^{2}}
\Delta_{f}(i(\omega_{q}-\omega_{r});E_{p})\Delta_{f}(i\omega_{r};E_{\ell})[\Xi_{\mu\nu}^{(q)}]_{ij}^{p\ell}.
\end{eqnarray}
In \eqref{D14}, the coefficients ${\cal{K}}_{ij}^{(q)}$ arises from the integration over $\ell_{2}, x_{1}$ and $y_{1}$ in (\ref{D3}), as described before. The final results for ${\cal{K}}_{ij}^{(q)}$ are presented in (\ref{D39}), (\ref{D40}) and in \eqref{appA1}-\eqref{appA4}. Moreover, the bases $[\Xi^{(q)}]_{ij}^{p\ell}$, appearing in (\ref{D15}), are given in (\ref{D10}).
\par
Let us notice at this stage that, according to our notations, the indices $p$ and $\ell$ label the Landau levels corresponding to the internal fermion propagators in figure \ref{fig1}. In \eqref{A33} and \eqref{A34}, however, $\mbox{Im}[\Pi_{\mu\nu}]$ shall depend on the indices $n$ and $k$, which label the Landau levels corresponding to the external lepton-antilepton legs
[see \eqref{A17}]. To determine the relation between the external $(n,k)$ and internal $(p,\ell)$ indices, we use the conservation relation $\ell^{(2)}-p^{(2)}=q^{(2)}=p_{1}^{(2)}+p_{2}^{(2)}$, where the superscripts $(2)$ denote to the second component of the corresponding four-momenta. Discretizing the momenta of fermionic fields according to the Ritus prescription, we obtain
\begin{eqnarray}\label{D16}
\sqrt{n}-\sqrt{k}=-s_{q}\sqrt{|q_{f}|}\left(\sqrt{\ell}-\sqrt{p}\right).
\end{eqnarray}
Later, we will use
\begin{eqnarray}\label{D17}
[\Pi_{\mu\nu}]^{nk}=\sum\limits_{p,\ell=0}^{\infty}\eta_{nkp\ell}^{(q)}[\Pi_{\mu\nu}]^{p\ell},
\end{eqnarray}
for each flavor $q_{f}$. Here,  $$\eta_{nkp\ell}^{(q)}\equiv \delta_{\sqrt{n}-\sqrt{k}+s_{q}\sqrt{|q_{f}|}\left(\sqrt{\ell}-\sqrt{p}\right),0}.$$
\par
Let us also notice that the general structure of $\Pi^{\mu\nu}$ in the presence of external magnetic fields is studied previously in a number of papers \cite{vacuum-old-1, vacuum-old-2,vacuum-new-1, vacuum-new-2, vacuum-new-3, vacuum-new-4}. In \cite{vacuum-old-1, vacuum-old-2}, it is in particular shown that $\Pi^{\mu\nu}$ can be given as a linear combination of fourteen bases. Using the Ward identity $q_{\mu}\Pi^{\mu\nu}(q)=0$, $\Pi^{\mu\nu}$ is then diagonalized and its eigenvalues and eigenfunctions are determined. In \cite{vacuum-new-1, vacuum-new-2, vacuum-new-3, vacuum-new-4},  $\Pi^{\mu\nu}$ is brought into the following form:
\begin{eqnarray}\label{x1}
\Pi^{\mu\nu}=-(\chi_{0}P_{0}^{\mu\nu}+\chi_{1}P_{1}^{\mu\nu}+
\chi_{2}P_{2}^{\mu\nu}),
\end{eqnarray}
with the projectors $P_{i}^{\mu\nu}, i=0,1,2$ defined by
\begin{eqnarray}\label{x2}
P_{0}^{\mu\nu}&=&q^{2}g^{\mu\nu}-q^{\mu}q^{\nu}, \nonumber\\
P_{1}^{\mu\nu}&=&q_{\|}^{2}g^{\mu_{\|}\nu_{\|}}-q^{\mu_{\|}}q^{\nu_{\|}},\nonumber\\
P_{2}^{\mu\nu}&=&q_{\perp}^{2}g^{\mu_{\perp}\nu_{\perp}}-q^{\mu_{\perp}}q^{\nu_{\perp}}.
\end{eqnarray}
The coefficients $\chi_{i}, i=1,2,3$ are then determined using Schwinger proper-time formalism \cite{schwinger1951} at $B\neq 0, T=0$. In \cite{alexandre2000}, the one-loop vacuum polarization tensor of hot QED for nonzero magnetic field and temperature is computed using the method of Schwinger proper-time. In the present paper, however, we have used the Ritus eigenfunction method to compute $\Pi^{\mu\nu}$ at $T\neq 0$ and $eB\neq 0$. The fact that $\Pi^{\mu\nu}$ presented in (\ref{D14}) can be given as a linear combination of $\Xi^{(q)\mu\nu}_{ij}$ from (\ref{D10}), and can therefore be separated into four groups of $(\mu,\nu)=(\|,\|)$, $(\mu,\nu)=(\|,\perp)$, $(\mu,\nu)=(\perp,\|)$ and $(\mu,\nu)=(\perp,\perp)$, strongly indicates that its general form can be brought into the form (\ref{x1}), presented in \cite{vacuum-new-1, vacuum-new-2, vacuum-new-3, vacuum-new-4}. On the other hand, it is possible to check the gauge invariance of our result. To show this, it is enough to prove the Ward identity, $q_{\mu}\Pi^{\mu\nu}=0$. Using $q_{\mu}=q_{\mu_{\|}}+q_{\mu_{\perp}}$, the product $q_{\mu}\Pi^{\mu\nu}$, can be first brought into the following form:
\begin{eqnarray}\label{x3}
q_{\mu}\Pi^{\mu\nu}(q)=q_{\mu_{\|}}\Pi^{\mu_{\|}\nu_{\|}}+q_{\mu_{\|}}
\Pi^{\mu_{\|}\nu_{\perp}}+q_{\mu_{\perp}}\Pi^{\mu_{\perp}\nu_{\|}}+q_{\mu_{\perp}}
\Pi^{\mu_{\perp}\nu_{\perp}},
\end{eqnarray}
where, according to our notations, the components $\Pi^{\mu_{\|}\nu_{\|}}, \Pi^{\mu_{\|}\nu_{\perp}}, \Pi^{\mu_{\perp}\nu_{\|}}$ and $\Pi^{\mu_{\perp}\nu_{\perp}}$ are defined by
\begin{eqnarray}\label{x4}
\begin{array}{rclcrcl}
\hspace{-0.8cm}\Pi^{\mu_{\|}\nu_{\|}}&\equiv&N_{c}\sum\limits_{q_{f}=\{u,d\}}\sum\limits_{p,\ell=0}^{\infty}\sum\limits_{j=1}^{2}{\cal{K}}_{1j}^{(q)}\tilde{\Xi}^{(q)\mu\nu}_{1j},
&\qquad&
\Pi^{\mu_{\|}\nu_{\perp}}&\equiv&N_{c}\sum\limits_{q_{f}=\{u,d\}}\sum\limits_{p,\ell=0}^{\infty}\sum\limits_{j=1}^{4}{\cal{K}}_{2j}^{(q)}\tilde{\Xi}^{(q)\mu\nu}_{2j},\\
\hspace{-0.8cm}\Pi^{\mu_{\perp}\nu_{\|}}&\equiv&N_{c}\sum\limits_{q_{f}=\{u,d\}}\sum\limits_{p,\ell=0}^{\infty}\sum\limits_{j=1}^{4}{\cal{K}}_{3j}^{(q)}\tilde{\Xi}^{(q)\mu\nu}_{3j},&\qquad&
\Pi^{\mu_{\perp}\nu_{\perp}}&\equiv&N_{c}\sum\limits_{q_{f}=\{u,d\}}\sum\limits_{p,\ell=0}^{\infty}\sum\limits_{j=1}^{4}{\cal{K}}_{4j}^{(q)}\tilde{\Xi}^{(q)\mu\nu}_{4j}.\\
\end{array}
\end{eqnarray}
We have performed the above computation with our ${\cal{K}}_{ij}^{(q)}$, from (\ref{D30})-(\ref{D31}) as well as (\ref{appA1})-(\ref{appA4}) and the bases $\tilde{\Xi}^{(q)\mu\nu}_{ij}$ from (\ref{D10}). We have shown that after an appropriate renormalization the Ward identity $q_{\mu}\Pi^{\mu\nu}=0$ is valid, and our result for $\Pi^{\mu\nu}$ is therefore gauge invariant.\footnote{We prefer to postpone the presentation of the details of this specific computation to our future work, because, in the computation of dilepton production rate it is enough to only focus on  $\Pi^{\mu_{\|}\nu_{\|}}$ and $\Pi^{\mu_{\perp}\nu_{\perp}}$ from (\ref{x4}), whose imaginary parts are to be determined to compute the dilepton production rate $\Delta_{B}$ from (\ref{A34}).}
\par
To determine the imaginary part of $[\Pi_{\mu\nu}]_{ij}^{nk}$, appearing in (\ref{A34}), let us first determine $\mbox{Im}[\tilde{\Xi}_{\mu\nu}^{(q)}]_{ij}^{p\ell}$ from (\ref{D15}) by making use of
\begin{eqnarray}\label{ِD18}
\lefteqn{\hspace{-2cm}\mbox{Im}\big[ T\sum\limits_{r}(i\omega_{r})^{s}\Delta_{f}(i\omega_{r},E_{\ell})\Delta_{f}(i(\omega_{q}-\omega_{r}), E_{p})\big]
=\pi
(e^{\beta q_{0}}-1)
}\nonumber\\
&&\times\int\frac{dk_{0}dk'_{0}}{(2\pi)^{2}}k_{0}^{s}N_{f}(k_{0})N_{f}(k'_{0})\rho_{f}(k_{0},E_{\ell})\rho_{f}(k'_{0},E_{p})\delta(q_{0}-k_{0}-k'_{0}),
\end{eqnarray}
where $N_{f}(E)\equiv (e^{\beta E}+1)^{-1}$ and $\rho_{f}(k_{0}, E_{k})$ is the free on the mass-shell fermionic spectral density function \cite{lebellac}.
Following this standard method, we obtain
\begin{eqnarray}\label{D19}
\mbox{Im}[\tilde{\Xi}_{\mu\nu}^{(q)}]_{ij}^{p\ell}=-\frac{\alpha q_{f}^{2}}{2\xi_{q}}\left(e^{\beta q_{0}}-1\right)
\bigg\{[\Xi_{\mu\nu}^{(q)}(E_{\ell}^{+},\ell_{3}^{+})]_{ij}^{p\ell}{\cal{N}}_{+}^{(q)}+[\Xi_{\mu\nu}^{(q)}(E_{\ell}^{-},\ell_{3}^{-})]_{ij}^{p\ell}{\cal{N}}_{-}^{(q)}
\bigg\},
\end{eqnarray}
where $\Xi_{\mu\nu}^{(q)}$ are given in (\ref{D10}), and for $i=\pm$
\begin{eqnarray}\label{D20}
{\cal{N}}_{i}^{(q)}&\equiv&\frac{\xi_{q}}{2}\bigg[\frac{N_{f}(E_{\ell}^{i})N_{f}(E_{p}^{i})}{|-\ell_{3}^{i}q_{0}+E_{\ell}^{i}q_{3}|}-\frac{N_{f}(E_{\ell'}^{i})N_{f}(E_{p'}^{i})}{|\ell_{3}^{i}q_{0}+E_{\ell}^{i}q_{3}|}\bigg].
\end{eqnarray}
Here, $\ell_{3}^{\pm}\equiv\frac{\xi_{q}\left(q_{3}\pm q_{0}{\cal{R}}_{q}\right)}{2{\mathbf{q}}_{\|}^{2}}$,
\begin{eqnarray}\label{D21}
E_{\ell}^{\pm}&\equiv&\frac{\xi_{q}\left(q_{0}\pm q_{3}{\cal{R}}_{q}\right)}{2{\mathbf{q}}_{\|}^{2}},\qquad
E_{p}^{\pm}\equiv\frac{q_{0}\xi_{q}'\mp q_{3}\xi_{q}{\cal{R}}_{q}}{2{\mathbf{q}}_{\|}^{2}},\qquad
E_{\ell'}^{\pm}\equiv -\frac{\xi_{q}\left(q_{0}\pm q_{3}{\cal{R}}_{q}\right)}{2{\mathbf{q}}_{\|}^{2}},\nonumber\\
E_{p'}^{\pm}&\equiv&\frac{q_{0}\xi_{q}''\pm q_{3}\xi_{q}{\cal{R}}_{q}}{2\qpar},
\end{eqnarray}
and
\begin{eqnarray}\label{D22}
\xi_{q}\equiv {\mathbf{q}}_{\|}^{2}+(M_{\ell}^{2}-M_{p}^{2}),\qquad
\xi_{q}'\equiv {\mathbf{q}}_{\|}^{2}-(M_{\ell}^{2}-M_{p}^{2}),\qquad\mbox{and}\qquad
\xi_{q}''\equiv 3\qpar+(M_{\ell}^{2}-M_{p}^{2}),
\end{eqnarray}
as well as
\begin{eqnarray}\label{D23}
{\cal{R}}_{q}\equiv \left(1-\frac{4M_{\ell}^{2}{\mathbf{q}}_{\|}^{2}}
{\xi_{q}^{2}}\right)^{1/2},
\end{eqnarray}
with $M_{\ell}^{2}=m_{q}^{2}+2\ell |q_{f}eB|$, and $m_{q}$ the bare mass of quarks with charge $q_{f}$. Here, $\ell$ labels the Landau levels. Let us notice that for $eB=0$, the  second term appearing in (\ref{D20}) does not arise. Its appearance is mainly because of the aforementioned dimensional reduction from $D=4$ to $D=2$ dimensions in the presence of external magnetic fields. In other words, for $eB=0$, apart from $\qpar=q_{0}^{2}-q_{3}^{2}$, the transverse components of $q^{\mu}$, i.e. $\mathbf{q}_{\perp}^{2}=q_{1}^{2}+q_{2}^{2}$ also appear in $E_{\ell}^{\pm}$ and $E_{p}^{\pm}$ as well as in $E_{\ell'}^{\pm}$ and $E_{p'}^{\pm}$, and this makes the condition $q_{0}=E_{p}-E_{\ell}$, from which the additional second term in (\ref{D20}) arises, invalid.\footnote{Let us notice that whereas $E_{p}^{\pm}=q_{0}-E_{\ell}^{\pm}$, $E_{p'}^{\pm}=q_{0}+E_{\ell}^{\pm}$.}
\par
Plugging at this stage (\ref{D19}) into (\ref{D14}), and using (\ref{D17}) to connect $\mbox{Im}[\Pi_{\mu\nu}]_{ij}^{p\ell}$ with $\mbox{Im}[\Pi_{\mu\nu}]_{ij}^{nk}$, we arrive at $\mbox{Im}[\Pi_{\mu\nu}]_{ij}^{nk}$
with $(\mu,\nu)=(\|,\|)$, i.e. with $(i=1, j=1,2)$, and $(\mu,\nu)=(\perp,\perp)$, i.e. with $(i=4, j=1,\cdots, 4)$.\footnote{
Because of the special structure of  $\Xi_{\mu\nu}^{(q)ij}$ from (\ref{D10}), the contributions from $(\mu,\nu)=(\perp,\|)$, i.e. $(i=2,j=1,\cdots,4)$, and $(\mu,\nu)=(\|,\perp)$, i.e. $(i=3,j=1,\cdots,4)$, once multiplied with the corresponding leptonic bases $\Psi_{ij, nk}^{\mu\nu}$ from (\ref{A24}), vanish. This multiplication is to be performed according to (\ref{A34}).} To keep the presentation in this section as short as possible, we present the final results for $\mbox{Im}[\Pi_{\mu\nu}]_{ij}^{nk}$ in appendix \ref{app2b}. In appendix \ref{appB}, we have presented $\mbox{Im}[\Pi_{\mu\nu}]$ arising from LLL.
\subsection{Leptonic part of ${\Delta_{B}}$}\label{sec4B}
\par\noindent
In what follows, we will present the results for the leptonic part of $\Delta_{B}$, ${\cal{L}}_{ij,nk}^{\mu\nu}$ from (\ref{A35}). After performing the integrations over $p_{1}^{(3)}$ and $p_{2}^{(3)}$, we arrive first at
\begin{eqnarray}\label{D28}
{\cal{L}}_{ij,nk}^{\mu\nu}={\cal{C}}_{ij,nk}^{\mu\nu}+
{\cal{D}}_{ij,nk}^{\mu\nu},
\end{eqnarray}
with
\begin{eqnarray}\label{D29}
{\cal{C}}_{ij,nk}^{\mu\nu}\equiv\frac{\Psi_{ij,nk}^{\mu\nu}(p_{1}^{(0)+}, p_{2}^{(0)+}, p_{2}^{(3)+})}{\big|-p_{1}^{(3)+} q_{0}+q_{3}E_{k}^{+}\big|},\qquad\mbox{and}\qquad
{\cal{D}}_{ij,nk}^{\mu\nu}\equiv\frac{\Psi_{ij,nk}^{\mu\nu}(p_{1}^{(0)-}, p_{2}^{(0)-}, p_{2}^{(3)-})}{\big|-p_{1}^{(3)-} q_{0}+q_{3}E_{k}^{-}\big|}.
\end{eqnarray}
Here, $p_{1}^{(0)}=E_{k}^{\pm}$, $p_{2}^{(0)\pm}=q_{0}-E_{k}^{\pm}$,  and $p_{1}^{(3)\pm}=\frac{\eta\left(q^{3}\pm q^{0}{\cal{Q}}\right)}{2{\mathbf{q}}_{\|}^{2}}$. Moreover, we have
\begin{eqnarray}\label{D30}
E_{k}^{\pm}=\frac{\eta\left(q^{0}\pm q^{3}{\cal{Q}}\right)}{2{\mathbf{q}}_{\|}^{2}},\qquad\mbox{and}\qquad
E_{n}^{\pm}=\frac{\left(q^{0}\eta'\mp q^{3}\eta{\cal{Q}}\right)}{2{\mathbf{q}}_{\|}^{2}},
\end{eqnarray}
where following definitions are used:
\begin{eqnarray}\label{D31}
\eta\equiv {\mathbf{q}}_{\|}^{2}+(M_{k}^{2}-M_{n}^{2}), \qquad\mbox{and}\qquad
\eta'\equiv {\mathbf{q}}_{\|}^{2}-(M_{k}^{2}-M_{n}^{2}),
\end{eqnarray}
and
\begin{eqnarray}\label{D32}
{\cal{Q}}\equiv \left(1-\frac{4M_{k}^{2}{\mathbf{q}}_{\|}^{2}}{\eta^{2}}\right)^{1/2}.
\end{eqnarray}
Here, $M_{k}^{2}=m_{\ell}^{2}+2k |eB|$ for leptons with bare masses $m_{\ell}$ and charges $q_{f}=\pm 1$.
\par
Plugging now $\Psi_{ij,nk}^{\mu\nu}$ from (\ref{A24}) into (\ref{D29}), ${\cal{L}}_{ij,nk}^{\mu\nu}$ can explicitly be determined as matrices in the $\mu\nu$-space. According to (\ref{A34}), the resulting expressions are to be multiplied with $\mbox{Im}[\Pi_{\mu\nu}]_{ij}^{nk}$.  As it turns out, because of the special structure of $\Xi_{\mu\nu}^{(q)ij}$ from (\ref{D10}), the contributions from $(\mu,\nu)=(\perp,\|)$, i.e. $(i=2,j=1,\cdots,4)$, and $(\mu,\nu)=(\|,\perp)$, i.e. $(i=3,j=1,\cdots,4)$, once multiplied with the corresponding leptonic bases $\Psi_{ij, nk}^{\mu\nu}$ from (\ref{A24}), vanish. We will therefore only present the results for $(\mu,\nu)=(\|,\|)$ and $(\mu,\nu)=(\perp,\perp)$ [see appendix \ref{app2c}]. In appendix \ref{appB}, we have presented the result for ${\cal{L}}^{\mu\nu}$ arising from LLL.

\subsection{Final analytical result for ${\Delta_{B}}$}\label{sec4C}
\par\noindent
According to (\ref{A34}), we have to multiply  $\mbox{Im}[\Pi_{\mu\nu}]_{ij}^{nk}$ from (\ref{D24}) and (\ref{D27}) with ${\cal{L}}_{ij,nk}^{\mu\nu}$ from (\ref{D33}) and (\ref{D35}). As aforementioned, only the components $(i=1, j=1,2)$ and $(i=4, j=1,\cdots,4)$ contribute to this product. The final result for the differential multiplicity of dileptons in the presence of a constant magnetic field can then be separated into two parts
\begin{eqnarray}\label{D36}
\Delta_{B}=\Delta_{B}^{\|}+\Delta_{B}^{\perp},
\end{eqnarray}
with
\begin{eqnarray}\label{D37}
\Delta_{B}^{\|}\equiv\sum\limits_{n,k=0}^{\infty}\sum\limits_{j=1}^{2}
\Delta^{1j,nk}_{B},\qquad\mbox{and}\qquad\Delta_{B}^{\perp}\equiv\sum\limits_{n,k=0}^{\infty}
\sum\limits_{j=1}^{4}\Delta^{4j,nk}_{B},
\end{eqnarray}
arising from product of $(\mu,\nu)=(\|,\|)$ and $(\mu,\nu)=(\perp,\perp)$ contributions, as previously described. In (\ref{D37}), $\Delta_{B}^{ij,nk}$ are given by
\begin{eqnarray*}
\Delta^{11,nk}_{B}&=&-\frac{N_{c}eB\alpha^{2}}{2\eta{\cal{Q}}(\qpar)^{4}(q^{2})^{2}}\sum_{q_{f}=\{u,d\}}q_{f}^{2}\sum_{p,\ell=0}^{\infty}\frac{\eta_{nkp\ell}^{(q)}{\cal{K}}_{11}^{(q)}}{\xi_{q}}\mbox{tr}(XY^{T}),\nonumber\\
\Delta^{12,nk}_{B}&=&-\frac{8N_{c}eB\alpha^{2}\tilde{p}_{1}^{(2)}\tilde{p}_{2}^{(2)}}{\eta{\cal{Q}}(q^{2})^{2}}\sum_{q_{f}=\{u,d\}}q_{f}^{2}\sum_{p,\ell=0}^{\infty}\frac{\eta_{nkp\ell}^{(q)}{\cal{K}}_{12}^{(q)}\tilde{p}_{q}^{(2)}\tilde{\ell}_{q}^{(2)}}{\xi_{q}}{\cal{N}}^{(q)},\nonumber\\
\end{eqnarray*}
\begin{eqnarray}\label{D38}
\Delta^{41,nk}_{B}&=&-\frac{4N_{c}eB\alpha^{2}\tilde{p}_{1}^{(2)}\tilde{p}_{2}^{(2)}}{\eta{\cal{Q}}(q^{2})^{2}}\sum_{q_{f}=\{u,d\}}q_{f}^{2}\sum_{p,\ell=0}^{\infty}\frac{\eta_{nkp\ell}^{(q)}{\cal{K}}_{41}^{(q)}
\tilde{p}_{q}^{(2)}\tilde{\ell}_{q}^{(2)}(2-2s_{q})
}{\xi_{q}}{\cal{N}}^{(q)},\nonumber\\
\Delta^{42,nk}_{B}&=&-\frac{4N_{c}eB\alpha^{2}\tilde{p}_{1}^{(2)}\tilde{p}_{2}^{(2)}}{\eta{\cal{Q}}(q^{2})^{2}}\sum_{q_{f}=\{u,d\}}q_{f}^{2}\sum_{p,\ell=0}^{\infty}\frac{\eta_{nkp\ell}^{(q)}{\cal{K}}_{42}^{(q)}
\tilde{p}_{q}^{(2)}\tilde{\ell}_{q}^{(2)}(2-2s_{q})
}{\xi_{q}}{\cal{N}}^{(q)},\nonumber\\
\Delta^{43,nk}_{B}&=&+\frac{N_{c}eB\alpha^{2}(\eta-4keB)}{\eta{\cal{Q}}(q^{2})^{2}}\sum_{q_{f}=\{u,d\}}q_{f}^{2}\sum_{p,\ell=0}^{\infty}\frac{\eta_{nkp\ell}^{(q)}{\cal{K}}_{43}^{(q)}
(\xi_{q}-4\ell|q_{f}eB|)(2-2s_{q})
}{\xi_{q}} {\cal{N}}^{(q)},\nonumber\\
\Delta^{44,nk}_{B}&=&+\frac{N_{c}eB\alpha^{2}(\eta-4keB)}{\eta{\cal{Q}}(q^{2})^{2}}\sum_{q_{f}=\{u,d\}}q_{f}^{2}\sum_{p,\ell=0}^{\infty}\frac{\eta_{nkp\ell}^{(q)}{\cal{K}}_{44}^{(q)}
(\xi_{q}-4\ell|q_{f}eB|)(2-2s_{q})
}{\xi_{q}}{\cal{N}}^{(q)},
\end{eqnarray}
where $X$- and $Y$-matrices are given in (\ref{D25}) and (\ref{D34}), respectively. Moreover, ${\cal{N}}^{(q)}={\cal{N}}^{(q)}_{+}+{\cal{N}}^{(q)}_{-}$, where ${\cal{N}}_{\pm}^{(q)}$ are defined in (\ref{D26}). The coefficients ${\cal{K}}^{(q)}_{1j}, j=1,2$ and ${\cal{K}}^{(q)}_{4j}, j=1,\cdots,4$ for up (positively charged) and down (negatively charged) quarks are given by
\begin{eqnarray}\label{D39}
{\cal{K}}_{11}^{(+)}&=&+\frac{2}{\ell_{B+}^{2}}\frac{e^{-\kappa_{+}}}{p!\ell!}\bigg[\kappa_{+}^{M_{1}-m_{1}}[U_{M_{1}-m_{1}+1}^{-m_{1}}(\kappa_{+})]^{2}+p\ell\Pi_{p}\Pi_{\ell}\kappa_{+}^{M_{2}-m_{2}}[U_{M_{2}-m_{2}+1}^{-m_{2}}(\kappa_{+})]^{2}\bigg],\nonumber\\
{\cal{K}}_{12}^{(+)}&=&-\frac{4\Pi_{p}\Pi_{\ell}}{\ell_{B+}^{2}}\frac{\sqrt{p\ell}}{p!\ell!}\kappa_{+}^{M_{1}-m_{1}}e^{-\kappa_{+}}[U_{M_{1}-m_{1}+1}^{-m_{1}}(\kappa_{+})][U_{M_{1}-m_{1}+1}^{-m_{1}+1}(\kappa_{+})],\nonumber\\
{\cal{K}}_{41}^{(+)}&=&-\frac{\Pi_{p}\Pi_{\ell}\sqrt{p\ell}}{p!\ell!}\kappa_{+}^{p+\ell-2-m_{3}-m_{4}}e^{-\kappa_{+}}(q_{2}+iq_{1})^{2}[U_{M_{3}-m_{3}+1}^{-m_{3}}(\kappa_{+})][U_{M_{4}-m_{4}+1}^{-m_{4}}(\kappa_{+})],\nonumber\\
{\cal{K}}_{42}^{(+)}&=&-\frac{\Pi_{p}\Pi_{\ell}\sqrt{p\ell}}{p!\ell!}\kappa_{+}^{p+\ell-2-m_{3}-m_{4}}e^{-\kappa_{+}}(q_{2}-iq_{1})^{2}[U_{M_{3}-m_{3}+1}^{-m_{3}}(\kappa_{+})][U_{M_{4}-m_{4}+1}^{-m_{4}}(\kappa_{+})],\nonumber\\
{\cal{K}}_{43}^{(+)}&=&+\frac{2\Pi_{p}}{\ell_{B+}^{2}}\frac{p}{p!\ell!}\kappa_{+}^{M_{4}-m_{4}}e^{-\kappa_{+}}[U_{M_{4}-m_{4}+1}^{-m_{4}}(\kappa_{+})]^{2},\nonumber\\
{\cal{K}}_{44}^{(+)}&=&+\frac{2\Pi_{\ell}}{\ell_{B+}^{2}}\frac{\ell}{p!\ell!}\kappa_{+}^{M_{3}-m_{3}}e^{-\kappa_{+}}[U_{M_{3}-m_{3}+1}^{-m_{3}}(\kappa_{+})]^{2},
\end{eqnarray}
for positive charges, and
\begin{eqnarray}\label{D40}
{\cal{K}}_{11}^{(-)}&=&+\frac{2}{\ell_{B-}^{2}}\frac{e^{-\kappa_{-}}}{p!\ell!}
\bigg[\Pi_{p}\Pi_{\ell}
\kappa_{-}^{M_{1}-m_{1}}[U_{M_{1}-m_{1}+1}^{-m_{1}}(\kappa_{-})]^{2}+p\ell
\kappa_{-}^{M_{2}-m_{2}}[U_{M_{2}-m_{2}+1}^{-m_{2}}(\kappa_{-})]^{2}
\bigg],\nonumber\\
{\cal{K}}_{12}^{(-)}&=&-\frac{4\Pi_{p}\Pi_{\ell}}{\ell_{B-}^{2}}\frac{\sqrt{p\ell}}{p!\ell!}\kappa_{-}^{M_{1}-m_{1}}e^{-\kappa_{-}}[U_{M_{1}-m_{1}+1}^{-m_{1}}(\kappa_{-})][U_{M_{1}-m_{1}+1}^{-m_{1}+1}(\kappa_{-})],\nonumber\\
{\cal{K}}_{41}^{(-)}&=&-\frac{\Pi_{p}\Pi_{\ell}\sqrt{p\ell}}{p!\ell!}\kappa_{-}^{p+\ell-2-m_{3}-m_{4}}e^{-\kappa_{-}}(q_{2}+iq_{1})^{2}[U_{M_{3}-m_{3}+1}^{-m_{3}}(\kappa_{-})][U_{M_{4}-m_{4}+1}^{-m_{4}}(\kappa_{-})],\nonumber\\
{\cal{K}}_{42}^{(-)}&=&-\frac{\Pi_{p}\Pi_{\ell}\sqrt{p\ell}}{p!\ell!}\kappa_{-}^{p+\ell-2-m_{3}-m_{4}}e^{-\kappa_{-}}(q_{2}-iq_{1})^{2}[U_{M_{3}-m_{3}+1}^{-m_{3}}(\kappa_{-})][U_{M_{4}-m_{4}+1}^{-m_{4}}(\kappa_{-})],\nonumber\\
{\cal{K}}_{43}^{(-)}&=&+\frac{2\Pi_{p}}{\ell_{B-}^{2}}\frac{\ell}{p!\ell!}\kappa_{-}^{M_{3}-m_{3}}e^{-\kappa_{-}}[U_{M_{3}-m_{3}+1}^{-m_{3}}(\kappa_{-})]^{2},\nonumber\\
{\cal{K}}_{44}^{(-)}&=&+\frac{2\Pi_{\ell}}{\ell_{B-}^{2}}\frac{p}{p!\ell!}\kappa_{-}^{M_{4}-m_{4}}e^{-\kappa_{-}}[U_{M_{4}-m_{4}+1}^{-m_{4}}(\kappa_{-})]^{2},
\end{eqnarray}
for negative charges. In the above expressions, $\kappa_{q}\equiv \frac{\ell_{B_{q}}^{2}{\mathbf{q}}_{\perp}^{2}}{2}$ with $\ell_{B_{q}}=|q_{f}eB|^{-1/2}$ and ${\mathbf{q}}_{\perp}^{2}=q_{1}^{2}+q_{2}^{2}$. Moreover, we have
\begin{eqnarray}\label{D41}
\begin{array}{rclcrcl}
m_{1}&\equiv&\mbox{min}(p,\ell),&\qquad&M_{1}&\equiv&\mbox{max}(p,\ell),\\
m_{2}&\equiv&\mbox{min}(p-1,\ell-1),&\qquad&M_{2}&\equiv&\mbox{max}(p-1,\ell-1),\\
m_{3}&\equiv&\mbox{min}(p,\ell-1),&\qquad&M_{3}&\equiv&\mbox{max}(p,\ell-1),\\
m_{4}&\equiv&\mbox{min}(p-1,\ell),&\qquad&M_{4}&\equiv&\mbox{max}(p-1,\ell).
\end{array}
\end{eqnarray}
In appendix \ref{appB}, the analytical expression for $\Delta_{B}$ arising from LLL is presented.

Let us notice that the above results for $\Delta_{B}$ are invariant under $\mathbf{B}\to -\mathbf{B}$. This is because electromagnetic processes
are invariant under the operation of charge conjugation operator. Our analytical results are
thus (theoretically) invariant under the inversion of $B$, and should be suitable for further
phenomenological studies once the direction of the magnetic field is fixed.\footnote{In general, $eB$ appearing before the summations over $q_{f}$  in (\ref{D38}) are to be replaced by $|eB|$. See also
footnote 7.} However, in order to study the implication of our results for more realistic scenarios of HICs, where the magnitude and direction of the background magnetic field 
fluctuate from event to event, $B$ in the above relations should probably be replaced by its average over several events \cite{skokov2012, skokov2009}.
Moreover, it is worth to remind that, as in the case of a zero magnetic field, in the case of
a nonvanishing magnetic field, the general expression (III.33) for $\Delta_{B}$ is derived by starting with the thermally averaged multiplicity $N_{B}$ from 
(\ref{A15}), which is defined in the local rest frame of QGP. Hence, in order to bring the corresponding results of $\Delta_{B}$ from (\ref{D36})-(\ref{D41})
into connection with the experimental results from HICs, where QGP is not at rest, it is necessary to replace $q_0$ in the above results for the differential multiplicity 
$\Delta_{B}=\frac{dN_{B}}{d^{4}x d^{4}q}$ by $q\cdot u$, where $u_{\mu}$
is the four-velocity of the plasma. Then, after choosing an appropriate parametrization \cite{hatsuda2005},
and assuming an appropriate proper-time dependence of $T$ and $eB$, it is possible to use $\Delta_{B}$ to
determine, e.g., the flow coefficients $v_{n}$ of QGP from $\Delta_{B}$ as functions of the energy of virtual photons. We will postpone this rather involved computation to our future publications, and will only numerically evaluate the above results for $\Delta_{B}$ in the next section. We will, in particular, focus on the effects of $T$ and $B$ on $\Delta_{B}, \Delta_{B}^{\|}$ and $\Delta_{B}^{\perp}$.
\section{Dilepton production rate in a hot and magnetized quark-gluon plasma: Numerical  results}\label{sec5}
\setcounter{equation}{0}
\subsection{General considerations}\label{sec5a}
\par\noindent
In this section, we will use the analytical results presented in previous sections to study the effect of constant magnetic fields on the production rates of dileptons in a hot and magnetized QCD plasma. We are mainly interested in the effect of magnetic fields and temperatures on the ratio $\Delta_{B}/\Delta_{0}$ and a certain anisotropy factor $\nu_{B}$ between $\Delta_{B}^{\|}$ and $\Delta_{B}^{\perp}$, appearing in (\ref{D37}). Here, $\Delta_{0}$ and $\Delta_{B}$ are dilepton multiplicities for vanishing and nonvanishing magnetic fields from \eqref{A14} and \eqref{D36}-\eqref{D41}. Moreover, the ratio $\Delta_{B}/\Delta_{0}$ is defined by
\begin{eqnarray}\label{O1}
\frac{\Delta_{B}}{\Delta_{0}}\equiv \bigg[\frac{\Delta_{B}}{\Delta_{0}}\bigg]_{u}+\bigg[\frac{\Delta_{B}}{\Delta_{0}}\bigg]_{d},
\end{eqnarray}
where $[\Delta_{B}/\Delta_{0}]_{u}$ and $[\Delta_{B}/\Delta_{0}]_{d}$ are the contributions of up ($u$) and down ($d$) quarks to $\Delta_{B}/\Delta_{0}$, respectively.\footnote{Later, we will replace the quark masses appearing in $\Delta_{0}$ from (\ref{A14}) and in $\Delta_{B}$ from (\ref{D36})-(\ref{D40}) with thermal masses given in (\ref{O5}). Because of the factor $eq_{f}$ appearing on the r.h.s. of (\ref{O5}), the thermal contribution to quark mass breaks the isospin symmetry. It is therefore necessary to consider the contributions of up and down quarks to the ratio $\Delta_{B}/\Delta_{0}$ separately, as it is done in (\ref{O1}).} This ratio and $\nu_{B}$ will be plotted as functions of $q_{0}/T$ for $q_{3}=0$,  and fixed but small ${\mathbf{q}}_{\perp}=(q_{1},q_{2})$ for $eB=0.02, 0.03$ GeV$^{2}$ and $eB=0.2, 0.3$ GeV$^{2}$ at $T=200$ and $T=400$ MeV.
Our specific choices for $q_{i}/T, i=1,2,3$ are $(q_{1},q_{2},q_{3})=(0.1 T,0.1 T,0)$. Our findings for $eB=0.02, 0.03$ GeV$^{2}$ and $eB=0.2, 0.3$ GeV$^{2}$ may be relevant for the physics of heavy ion collisions, because, as it is pointed out in section \ref{sec1},  these magnetic field strengths are in the same order of magnitude of magnetic fields created in early stage of noncentral HICs at RHIC ($eB\sim 1.5 m_{\pi}^{2}$) and LHC ($eB\sim 15 m_{\pi}^{2}$)\footnote{$m_{\pi}\sim 140$ MeV.} \cite{skokov2009, kharzeev2007}. Moreover, whereas the results for $T=200$ MeV are relevant for a temperature near the QCD phase transition point ($\sim 150-170$ MeV), $T=400$ MeV is high enough to correspond to a temperature at the beginning of the formation of QGP, where the thermodynamic equilibrium is assumed to be built after the collision. Our perturbative approach is assumed to be relevant for both these temperatures. Before presenting our numerical results, a couple of remarks on the specific features of our method are in order:

\par
\textit{i) The choice for the upper limit in the summation over Landau levels:} This specific summation appears in our analytical results for $\Delta_{B}$ from (\ref{D37})  and (\ref{D38}). As we have discussed before, the corresponding relations for $\Delta_{B}$ include a certain factor $\eta_{nkp\ell}^{(q)}$ defined by $\eta_{nkp\ell}^{(q)}= \delta_{\sqrt{n}-\sqrt{k}+s_{q}\sqrt{|q_{f}|}\left(\sqrt{\ell}-\sqrt{p}\right),0}$ [see also (\ref{D16})]. This factor connects the external Landau levels, $(n,k)$, with the internal (loop) Landau levels, $(p,\ell)$. Having in mind that $n,k,p,\ell$ are non-negative integers, only a limited number of them satisfies the constraint (\ref{D16}). For $n,k,p,\ell\leq 10$, they are given by
\begin{eqnarray}\label{O2}
n=k=0&\longrightarrow&p=\ell=0,\cdots, 10,\nonumber\\
n=k=1&\longrightarrow&p=\ell=0,\cdots, 10,\nonumber\\
\cdots&\longrightarrow&\cdots,\nonumber\\
n=k=10&\longrightarrow&p=\ell=0,\cdots, 10,
\end{eqnarray}
for both positive and negative charges, and
\begin{eqnarray}\label{O3}
n= 2,k = 0,p = 3,\ell = 0,\qquad
n= 4,k = 0,p = 6,\ell = 0,\qquad
n= 4,k = 2,p = 6,\ell = 3,
\end{eqnarray}
for positive charges, as well as
\begin{eqnarray}\label{O4}
\begin{array}{cccccccccccccc}
\hspace{-0.5cm}n= 1,& k = 0,& p = 0,&\ell = 3,&\qquad&n= 1,& k = 2,& p = 6,&\ell = 3,&&&&&\\
\hspace{-0.5cm}n= 2,& k = 0,& p = 0,&\ell = 6,&\qquad&n= 2,& k = 1,& p = 3,&\ell = 6,&\qquad&
n =2,& k = 3,& p = 9,&\ell = 6,\\
\hspace{-0.5cm}n =3,& k = 0,& p = 0,&\ell = 9,&\qquad&n =3,& k = 1,& p = 3,&\ell = 9,&\qquad&
n =3,& k = 2,& p = 6,&\ell = 9,\\
\hspace{-0.5cm}n= 4,& k = 1,& p = 0,&\ell = 3,&&&&&&&&&&
\end{array}
\end{eqnarray}
for negative charges. In what follows, we will only consider the sets with $n=k$ and $p=\ell$ presented in (\ref{O2}). The sets given in (\ref{O3}) and (\ref{O4}) will be ignored, because they do not obey $n=k$ and $p=\ell$. Let us notice that, $p=\ell$ is required by gauge invariance, and, according to (\ref{O2}), this fixes $n$ to be equal to $k$. To describe why we have worked only with Landau levels up to $n,k,p,\ell\leq 10$, let us remind that the upper limit of the summation over Landau levels is indeed related to $\lfloor \frac{\Lambda^2}{eB}\rfloor$,\footnote{The floor function $\lfloor x\rfloor$, also called the greatest integer function or integer value, gives the largest integer less than or equal to $x$.} where $\Lambda$ is a characteristic energy scale of the theory, e.g. the energy cutoff in a Nambu--Jona-Lasinio model \cite{fayazbakhsh2013, fayazbakhsh2012, fayazbakhsh2011, fayazbakhsh2010}. For hot QCD, because of the lack of a natural cutoff, the temperature $T$ seems to be an appropriate energy scale. In the present work, we have used $\lfloor \frac{T^2}{eB}\rfloor$ to determine the upper limit of the summation over Landau levels (see also \cite{fukushima2015, warringa2009} for similar arguments). Hence, for our specific choice of magnetic field strengths $eB=0.2, 0.3$ GeV$^{2}$ (moderate magnetic fields) and temperatures $T=200$ MeV and $400$ MeV, it is probably enough to consider only the LLL for both external and internal Landau levels, $n,k$ and $p,\ell$. But, according to the above argument, for weak magnetic fields $eB=0.02, 0.03$ GeV$^{2}$, we have to consider higher Landau levels up to 8. We will, however, work with $n,k,p,\ell\leq 10$ to guarantee the reliability of our qualitative conclusions. Later, we will compare the results for $n,k,p,\ell\leq 10$ with the results corresponding to $n,k,p,\ell\leq 5$ and $n,k,p,\ell\leq 8$, and will discuss the impact of increasing the upper limit of the summation over Landau levels on some specific quantities related to the production rates of dileptons.
\par
\textit{ii) Explicit dependence on quark and lepton masses, $m_{q}$ and $m_{\ell}$:} In the previous section, we assumed that the hot quark-gluon plasma created after the collision consists of up and down quarks. At high enough temperatures, the small bare masses of these quarks are indeed negligible.  Instead, they receive significant thermal corrections, given by
\begin{eqnarray}\label{O5}
{\cal{M}}_{q}^{2}=m_{q}^{2}+\frac{e^{2}q_{f}^{2}T^{2}}{8}.
\end{eqnarray}
The second term arises from the standard Hard Thermal Loop (HTL) approximation arising from the QED coupling of quarks with photons \cite{lebellac}.\footnote{
Let us notice that magnetic fields can principally correct quarks and leptons bare masses too. As we have argued in section \ref{sec1}, the presence of hot and/or magnetized plasminos \cite{taghinavaz2015} can also affect ${\cal{M}}_{q}$ from (\ref{O5}). These kinds of corrections are not considered in the present paper to avoid additional complications. Apart from this correction, in principal, the thermal mass of quarks receives contributions from QCD coupling. Although, comparing with QED, the QCD mass correction of quarks is larger, but, as it turns out, considering these kinds of corrections has no significant impact on the ratio $\Delta_{B}/\Delta_{0}$ demonstrated in the present section, where QED mass correction to quarks is solely considered.}
In \eqref{O5}, the factor $eq_{f}$ is the QED coupling of quarks with photons. Because of the explicit dependence of ${\cal{M}}_{q}$ on $q_{f}$, with $q_{u}=2/3$ and $q_{d}=-1/3$, the assumed isospin symmetry $m_{u}=m_{d}$ is broken by these thermal corrections. In what follows, we have worked with $m_{q}=3$ MeV for both quark flavors, $e^{2}=4\pi\alpha$ with QED fine structure constant $\alpha=1/137$. An appropriate thermal mass correction is also considered for leptons. Their HTL corrected thermal masses, arising from the QED coupling of leptons with photons,  are similarly given by\footnote{Let us notice that second terms appearing on the r.h.s. of (\ref{O5}) and (\ref{O6}) are QED thermal (Debye) mass corrections of quarks and leptons for small (zero) momenta. At large momenta and high enough temperatures, where the bare masses of quarks and leptons, $m_{q}$ and $m_{\ell}$, can be neglected, ${\cal{M}}_{q}^{2}$ and ${\cal{M}}_{\ell}^{2}$ from (\ref{O5}) and (\ref{O6}) are to be replaced by ${\cal{M}}_{q}^{2}\simeq 2q_{f}^{2}m_{D}^{2}$ and ${\cal{M}}_{\ell}^{2}\simeq 2m_{D}^{2}$, where $m_{D}^{2}\equiv e^{2}T^{2}/8$ is the Debye mass of fermions. It is worth to notice also that the additional factor two has no significant impact on the numerical results demonstrated in the present section. It arises because the dispersion relation of fermions in the large momentum limit $|\mathbf{k}|\gg m_{D}$ is given by $k_{0}\simeq |\mathbf{k}|+\frac{m_{D}^{2}}{|\mathbf{k}|}$ \cite{lebellac}.}
\begin{eqnarray}\label{O6}
{\cal{M}}_{\ell}^{2}=m_{\ell}^{2}+\frac{e^{2}T^{2}}{8}.
\end{eqnarray}
To compare $\Delta_{B}/\Delta_{0}$ for electron-positron and muon-antimuon pairs, we will use the electron and muon (bare) masses $m_{e}=0.5$ MeV and $m_{\mu}= 105$ MeV.
\par
\textit{iii) Threshold energy of photons $[q_{0}/T]_{th}$:} Similar to the case of vanishing magnetic fields, where the appearance of $R_{0}$ and $Q_{0}$ in (\ref{A14}), leads to a certain ``minimum'' energy threshold $q^{2}= 4\widetilde{m}^{2}$ with $\widetilde{m}\equiv \mbox{max}(m_{q},m_{\ell})$ for a dilepton pair to be produced,\footnote{
It is obvious that before inserting the $\Theta$-function into the r.h.s. of (\ref{A14}), the multiplicity $\Delta_{0}$ is imaginary for all $q^{2}<4\tilde{m}^{2}$.} in the case of nonvanishing magnetic fields, $\Delta_{B}$ from (\ref{D36})-(\ref{D41}), exhibits also a certain ``minimum'' energy threshold which is necessary for the photons to be converted into a dilepton pair. As it turns out, this minimum production threshold for $B\neq 0$ case is determined by the LLL, and, is independent of $eB$ and $T$.
In this section, however, we will define another threshold energy for virtual photons, which seems to be more appropriate for comparison of $\Delta_{B}$ with $\Delta_{0}$.\footnote{In what follows, the word ``threshold'' is used in the most general sense, and is not to be confused with the aforementioned ``minimum energy threshold for dilepton production''. In general, a threshold is  the magnitude or intensity that must be exceeded  for a certain reaction, phenomenon, result or condition to occur or to be manifested. See the main text for the condition which defines the threshold energy $[q_{0}]_{th}$ in the present work.} To do this, let us choose, as aforementioned, $(q_{1},q_{2},q_{3})=(0.1T,0.1T, 0)$, and consider $\Delta_{B}$ and $\Delta_{0}$ solely as a function of $q_{0}$, $T$ and $eB$. The new threshold of $q_{0}$ (or $q_{0}/T$), $[q_{0}]_{th}$ (or $[q_{0}/T]_{th}$), is then defined by the specific value of $q_{0}$ below which the ratio $\Delta_{B}/\Delta_{0}$ is either imaginary or negative and above which this ratio is positive and real (see below). Later, we will numerically determine $[q_{0}/T]_{th}$ for various fixed $T$ and $eB$. We will then plot $\Delta_{B}/\Delta_{0}$ for $q_{0}/T\geq [q_{0}/T]_{th}$.
\par
\begin{table}[tbp]
\centering
\begin{tabular}{ccccccccccccccc}
\hline
$eB$ [GeV$^{2}]$&&
 $T$ [MeV]&
&
$[\frac{q_{0}}{T}]_{th}^{(5)}$&
&
$[\frac{\Delta_{B}}{\Delta_{0}}]_{th}^{(5)}$&
&
$[\frac{q_{0}}{T}]_{th}^{(8)}$&
&
$[\frac{\Delta_{B}}{\Delta_{0}}]_{th}^{(8)}$&
&
$[\frac{q_{0}}{T}]_{th}^{(10)}$&
&
$[\frac{\Delta_{B}}{\Delta_{0}}]_{th}^{(10)}$\\
\hline\hline
$0.02$&&$200$&&$4.5$&&$45.5$&&$5.7$&&$46.6$&&$6.4$&&$48.5$\\
$0.03$&&$200$&&$5.5$&&$38.6$&&$7.0$&&$30.7$&&$7.8$&&$36.7$\\
$0.02$&&$400$&&$2.3$&&$48.0$&&2.9&&$47.6$&&$3.2$&&$49.9$\\
$0.03$&&$400$&&$2.8$&&$52.1$&&$3.5$&&$65.8$&&$3.9$&&$69.5$\\
$0.2$&&$200$&&$14.2$&&$35.1$&&$18.1$&&$0.6$&&$20.6$&&$0.05$\\
$0.3$&&$200$&&$17.4$&&$40.7$&&$22.2$&&$0.03$&&$26.0$&&$0.05$\\
$0.2$&&$400$&&$7.1$&&$29.3$&&$9.0$&&$17.3$&&$10.1$&&$19.2$\\
$0.3$&&$400$&&$8.7$&&$26.0$&&$11.0$&&$8.5$&&$12.3$&&$9.2$\\
\hline
\end{tabular}
\caption{The values of $[\frac{q_{0}}{T}]_{th}^{(i)}, i=5,8,10$ and $[\frac{\Delta_{B}}{\Delta_{0}}]_{th}^{(i)}, i=5,8,10$ for $eB=0.02, 0.03$ GeV$^{2}$ and  $eB=0.2, 0.3$ GeV$^{2}$ at $T=200, 400$ MeV. The superscripts $(5), (8)$ and $(10)$ correspond to the upper limits for the summation over external  and internal Landau levels,  $(n,k)$ and $(p,\ell)$  (see the main text).}\label{tab1}
\end{table}
\par
To explain why the definition of this new threshold seems to be necessary, let us first notice that the fact that $\Delta_{B}$ from (\ref{D36})-(\ref{D38}) is imaginary for $q_{0}<[q_{0}]_{th}$, is mainly related to the dependence of $\Delta_{B}$ on $\xi_{q}, {\cal{R}}_{q}$ as well as $\eta$ and ${\cal{Q}}$ from (\ref{D22}), (\ref{D23}) as well as  (\ref{D31}) and (\ref{D32}), respectively. As in the $B=0$ case, these imaginary values can be discarded by inserting certain Heaviside $\Theta$-functions, $\Theta(q_{0}^{2}-\widetilde{M}_{nk,p\ell}^{2})$, term by term for each fixed $n,k$ (external Landau levels)  and $p,\ell$ (internal Landau levels), into the r.h.s. of the final result for $\Delta_{B}$ from (\ref{D38}). Here,
\begin{eqnarray}\label{O7}
\widetilde{M}_{nk,p\ell}\equiv \mbox{max}(M_{n}+M_{k}, M_{p}+M_{\ell}),
\end{eqnarray}
with $M_{i}, i=n,k,p,\ell$ magnetic masses, including thermal quark and lepton masses ${\cal{M}}_{q}$ and ${\cal{M}}_{\ell}$ from (\ref{O5}) and (\ref{O6}),
\begin{eqnarray}\label{O8}
\begin{array}{rclcrcl}
M_{n}^{2}&=&{\cal{M}}_{\ell}^{2}+2n|eB|, &\qquad&
M_{k}^{2}&=&{\cal{M}}_{\ell}^{2}+2k|eB|, \\
M_{p}^{2}&=&{\cal{M}}_{q}^{2}+2p|q_{f}eB|, &\qquad&
M_{\ell}^{2}&=&{\cal{M}}_{q}^{2}+2\ell|q_{f}eB|.
\end{array}
\end{eqnarray}
\begin{figure}[t]
\centering
\includegraphics[width=8cm,height=6cm]{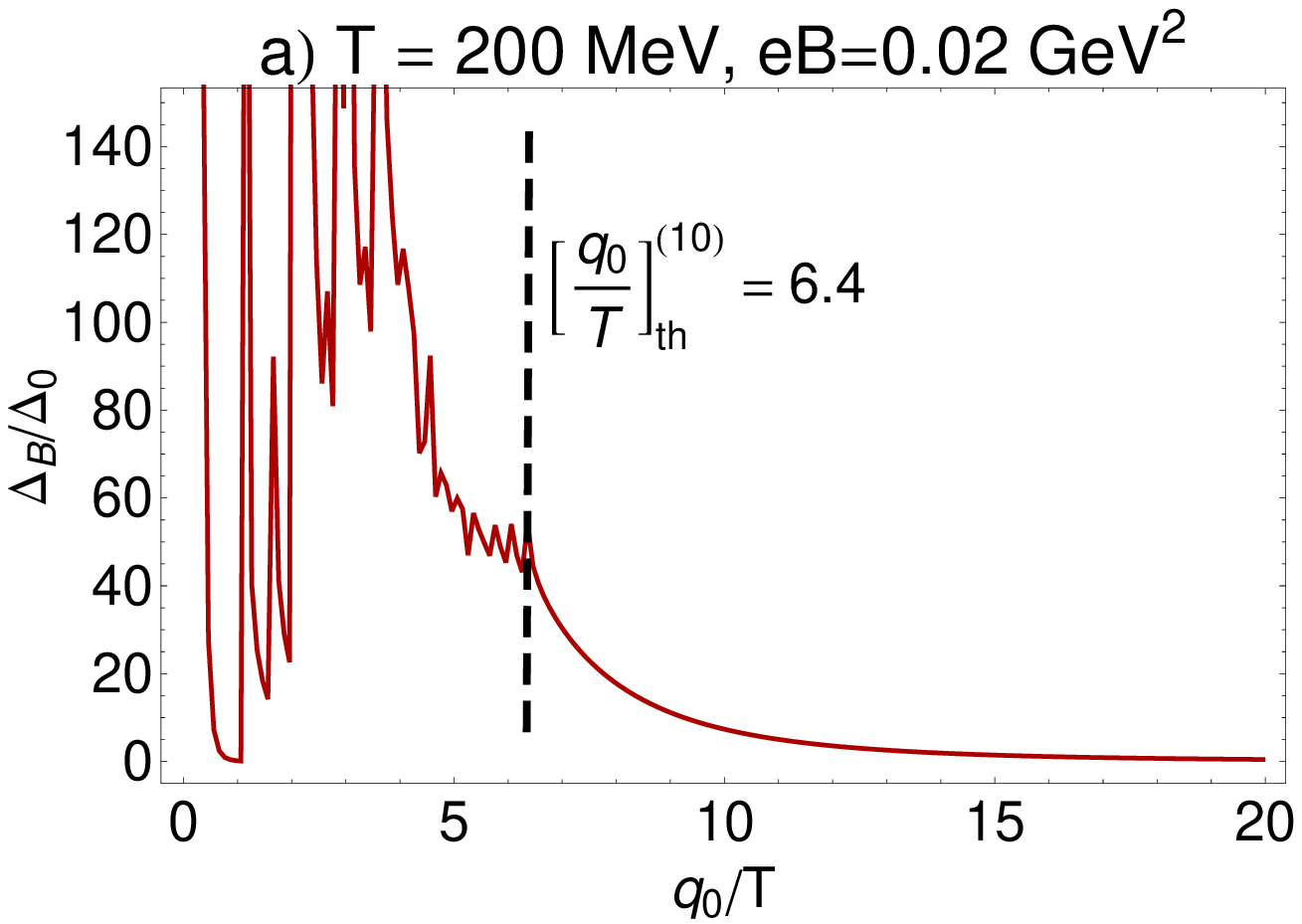}
\hfill
\includegraphics[width=8cm,height=6cm]{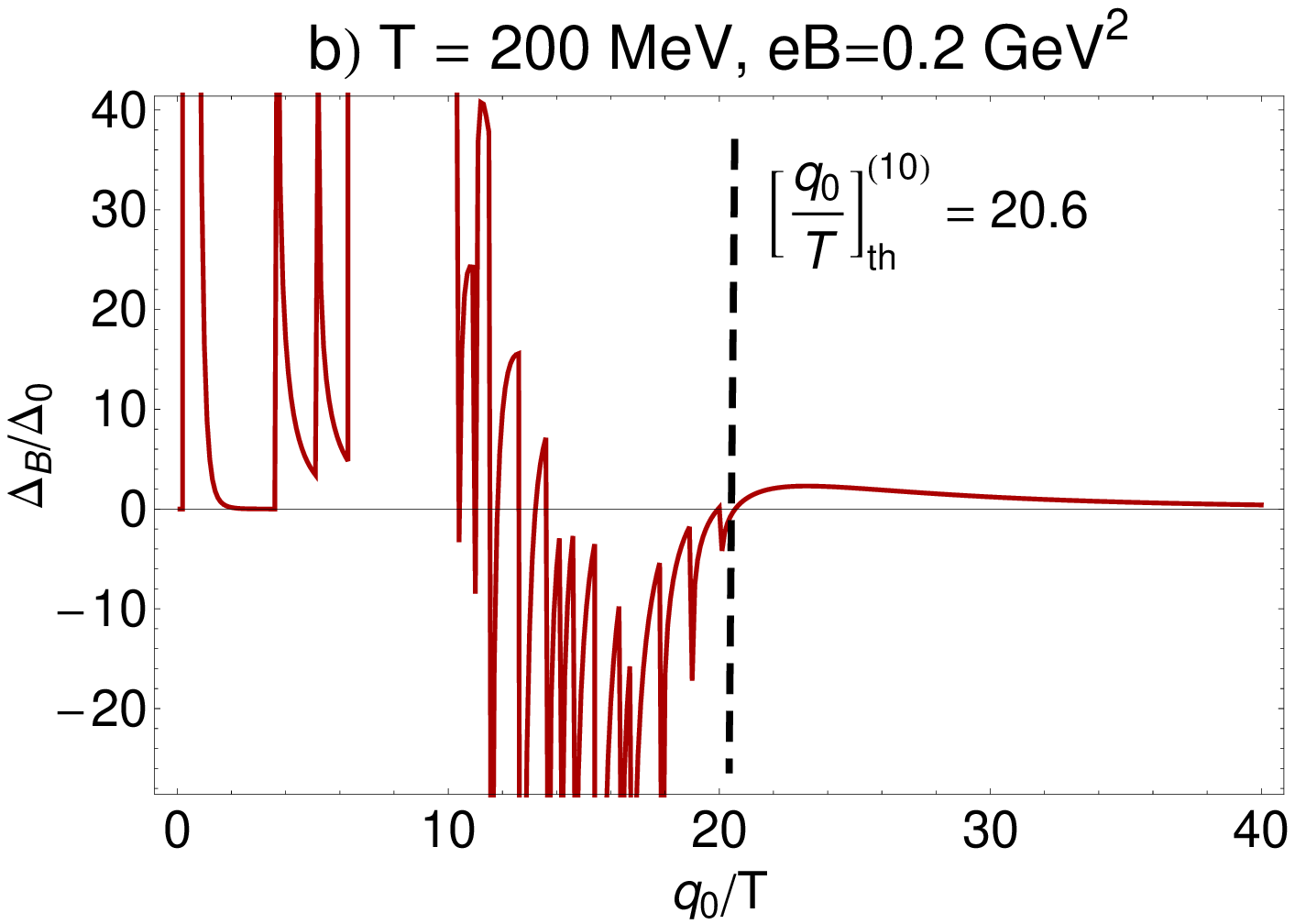}
\caption{(color online). Oscillatory pattern in the spectrum of $\Delta_{B}/\Delta_{0}$ arising below the thresholds $[q_{0}/T]_{th}^{(10)}=6.4$ for $T=200$ MeV and $eB=0.02$ GeV$^{2}$ and $[q_{0}/T]_{th}^{(10)}=20.6$ for $T=200$ MeV and $eB=0.2$ GeV$^{2}$ (dashed lines). The superscript $(10)$ indicates the upper limit in the summation over Landau levels. The minimum threshold for dilepton production does not depend on $eB$, as expected. For numerical values of $[q_{0}/T]_{th}^{(10)}$, see also table \ref{tab1}. Same qualitative picture arises also for $eB=0.03, 0.3$ GeV$^{2}$.}\label{fig2ab}
\end{figure}
\par\noindent
In this way, infinitely many production thresholds appear for each fixed $n,k$ and $p,\ell$. They are characterized by  $\widetilde{M}_{nk,p\ell}$ \cite{tuchin2013}. Because of these infinitely many thresholds for each fixed $n,k$ and $p,\ell$, the spectrum of $\Delta_{B}$ is expected to possess specific oscillatory pattern \cite{harding1983, baier2007} [see figures \ref{fig2ab}(a) and \ref{fig2ab}(b) as typical examples]. In our case, however, in contrast to $B=0$ case, another serious problem occurs. Because of the interplay of
different $\Delta_{B}^{ij,nk}$, which contribute to $\Delta_{B}$ with different positive and negative signs, there appear negative values in the spectrum of $\Delta_{B}(q_{0};T,eB)$, and, in particular, in the ratio $\Delta_{B}/\Delta_{0}$ [see figure \ref{fig2ab}(b)]. This makes the  comparison of $\Delta_{B}$ with $\Delta_{0}$ in the whole regime of $q_{0}$ rather difficult (even after the $\Theta$-functions are inserted).
The definition of a new threshold, which is different from ``the minimum energy threshold for dilepton production'', seems therefore to be necessary. Let us notice that since by using $[q_{0}]_{th}$, and focusing on the spectrum of $\Delta_{B}$ in the regime $q_{0}\geq [q_{0}]_{th}$,
we do not consider the aforementioned infinitely many thresholds for each fixed $n,k$ and $p,\ell$, no oscillatory pattern will appear in the spectrum of $\Delta_{B}/\Delta_{0}$ in the present work. Instead, the spectrum is only characterized with a single singularity which, by definition, occurs at the position of $[q_{0}/T]_{th}$. Later, we will show that the position of $[q_{0}/T]_{th}$ depends on the upper limit of the summation over $n,k$ and $p,\ell$. Moreover, a comparison of $\Delta_{B}$ with $\Delta_{0}$ in the regime $q_{0}\geq [q_{0}]_{th}$ shows that $\Delta_{B}\gg \Delta_{0}$ even in a regime where $\Delta_{0}$ is very small.
\par
In section \ref{sec5b}, we will separately study three different aspects of the dependence of $\Delta_{B}/\Delta_{0}$ on $q_{0}/T$. First, the dependence of the photon threshold energy $[q_{0}/T]_{th}$ on $eB$ and $T$ will be discussed. We will, in particular, focus on the interplay between magnetic field strengths and temperatures on these parameters. Then, the numerical results for $\Delta_{B}/\Delta_{0}$ of an electron-positron pair will be presented as a function of $q_{0}/T$ for fixed $T$ and different $eB$. We will finally compare the production rates of dielectrons and dimuons for different $eB$ at fixed $T$.
\par
In section \ref{sec5c}, we will then focus on possible effects of constant background magnetic fields on the anisotropy in the production rate of dileptons in the longitudinal and transverse directions with respect to these fields. To do this, we will use a novel anisotropy factor $\nu_{B}$, already defined in (\ref{O9}),
with $\Delta_{B}^{\|}$ and $\Delta_{B}^{\perp}$ from (\ref{D37}). We will study the dependence of $\nu_{B}$ on $q_{0}/T$ for $eB=0.02,0.03,0.2,0.3$ GeV$^{2}$ at various fixed temperatures $T=200,400$ MeV.
\subsection{Dependence of ${\Delta_{B}/\Delta_{0}}$ on ${q_{0}/T}$}\label{sec5b}
\subsubsection{$eB$ and $T$ dependence of the energy threshold of virtual photons}\label{sec5b1}
\begin{figure}[tbp]\centering
\includegraphics[width=8cm,height=6cm]{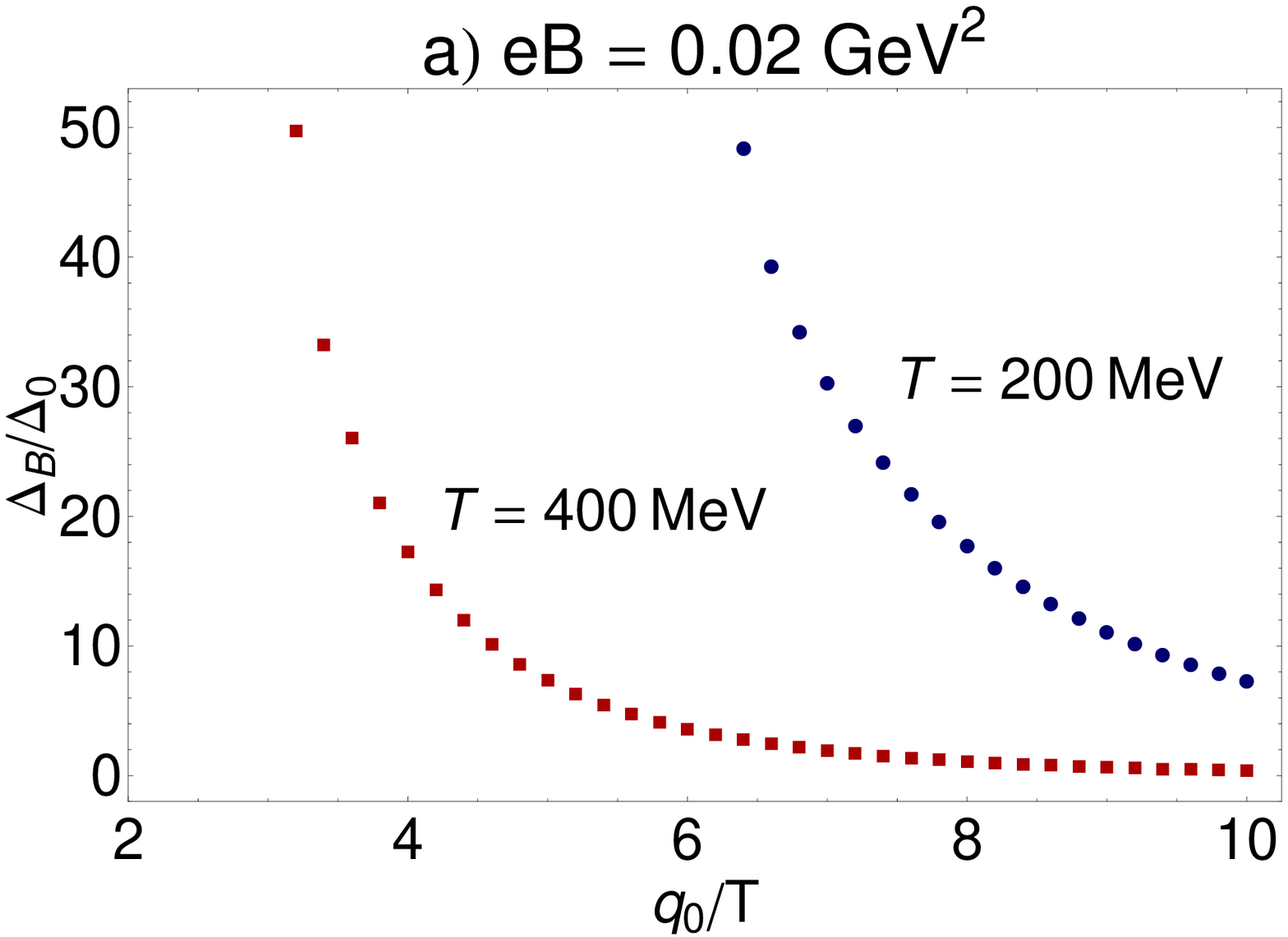}
\hfill
\includegraphics[width=8cm,height=6cm]{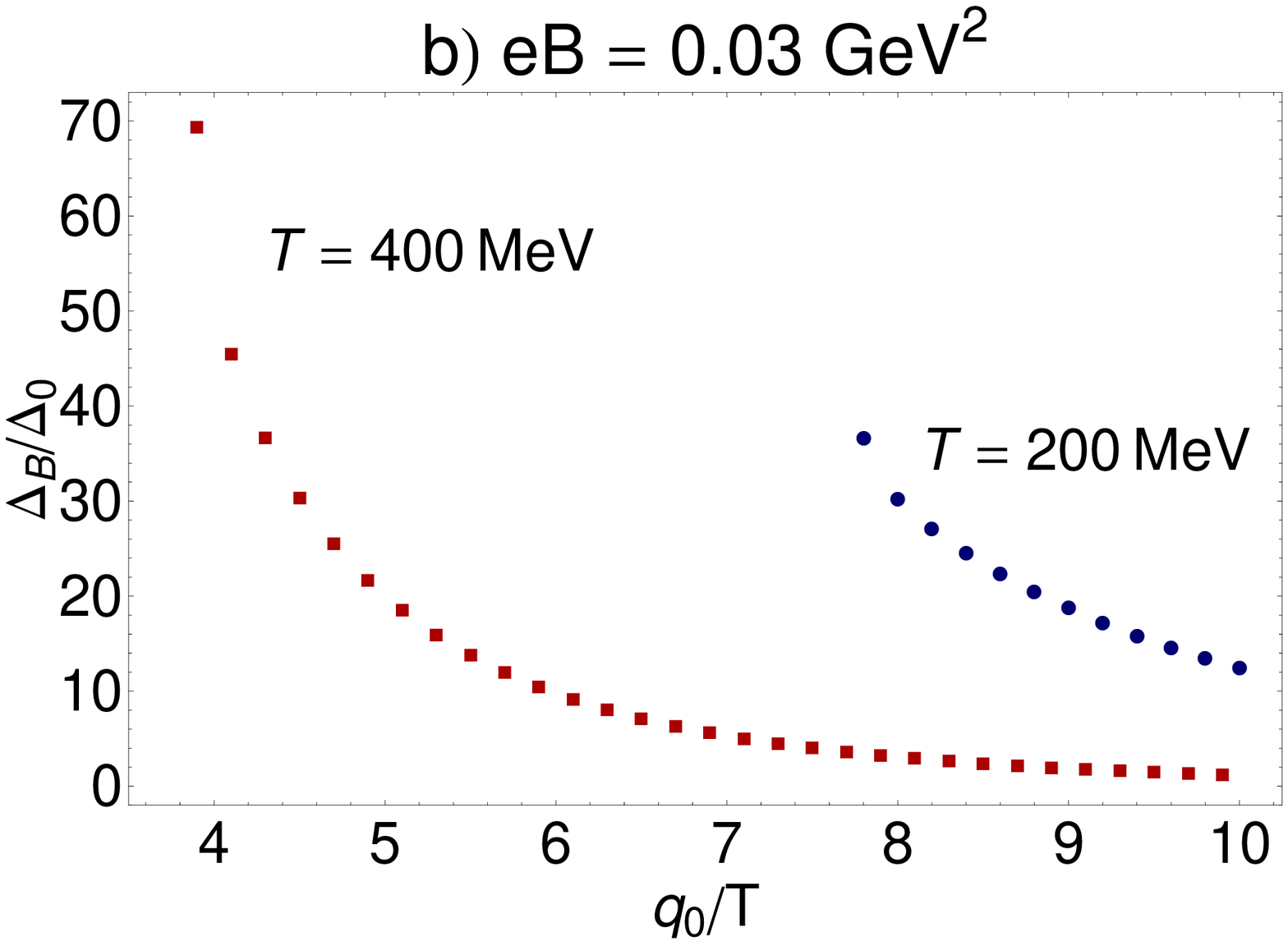}
\caption{(color online). The ratio $\Delta_{B}/\Delta_{0}$ for the production of electron-positron pairs is plotted as a function of rescaled photon energy $q_{0}/T$ for $eB=0.02$ GeV$^{2}$ (panel a) and  $eB=0.03$ GeV$^{2}$ (panel b) and at $T=200, 400$ MeV. The results correspond to $n,k,p,\ell\leq 10$. Blue circles and red squares correspond to $T=200$ MeV and $T=400$ MeV, respectively.}\label{fig2}
\end{figure}
\par\noindent
In figure \ref{fig2}, the ratio $\Delta_{B}/\Delta_{0}$ for the production of dielectrons is plotted as a function of rescaled photon energy $q_{0}/T$ for $eB=0.02$ GeV$^{2}$  and  $eB=0.03$ GeV$^{2}$ and at $T=200, 400$ MeV.
\begin{figure}[tbp]\centering
\includegraphics[width=7.5cm,height=6cm]{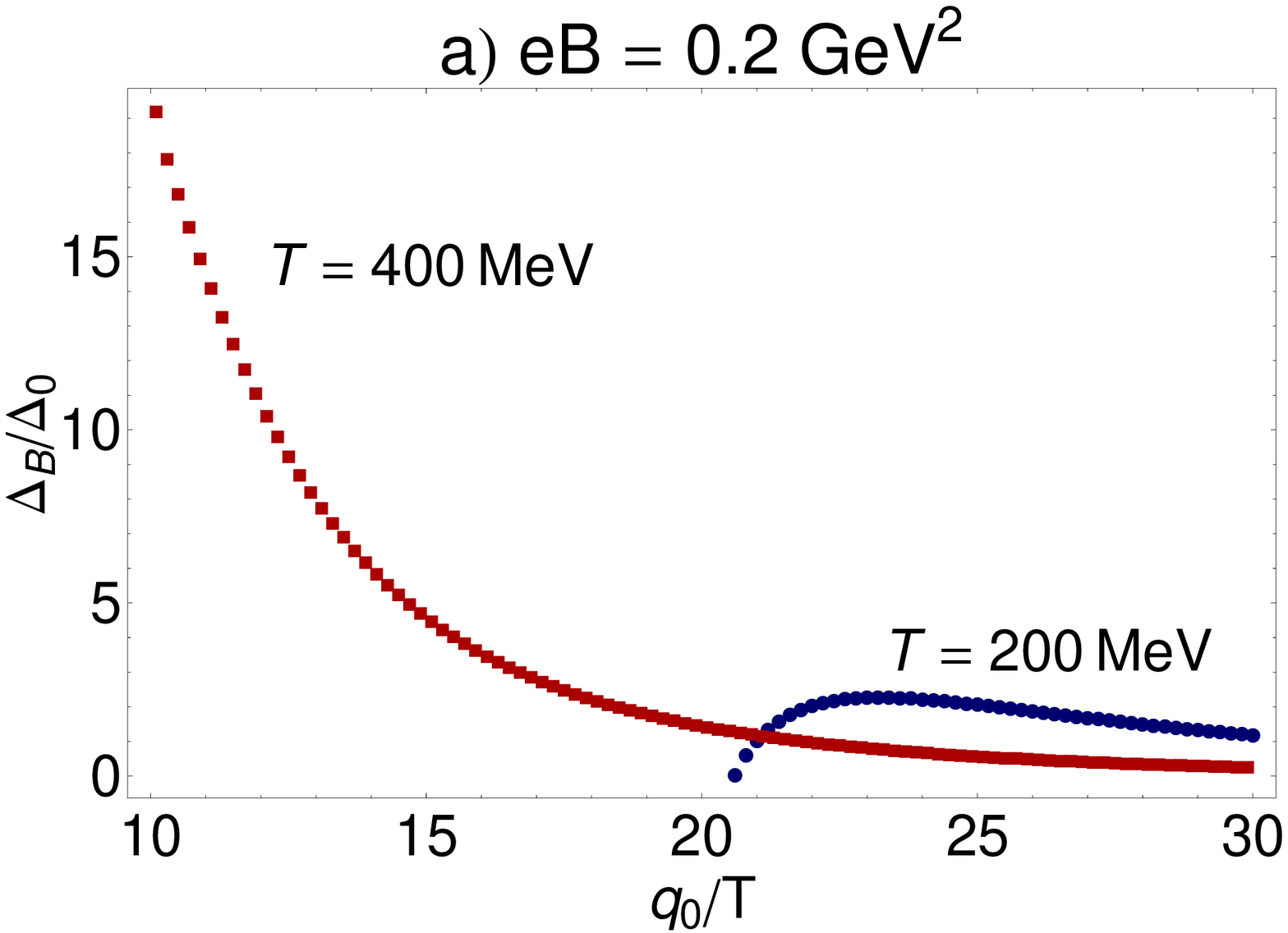}
\hfill
\includegraphics[width=7.5cm,height=6cm]{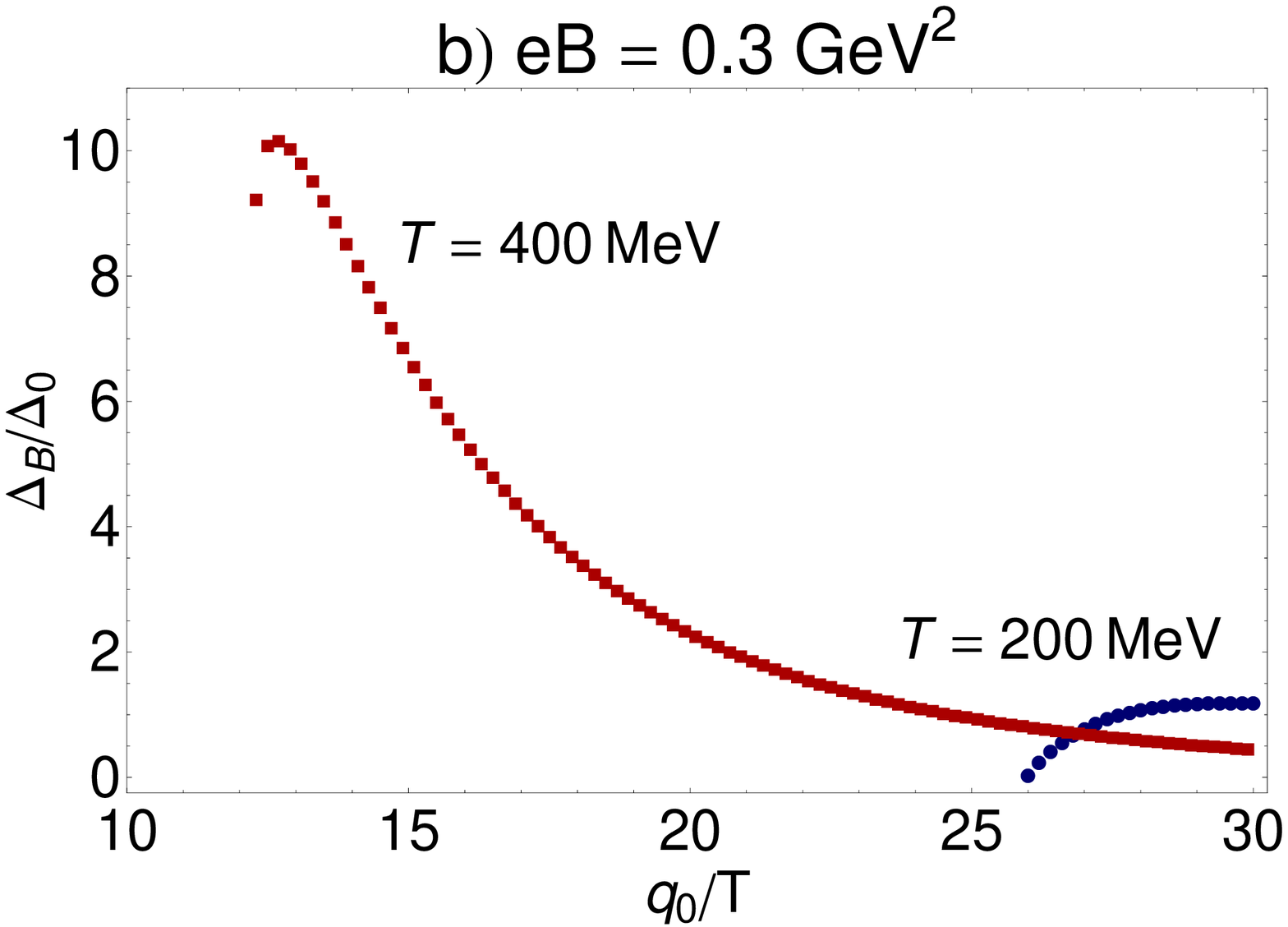}
\caption{(color online). The ratio $\Delta_{B}/\Delta_{0}$ for the production of electron-positron pairs is plotted as a function of rescaled photon energy $q_{0}/T$ for $eB=0.2$ GeV$^{2}$ (panel a) and  $eB=0.3$ GeV$^{2}$ (panel b) and at $T=200, 400$ MeV. The results correspond to $n,k,p,\ell\leq 10$. Blue circles and red squares correspond to $T=200$ MeV and $T=400$ MeV, respectively. }\label{fig3}
\end{figure}
In figure \ref{fig3}, the same dependence is plotted for $eB=0.2$ GeV$^{2}$ and  $eB=0.3$ GeV$^{2}$ and at $T=200, 400$ MeV. Blue circles and red squares correspond to $T=200$ MeV and $T=400$ MeV, respectively. The plots in figures \ref{fig2} and \ref{fig3} correspond to the choice $n,k,p,\ell\leq 10$ for the upper limit of the summation over Landau levels in (\ref{D38}). Same computation is also performed for $n,k,p,\ell\leq 5$ and $n,k,p,\ell\leq 8$. According to these results, independent of the choice of the upper limit of the summation over Landau levels, the rescaled photon threshold energy $[q_{0}/T]_{th}$ increases with increasing $eB$ at fixed $T$ (see table \ref{tab1} and figures \ref{fig5} and \ref{fig6}).
As concerns the $T$ dependence of $[q_{0}]_{th}$ for fixed $eB$, however, we have to distinguish between two cases of weak and moderate magnetic fields, $eB=0.02,0.03$ GeV$^{2}$ and $eB=0.2,0.3$ GeV$^{2}$. As it turns out, for weak magnetic fields $[q_{0}]_{th}$ does not significantly change with increasing $T$, while for moderate magnetic fields, it decreases once $T$ increases (see table \ref{tab1}).
Let us notice that the temperature dependence of $\Delta_{0}$ as well as $\Delta_{B}$ are mainly determined by Bose distribution function $(e^{\beta q_{0}}-1)^{-1}$, appearing in (\ref{A13}) as well as in (\ref{A34}). From here, it is expected that by increasing the temperature and by keeping $\Delta_{0}$ and/or $\Delta_{B}$ fixed, $q_{0}$ decreases. In the case of $\Delta_{B}$, however, the apparent interplay between $T$ and $B$ dependence of $\Delta_{B}$ leads to the above mentioned dependence of $[q_{0}]_{th}$ on $T$ for fixed $eB$.
\par
In table \ref{tab1}, we have presented numerical values for the energy thresholds, $[\frac{q_{0}}{T}]_{th}^{(i)}$ for different $eB$ and $T$ and different upper limits $(i), i=5,8,10$ for the summation over Landau levels. The superscript $(i)$ on  $[\frac{q_{0}}{T}]_{th}^{(i)}$ corresponds to these upper limits. The numerical values of $[\frac{q_{0}}{T}]_{th}^{(i)}$ are given with $0.1$ accuracy.\footnote{According to the results presented in table \ref{tab1}, the value of $[\frac{q_{0}}{T}]_{th}^{(10)}$ for $eB=0.02$ GeV$^{2}$ and at $T=200$ MeV is $6.4$. In this case, for instance, the accuracy $0.1$ means that by choosing $[\frac{q_{0}}{T}]_{th}^{(10)}=6.3$, the ratio $\Delta_{B}/\Delta_{0}$ would be imaginary (or negative). Let us notice that the accuracy $0.1$ in the numerical determination of the energy threshold leads to invisible errors in all our plots in section V.}
The numerical values of the ratio  $\frac{\Delta_{B}}{\Delta_{0}}$ at the threshold energy $[\frac{q_{0}}{T}]_{th}^{(i)}$, denoted by $[\frac{\Delta_{B}}{\Delta_{0}}]_{th}^{(i)}$ for $i=5,8,10$, are also presented in table \ref{tab1}. Let us consider the results for $n,k,p,\ell\leq 10$. In this case, for fixed $T$, $[\frac{\Delta_{B}}{\Delta_{0}}]_{th}^{(10)}$ decreases with increasing the magnetic field, e.g., from $eB=0.02$ GeV$^{2}$ to $eB=0.2$ GeV$^{2}$ and from $eB=0.03$ GeV$^{2}$ to $eB=0.3$ GeV$^{2}$.  The results for $[\frac{\Delta_{B}}{\Delta_{0}}]_{th}^{(i)}$ for $i=8$ show the same behavior. Moreover, a comparison between the results for $[\frac{\Delta_{B}}{\Delta_{0}}]_{th}^{(5)}$ with $[\frac{\Delta_{B}}{\Delta_{0}}]_{th}^{(8)}$ and $[\frac{\Delta_{B}}{\Delta_{0}}]_{th}^{(10)}$ shows that, for weak magnetic fields ($eB=0.02, 0.03$ GeV$^{2}$), increasing the upper limit for the summation over Landau levels does not significantly change the order of magnitude of $[\frac{\Delta_{B}}{\Delta_{0}}]_{th}^{(i)}$, while for strong magnetic fields ($eB=0.2, 0.3$ GeV$^{2}$), the values of $[\frac{\Delta_{B}}{\Delta_{0}}]_{th}^{(8)}$ and $[\frac{\Delta_{B}}{\Delta_{0}}]_{th}^{(10)}$ are much smaller than $[\frac{\Delta_{B}}{\Delta_{0}}]_{th}^{(5)}$. In contrast, changing the upper limit of the summation over Landau levels does not have such a drastic effects on $[\frac{q_{0}}{T}]_{th}^{(i)}$. To have a measure which quantifies this fact, let us introduce the following quantity:
\begin{eqnarray}\label{O10}
\eta_{i,j}\equiv\frac{[\frac{q_{0}}{T}]_{th}^{(j)}-[\frac{q_{0}}{T}]_{th}^{(i)}}{[\frac{q_{0}}{T}]_{th}^{(i)}}\qquad \mbox{in \%}.
\end{eqnarray}
Here, the superscripts $(i)$ indicate the upper limit of the summation over Landau levels, as before.

\begin{table}[tbp]\centering
\begin{tabular}{ccccccc}
\hline
$eB$ [GeV$^{2}]$&&
 $T$ [MeV]&
&
$\eta_{5,8} [\%]$&
&
$\eta_{8.10}[\%]$\\
\hline\hline
$0.02$&&$200$&&$26.7$&&$12.3$\\
$0.03$&&$200$&&$27.3$&&$11.4$\\
$0.02$&&$400$&&$26.0$&&$10.3$\\
$0.03$&&$400$&&$25.0$&&$11.4$\\
$0.2$&&$200$&&$27.5$&&$13.8$\\
$0.3$&&$200$&&$27.6$&&$17.1$\\
$0.2$&&$400$&&$26.8$&&$12.2$\\
$0.3$&&$400$&&$26.4$&&$11.8$\\
\hline
\end{tabular}
\caption{The values of $\eta_{i,j}$, defined in (\ref{O10}) for $eB=0.02, 0.03$ GeV$^{2}$ and  $eB=0.2, 0.3$ GeV$^{2}$ at $T=200, 400$ MeV.  A comparison between $\eta_{5,8}$ and $\eta_{8,10}$ shows that changes in the threshold values $[\frac{q_{0}}{T}]_{th}$ decreases with increasing the upper limit in the summation over Landau levels.}\label{tab2}
\end{table}
In table \ref{tab2}, we have listed $\eta_{5,8}$ and $\eta_{8,10}$ for different $eB$ and $T$. The results show that by increasing $(i)$ from $5$ to $8$, the threshold values for $[\frac{q_{0}}{T}]_{th}^{(5)}$ increases up to $\sim 27.6\%$, while by increasing $(i)$ from $8$ to $10$, we have
$10.3\% \leq \eta_{8,10}\leq 17.1\%$. For larger values of $(i)$, it is therefore expected that $\eta_{i,j}$ become smaller, and higher Landau levels become more and more irrelevant. Let us also notice that for $i>10$, we do not expect any qualitative changes in the final results for $[\frac{\Delta_{B}}{\Delta_{0}}]_{th}^{(i)}$. On the other hand, by assuming that for a realistic (experimental) setup, the relevant kinematical region for $q_{0}/T$ is at most $q_{0}/T\leq 40$ for $T=200$ MeV and $q_{0}/T\leq 20$ for $T=400$ MeV, and by having in mind that the threshold values $[\frac{q_{0}}{T}]_{th}^{(i)}$ increase by increasing the value of the upper limit to $n,k,p,\ell>10$ (see  table \ref{tab1}), the choice $(i)=10$ seems therefore to be acceptable. In the rest of this paper, we will only report the results for $n,k,p,\ell\leq 10$.
\par
Let us notice at this stage, that the above numerical analysis also shows that for small values of $q_{0}/T$, once the magnetic field is chosen to be very strong, only the lowest Landau level will contribute to $\Delta_{B}$. The question about the exact numerical values of $q_{0}/T$ and $eB/T^{2}$ that justify a LLL approximation remains open, and probably only after a rigorous comparison with experimental data, we will be able to decide about this issue.
\subsubsection{$eB$ and $T$ dependence of dielectron production rate}\label{sec5b2}
\par
\begin{figure}[tbp]\centering
\includegraphics[width=8.4cm,height=6cm]{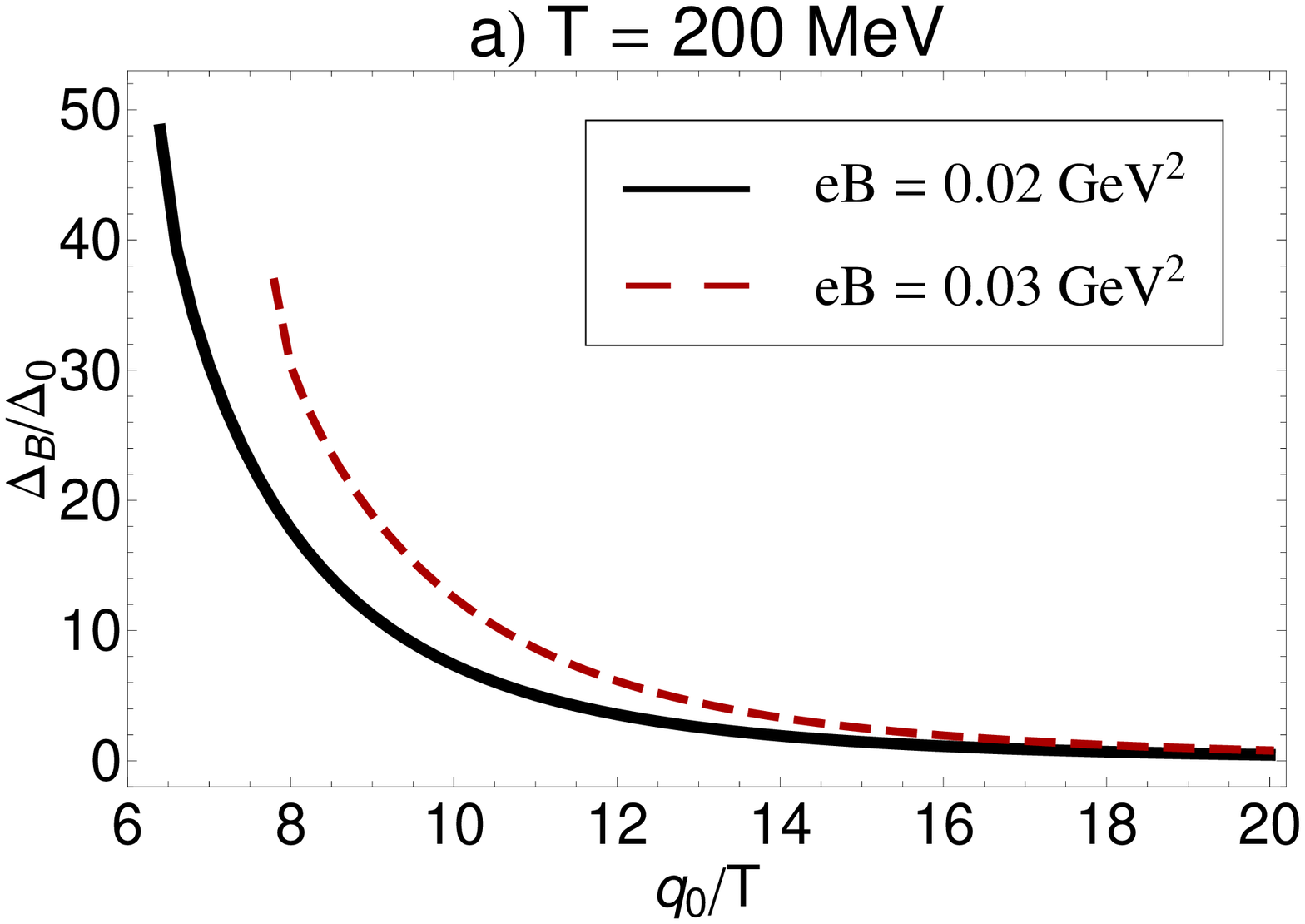}
\hfill
\includegraphics[width=8.4cm,height=6cm]{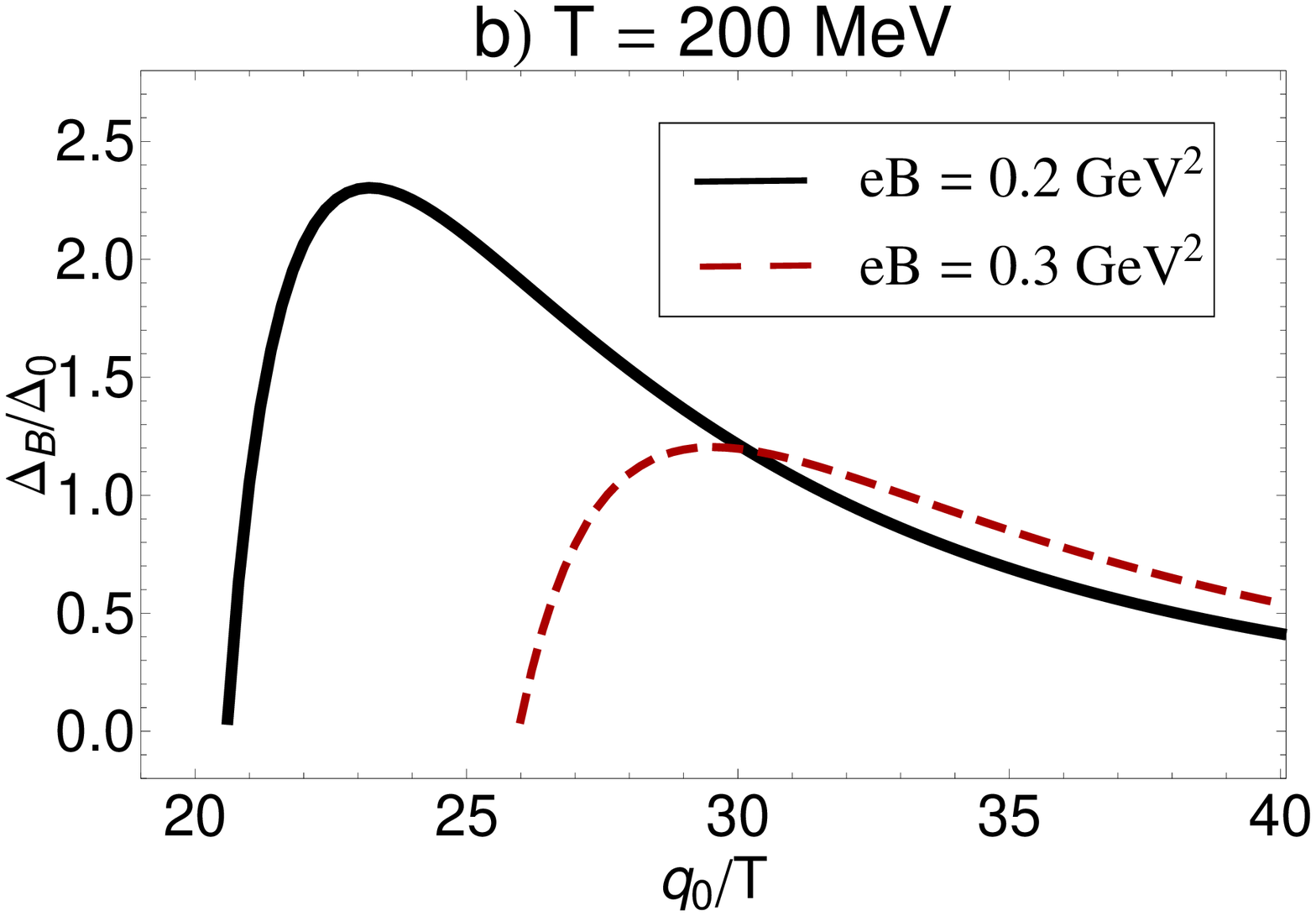}
\caption{(color online). (a) The ratio $\Delta_{B}/\Delta_{0}$ for the production of electron-positron pair is plotted as a function of rescaled photon energy $q_{0}/T$ at $T=200$ MeV for $eB=0.02$ GeV$^{2}$ (black solid line) and  $eB=0.03$ GeV$^{2}$ (red dashed line). (b)  The ratio $\Delta_{B}/\Delta_{0}$ for the production of electron-positron pair is plotted as a function of rescaled photon energy $q_{0}/T$ at $T=200$ MeV for $eB=0.2$ GeV$^{2}$ (black solid line) and  $eB=0.3$ GeV$^{2}$ (red dashed line). }\label{fig4}
\end{figure}

\begin{figure}[tbp]\centering
\includegraphics[width=8.4cm,height=6cm]{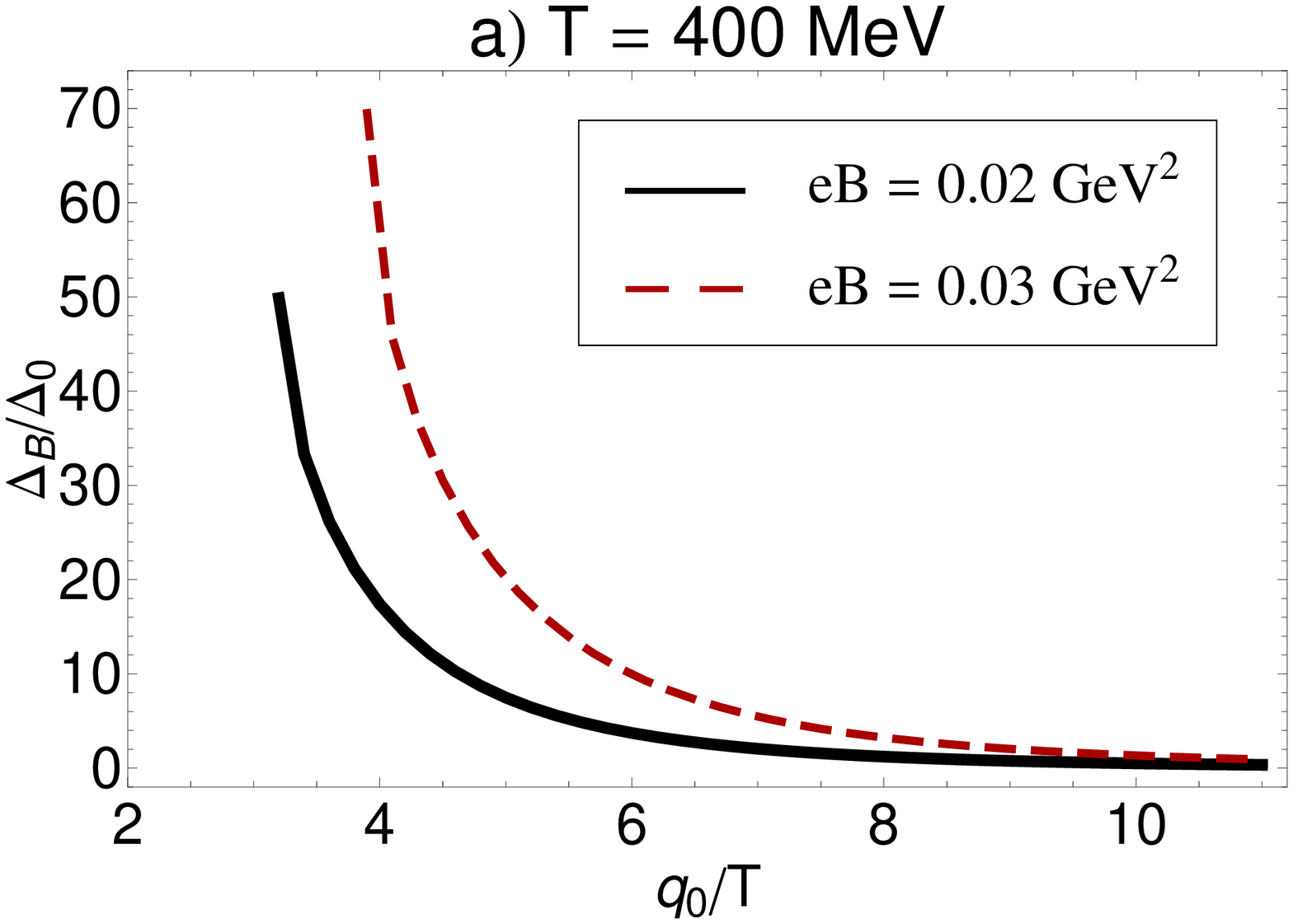}
\hfill
\includegraphics[width=8.4cm,height=6cm]{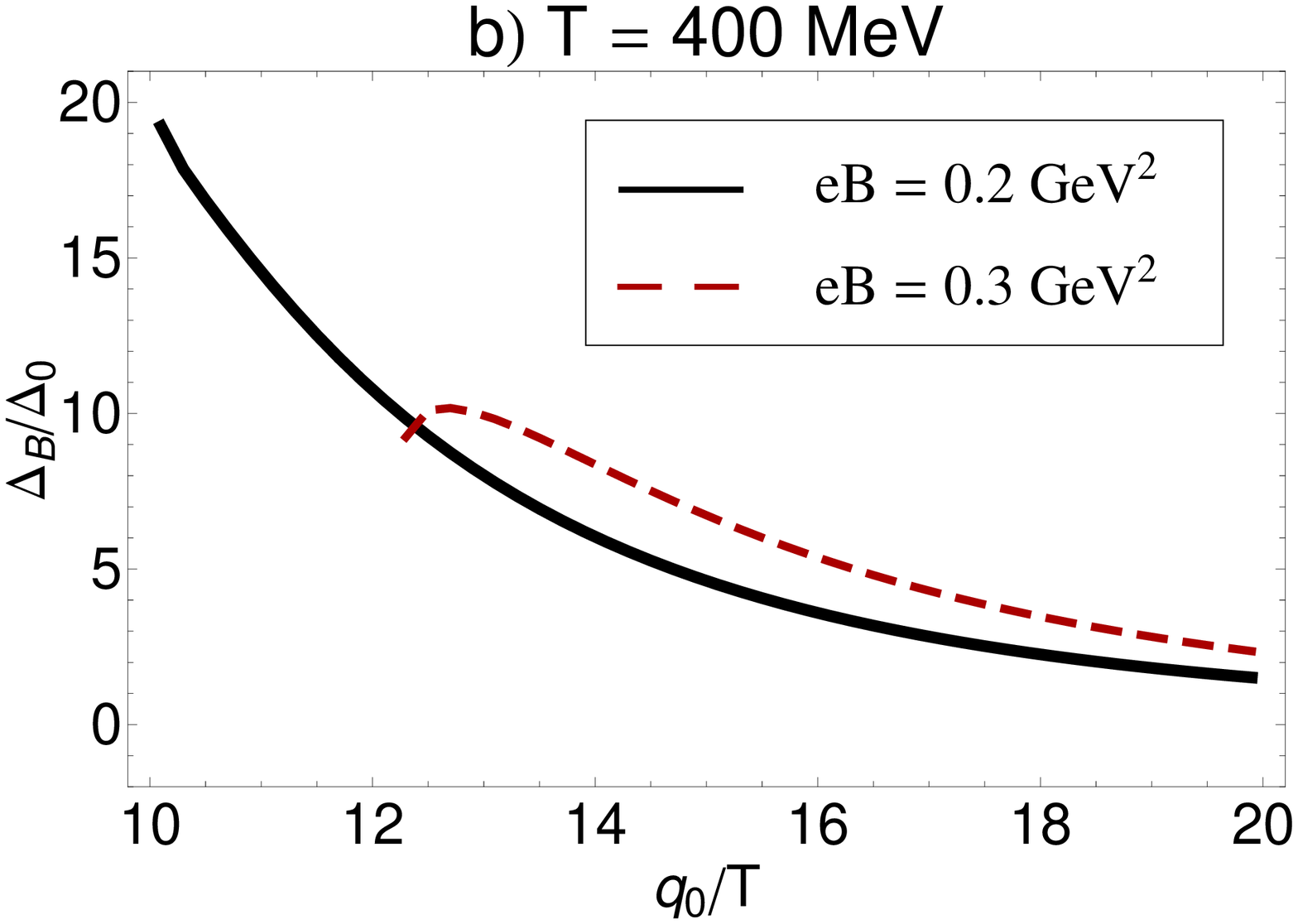}
\caption{(color online).  (a) The ratio $\Delta_{B}/\Delta_{0}$ for the production of electron-positron pair is plotted as a function of rescaled photon energy $q_{0}/T$ at $T=400$ MeV for $eB=0.02$ GeV$^{2}$ (black solid line) and  $eB=0.03$ GeV$^{2}$ (red dashed line). (b)  The ratio $\Delta_{B}/\Delta_{0}$ for the production of electron-positron pair is plotted as a function of rescaled photon energy $q_{0}/T$ at $T=400$ MeV for $eB=0.2$ GeV$^{2}$ (black solid line) and  $eB=0.3$ GeV$^{2}$ (red dashed line). }\label{fig5}
\end{figure}

\noindent
In the previous part, the ratio $\Delta_{B}/\Delta_{0}$ for electron-positron production rate was demonstrated as a function of $q_{0}/T$ for fixed $eB$ and different $T$. In contrast, in figures \ref{fig4} and \ref{fig5}, the same ratio $\Delta_{B}/\Delta_{0}$ is presented as a function of $q_{0}/T$ at fixed temperatures, $T=200$ MeV and $T=400$ MeV, for different magnetic field strengths $eB=0.02, 0.2$ GeV$^{2}$ (black solid lines) and $eB=0.03, 0.3$ GeV$^{2}$ (red dashed lines). Because of different thresholds $[q_{0}/T]_{th}$, the regime in which the results for two different values of $eB$ can be compared is different: For $T=200$ MeV, the relevant regime turns out to be $5<\frac{q_{0}}{T}< 40$, while for $T=400$ MeV this regime is given by $3< \frac{q_{0}}{T}< 20$.\footnote{We are looking for the regime of $q_{0}/T$, where $\Delta_{B}\gtrsim \Delta_{0}$, and will denote it as ``the relevant regime''.}
\\
According to the results demonstrated in figure \ref{fig4} and \ref{fig5}, except from $eB=0.2,0.3$ GeV$^{2}$ at $T=200$ MeV from figure \ref{fig4}(b) and from $eB=0.3$ GeV$^{2}$ from figure \ref{fig5}(b), the ratio $\Delta_{B}/\Delta_{0}$ has a maximum near the threshold energy $[\frac{q_{0}}{T}]_{th}$, and decreases with a relatively large slope to $\frac{\Delta_{B}}{\Delta_{0}}\leq 1$.
Same behavior can
also be observed separately in $\Delta_{B}$ and $\Delta_{0}$. For $\Delta_{0}$, this is related to the presence of Bose distribution factor $(e^{\beta q_0}-1)^{-1}$ in (\ref{A14}) (see the corresponding discussion in \cite{rapp-POS}). The same
factor appears also in ${\cal{N}}^{(q)}$ from (\ref{D38}). Let us notice that the specific behavior of the ratio $\Delta_{B}/\Delta_{0}$ in Figs. \ref{fig4}(a)-\ref{fig5}(a) for moderate $eB = 0.02, 0.03$ GeV$^{2}$, which mainly arises from
the interplay between $q_{0}, T$ and $eB$ in the final results for $\Delta_{B}$ and $\Delta_{0}$, confirms the expectation that magnetic fields enhance the production rate of particles in hot QGP \cite{mamo2012, munshi2016, ayala2016}. In the cases
of $eB = 0.2, 0.3$ GeV$^{2}$ at $T = 200$ MeV from figure \ref{fig4}(b) and from $eB = 0.3$ GeV$^2$ from figure \ref{fig5}(b), in contrast, the ratio $\Delta_{B}/\Delta_{0}$ has a maximum in certain $\frac{q_{0}}{T}>[\frac{q_{0}}{T}]_{th}$, and decreases with a moderate slope to $\frac{\Delta_{B}}{\Delta_{0}}\leq 1$. To elaborate the reason for the appearance of these maxima at certain $\frac{q_{0}}{T}\geq [\frac{q_{0}}{T}]_{th}$, let us consider $\Delta_{B}^{ij,nk}$ from (\ref{D38}). The corresponding expressions include, in particular, a summation over flavor index $q_{f}=\{u,d\}$. Because of the factor $(2-2s_{q})$ in $\Delta_{B}^{4j,nk}, j=1,\cdots 4$ with $s_{q}=\mbox{sign}(q_{f}eB)$, positive charges (up quarks) contribute only to  $\Delta^{1j,nk}, j=1,2$, while negative charges (down quarks) contribute to all $\Delta_{B}^{ij,nk}$ in (\ref{D38}). The total contribution of positive (negative) charges to $\Delta_{B}$ turns out to be always negative (positive). By adding the contributions from positive and negative charges, we arrive, depending on exact numerical values of $\Delta_{B}^{ij,nk}$, at different results: For large enough magnetic fields, for instance, the ratio $\Delta_{B}/\Delta_{0}$
possesses a maximum at certain value of $q_{0}/T$ greater than the threshold value $[q_{0}/T]_{th}$, while for weak and moderate field strengths and temperatures $\Delta_{B}/\Delta_{0}$ turns out to be maximum near $[q_{0}/T]_{th}$. These specific  features may be used to determine experimentally the (proper) time dependence of the magnetic fields created in HICs. The fact that positive and negative charges behaves differently is related to the fact that electromagnetic processes break the isospin symmetry of the original action.
\subsubsection{Comparison between dielectrons and dimuons production rates}\label{sec5b3}
\begin{figure}[tbp]\centering
\includegraphics[width=8.4cm,height=6cm]{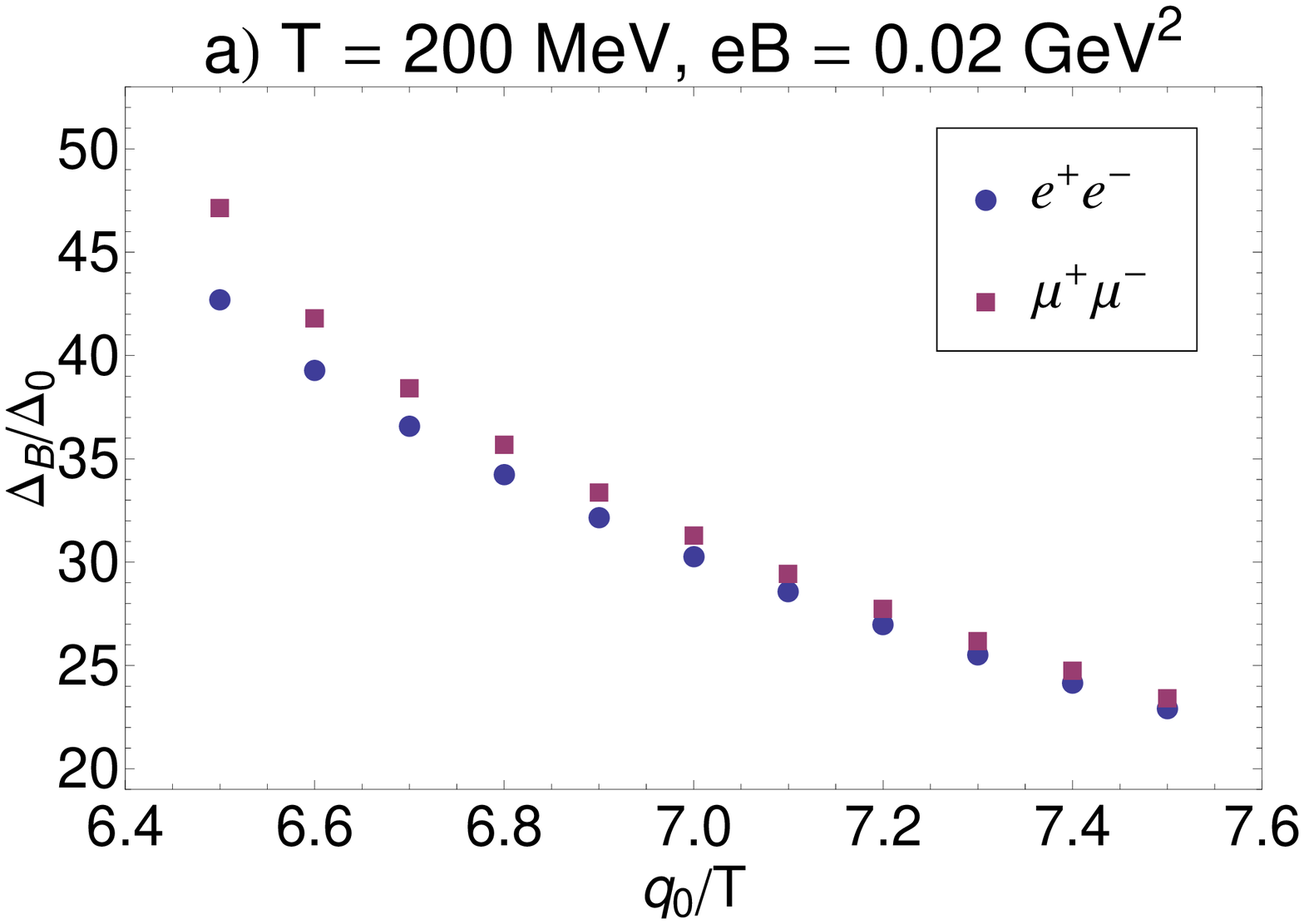}
\hfill
\includegraphics[width=8.4cm,height=6cm]{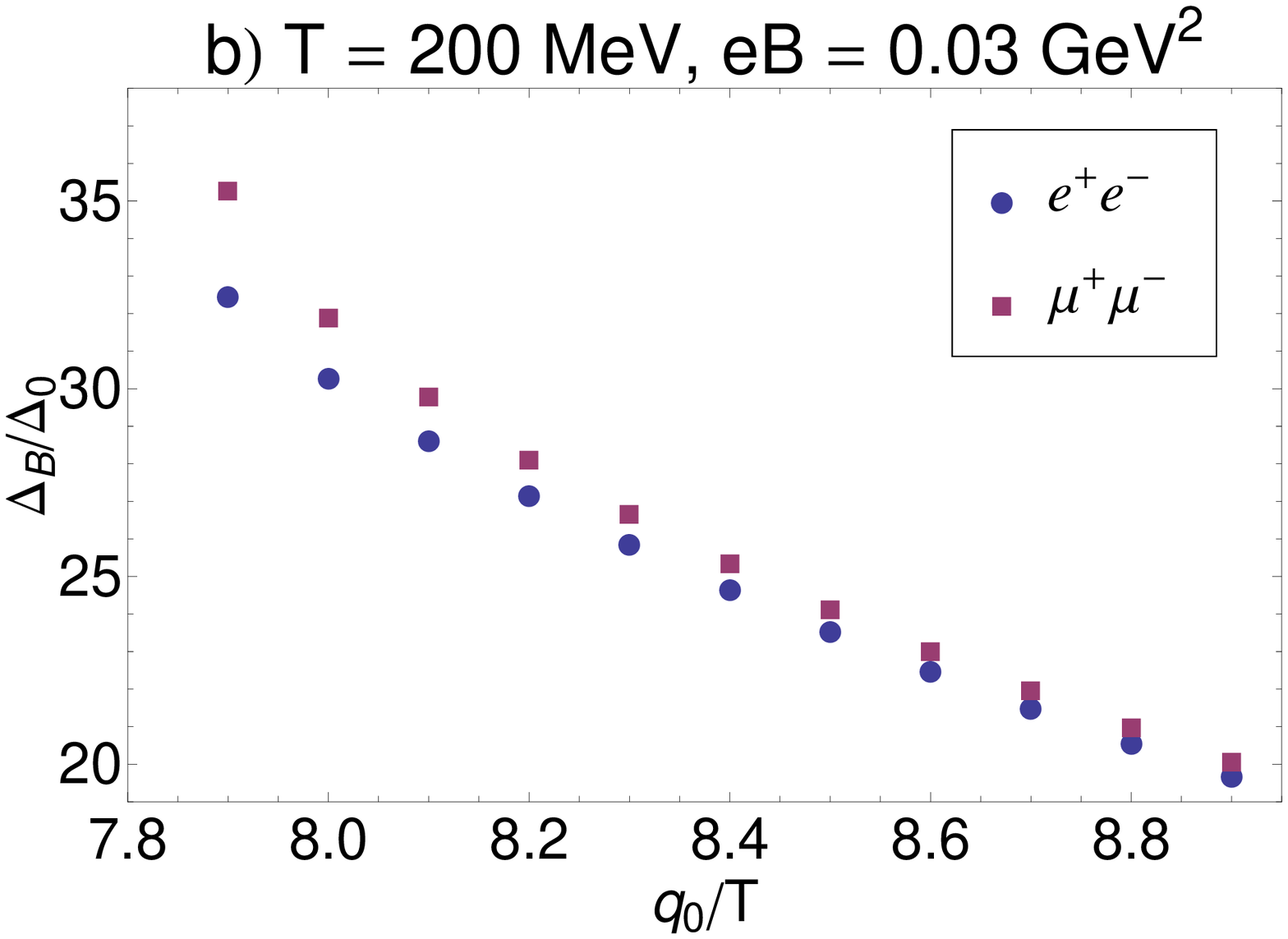}
\caption{(color online). The ratio $\Delta_{B}/\Delta_{0}$ for the production of dielectrons (blue circles) and dimuons (purple squares) are compared for $T=200$ MeV and $eB=0.02$ GeV$^{2}$ (panel a) and $eB=0.03$ GeV$^{2}$ (panel b). Only for $\frac{q_{0}}{T}\sim[\frac{q_{0}}{T}]_{th}$, i.e. at the beginning of the interval for which the curves are plotted, the results are slightly different (for more details see section \ref{sec5b3}).}\label{fig6}
\end{figure}
\begin{figure}[tbp]\centering
\includegraphics[width=8.4cm,height=6cm]{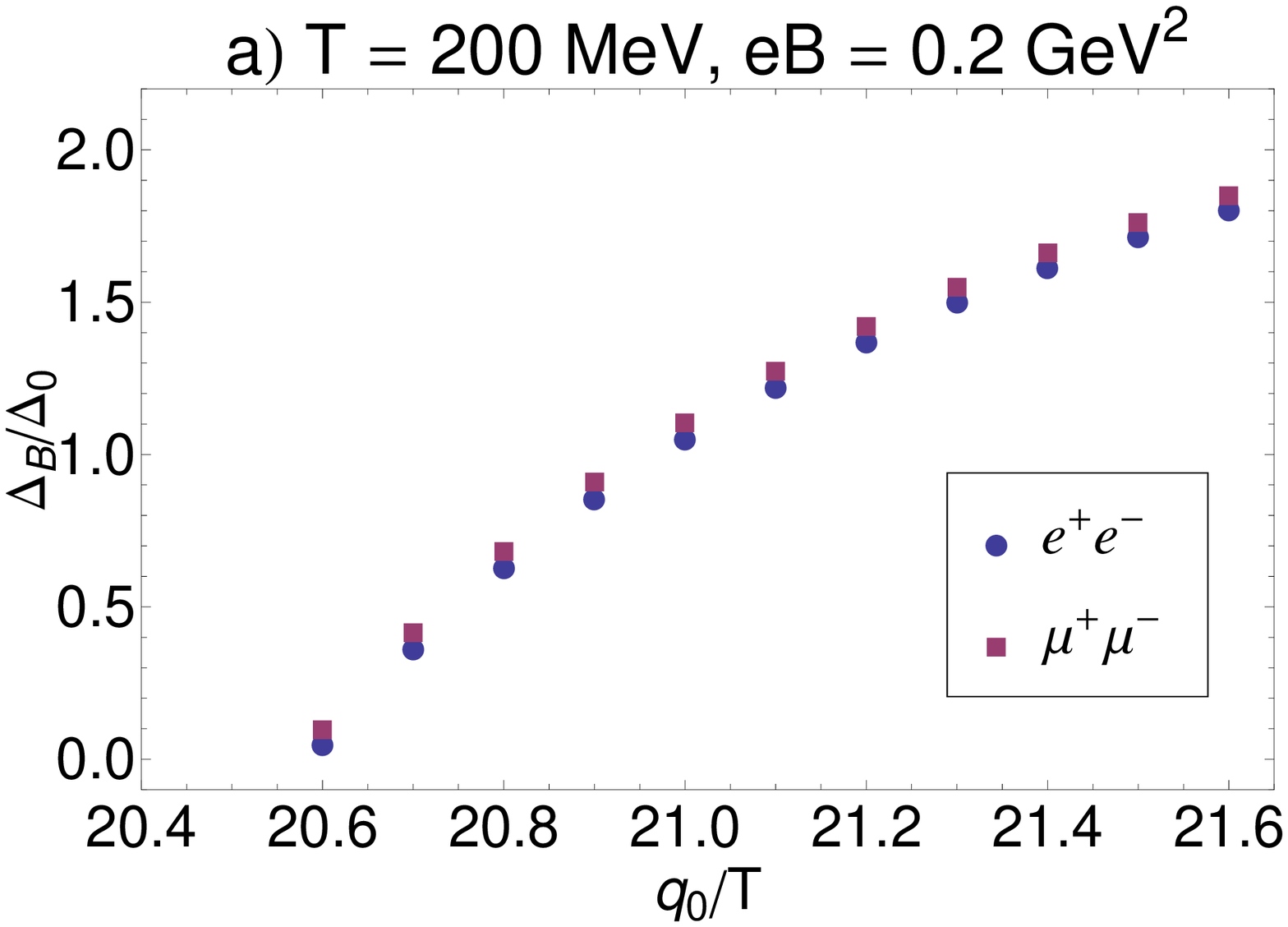}
\hfill
\includegraphics[width=8.4cm,height=6cm]{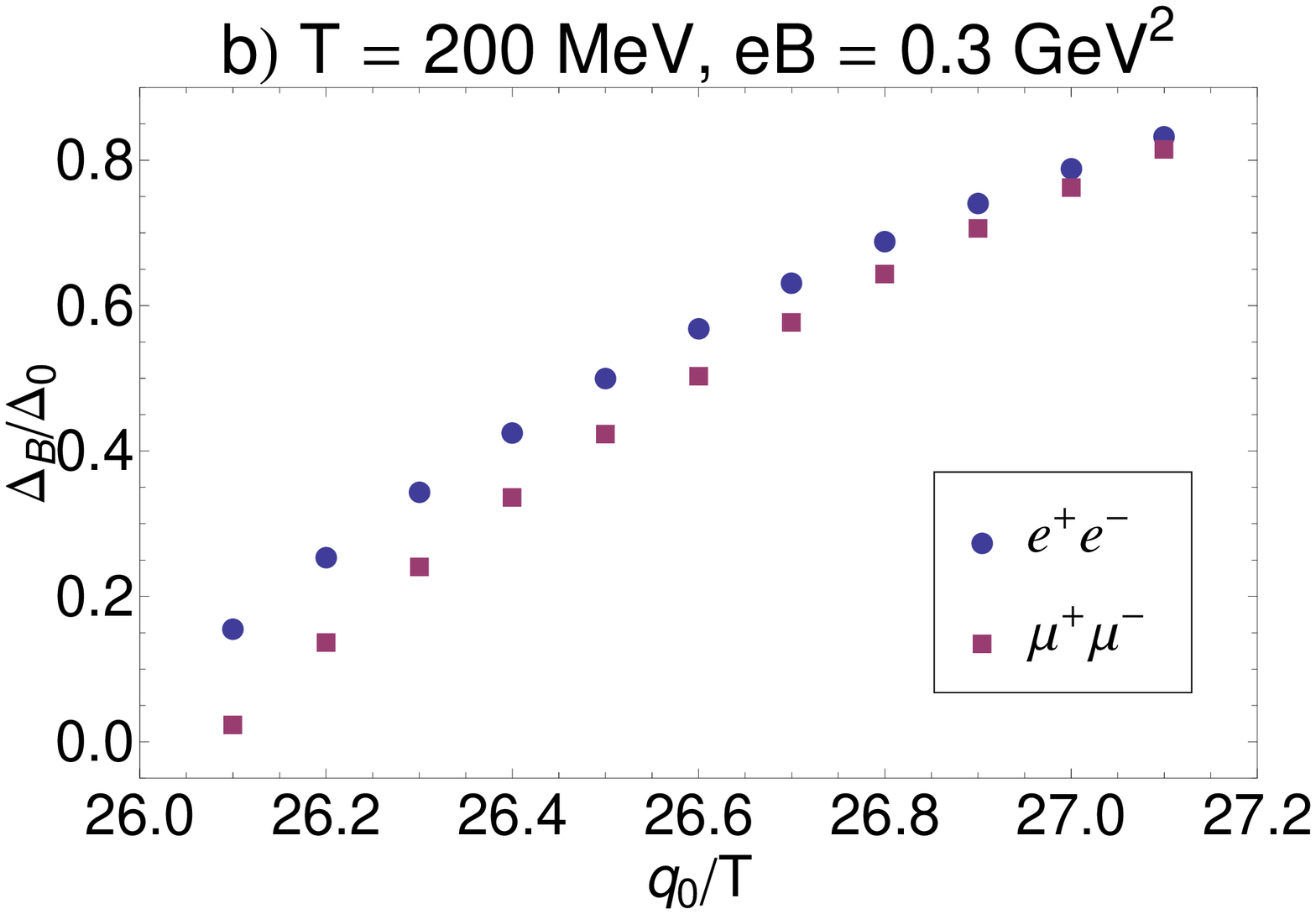}
\caption{(color online). The ratio $\Delta_{B}/\Delta_{0}$ for the production of dielectrons (blue circles) and dimuons (purple squares) are compared for $T=200$ MeV and $eB=0.2$ GeV$^{2}$ (panel a) and $eB=0.3$ GeV$^{2}$ (panel b). Only for $\frac{q_{0}}{T}\sim[\frac{q_{0}}{T}]_{th}$, i.e. at the beginning of the interval for which the curves are plotted, the results are slightly different (for more details see section \ref{sec5b3}).}\label{fig7}
\end{figure}
\par\noindent
As aforementioned, the ratio $\Delta_{B}/\Delta_{0}$ depends explicitly on the bare quark and lepton masses $m_{q}$ and $m_{\ell}$. To visualize the possible effect of different lepton (bare) masses on the ratio $\Delta_{B}/\Delta_{0}$, we have performed a similar analysis for dimuon pair production, as was previously carried out for dielectrons. In figures \ref{fig6} and \ref{fig7}, we have compared the $q_{0}/T$ dependence of $\Delta_{B}/\Delta_{0}$ for dielectrons (blue circles) and dimuons (purple squares) at $T=200$ MeV and
for $eB=0.02, 0.03$ GeV$^{2}$ and $eB=0.2,0.3$ GeV$^{2}$. As it turns out, the difference between the data corresponding to $e^{+}e^{-}$ and $\mu^{+}\mu^{-}$ maximizes in the vicinity of the threshold energies $[\frac{q_{0}}{T}]_{th}$, and quickly decreases with increasing $q_{0}/T$.
Quantitatively, these differences for all values of $eB$ and at $T=200$ MeV are, in general, between $\sim 10\%$ at the beginning, and $\sim 2\%$ at the end of the plotted interval. The same is also true for $T=400$ MeV for all values of $eB$. For large $eB=0.3$ GeV$^{2}$, in contrast to all the other cases, $\Delta_{B}/\Delta_{0}$ for dimuons is smaller than that of dielectrons.
\subsection{Dependence of the anisotropy factor ${\nu_{B}}$ on ${q_{0}/T}$}\label{sec5c}
\begin{figure}[tbp]\centering
\includegraphics[width=8.4cm,height=6cm]{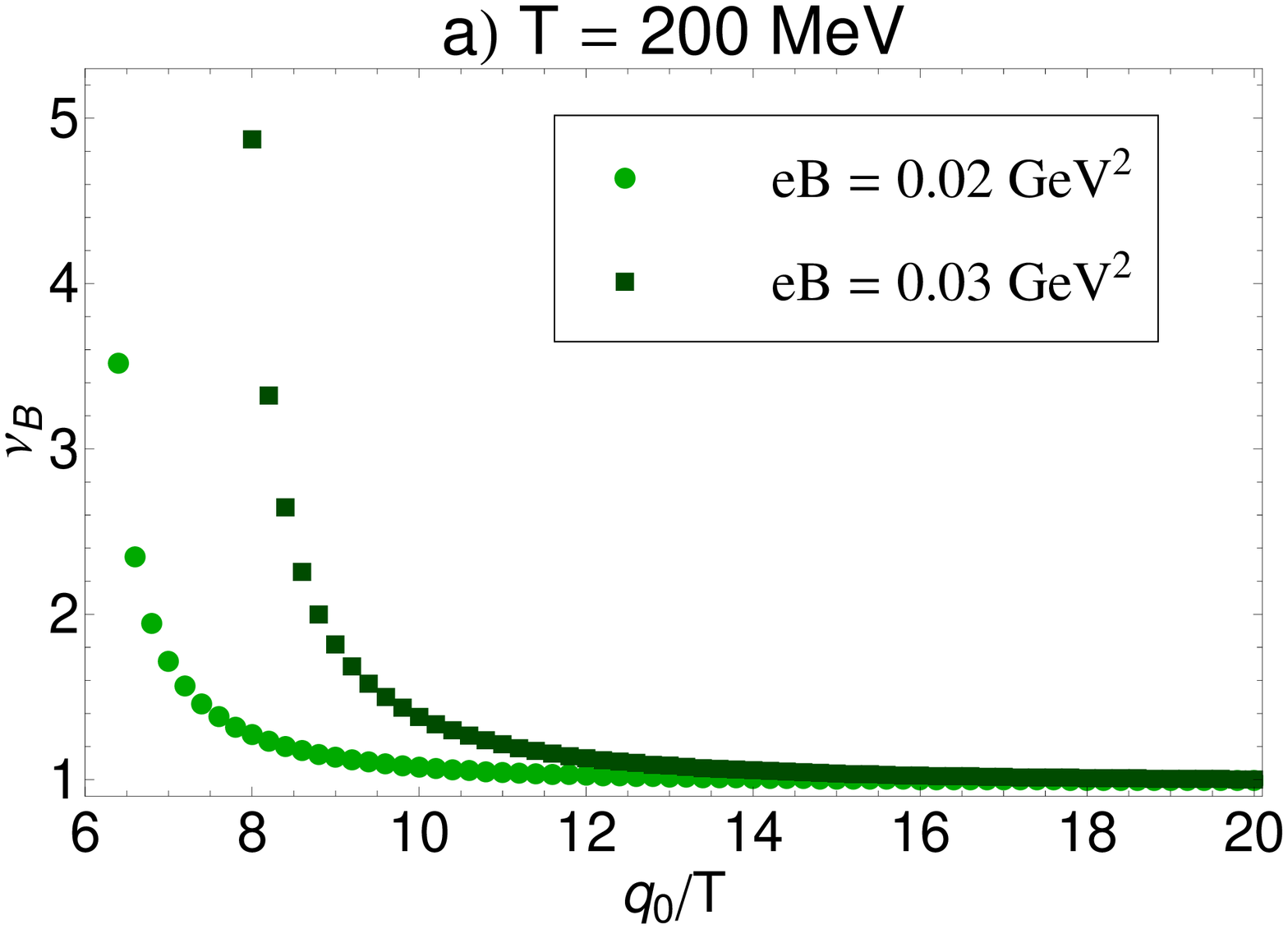}
\hfill
\includegraphics[width=8.4cm,height=6cm]{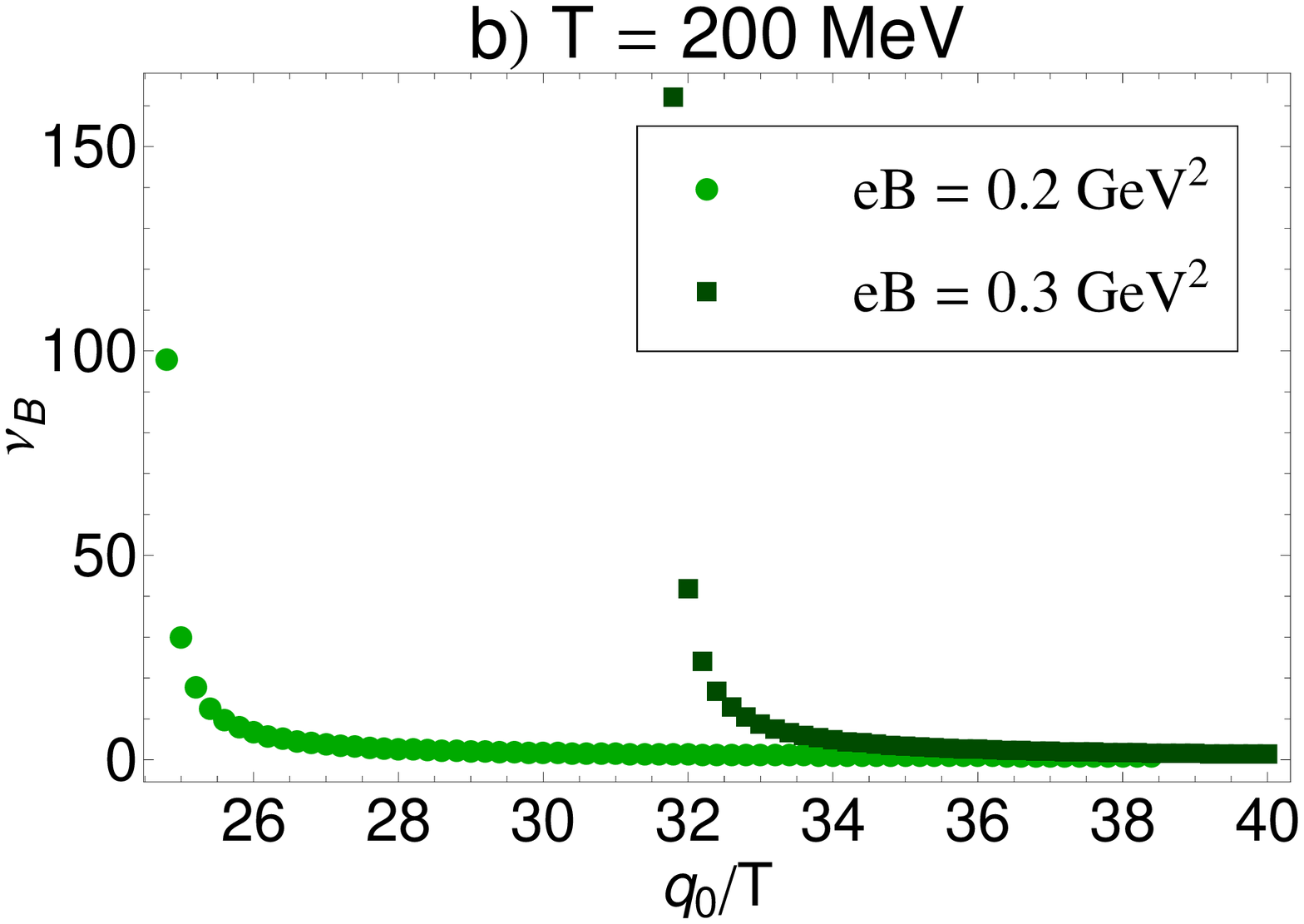}
\caption{(color online).  The anisotropy factor $\nu_{B}$ from (\ref{O9}) is plotted as a function of $q_{0}/T$ at $T=200$ MeV and $eB=0.02,0.03$ GeV$^{2}$ (panel a) and $eB=0.2,0.3$ GeV$^{2}$ (panel b). Light (dark) green circles (squares) correspond to $eB=0.02,0.2$ GeV$^{2}$ ($eB=0.03,0.3$ GeV$^{2}$). The results show that $\Delta_{B}^{\perp}>\Delta_{B}^{\|}$ as well as $|\Delta_{B}^{\perp}|>|\Delta_{B}^{\|}|$.}\label{fig8}
\end{figure}
\begin{figure}[tbp]\centering
\includegraphics[width=8.4cm,height=6cm]{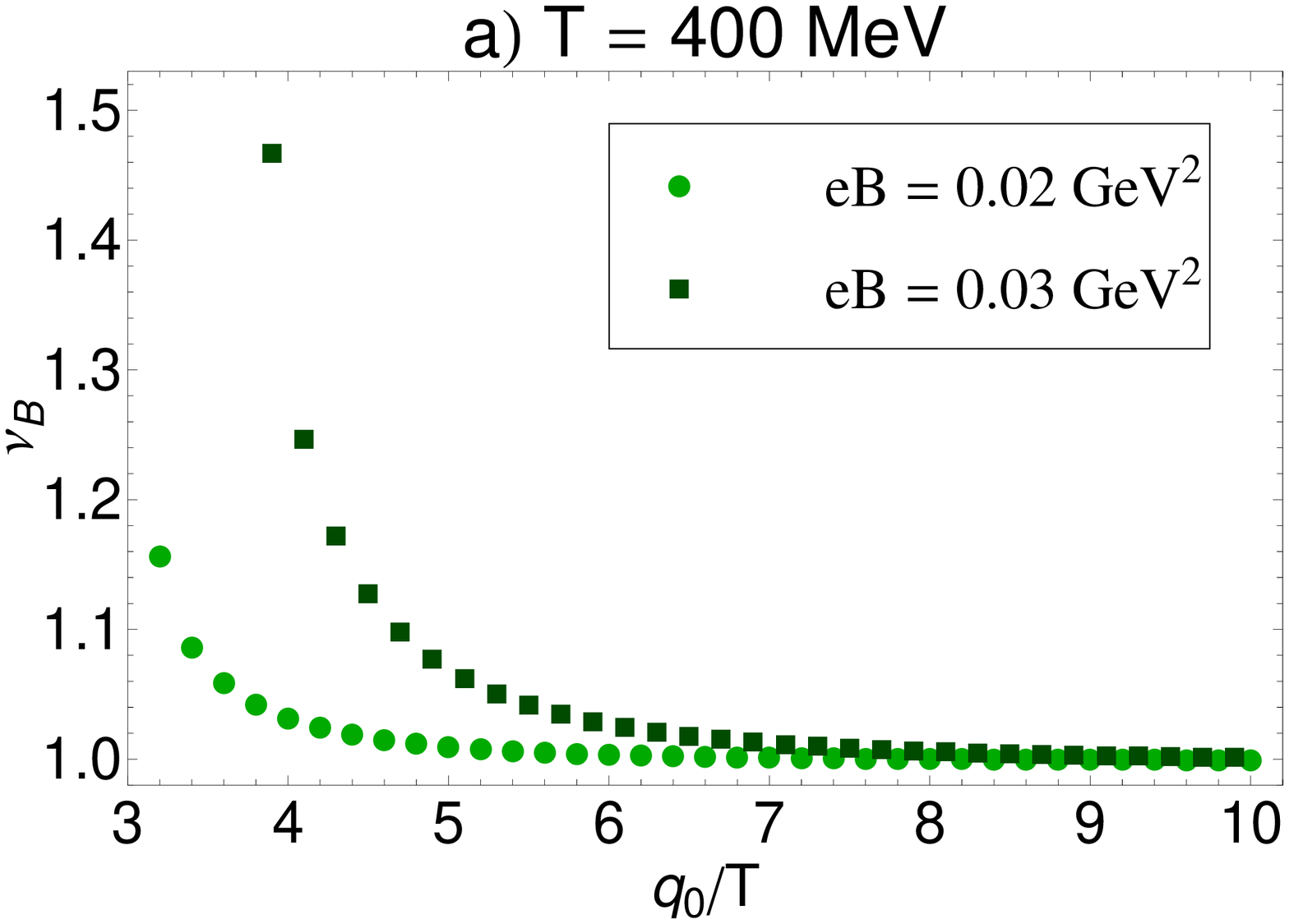}
\hfill
\includegraphics[width=8.4cm,height=6cm]{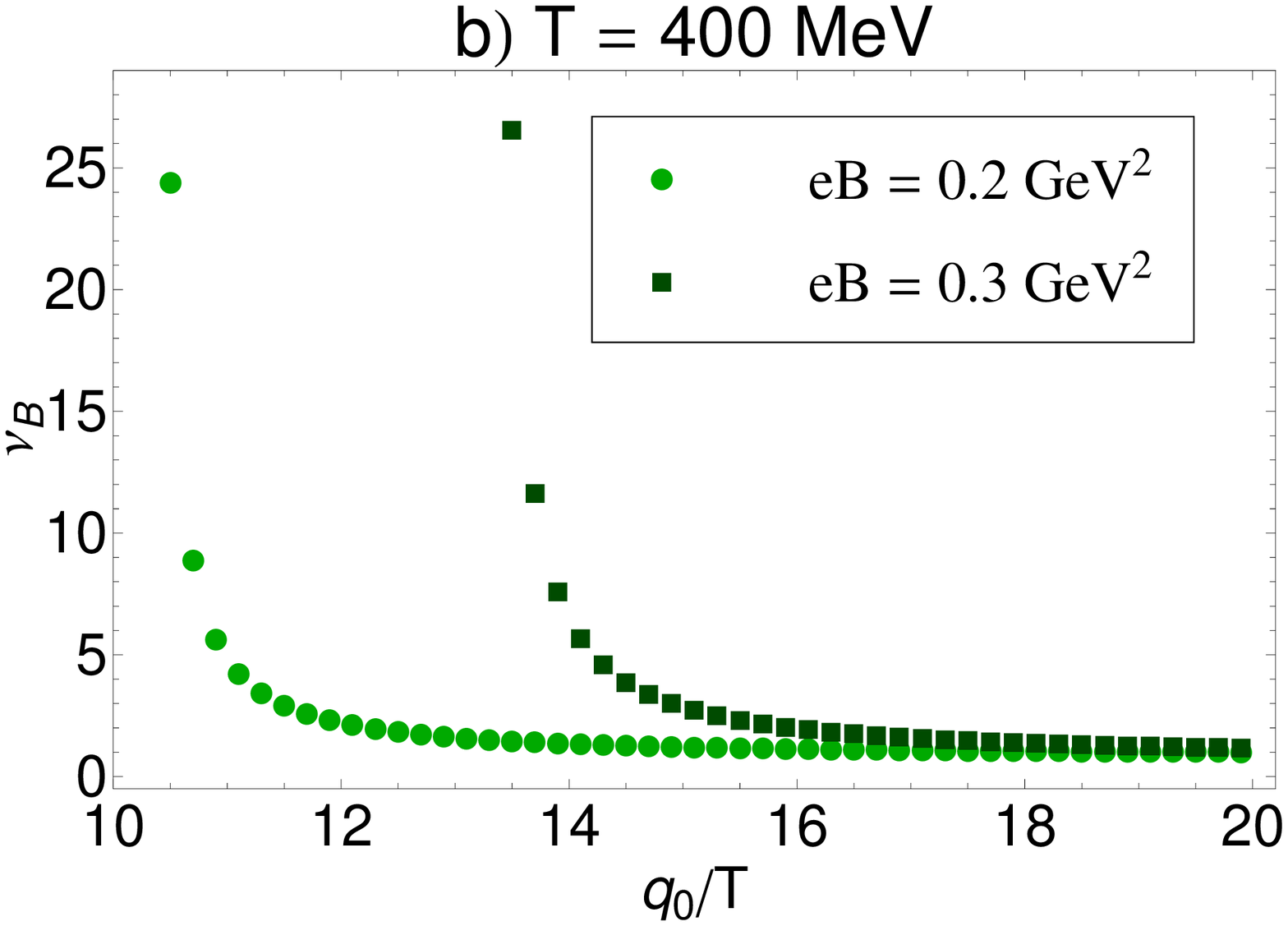}
\caption{(color online).  The anisotropy factor $\nu_{B}$ from (\ref{O9}) is plotted as a function of $q_{0}/T$ at $T=400$ MeV and $eB=0.02,0.03$ GeV$^{2}$ (panel a) and $eB=0.2,0.3$ GeV$^{2}$ (panel b). Light (dark) green circles (squares) correspond to $eB=0.02,0.2$ GeV$^{2}$ ($eB=0.03,0.3$ GeV$^{2}$). The results show that $\Delta_{B}^{\perp}>\Delta_{B}^{\|}$ as well as $|\Delta_{B}^{\perp}|>|\Delta_{B}^{\|}|$. }\label{fig9}
\end{figure}
In section \ref{sec4}, the analytical expression for $\Delta_{B}$ is presented in (\ref{D36})-(\ref{D41}). We have, in particular, shown that $\Delta_{B}$ receives contributions from two parts, $\Delta_{B}^{\|}$ and $\Delta_{B}^{\perp}$. As concerns the origin of these two contributions, let us remind that  $\Delta_{B}^{\|}$ and $\Delta_{B}^{\perp}$ arise from the $(\mu,\nu)=(\|,\|)$ and $(\mu,\nu)=(\perp,\perp)$ in the product of the photonic part $\mbox{Im}[\Pi_{\mu\nu}]$ and the leptonic part ${\cal{L}}^{\mu\nu}$ in (\ref{A34}). They are used to introduce the anisotropy factor $\nu_{B}$ in (\ref{O9}). The latter can presumably be brought in relation with the elliptic flow $v_{2}$, which is, by its part, a measure for the anisotropy in the particle distribution in the momentum space averaged over the whole volume in which the heavy ion experiment occurs.\footnote{Let us notice that, since the integration over this volume element is not performed in the present work, the anisotropy factor $\nu_{B}$ is defined for each volume element $d^{4}x$ separately. }
\par
\begin{figure}[tbp]\centering
\includegraphics[width=9cm,height=7cm]{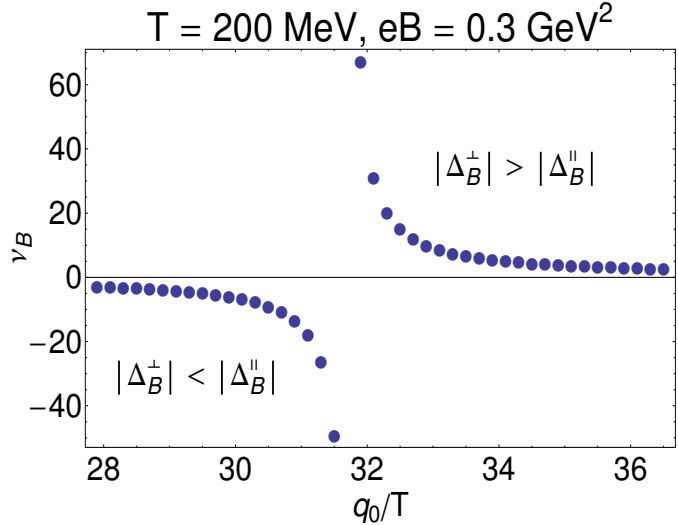}
\caption{(color online). The anisotropy factor $\nu_{B}$ from (\ref{O9}) is plotted at $T=200$ MeV and for $eB=0.3$ GeV$^{2}$ as a function of $q_{0}/T$. It turns out that in the interval $24.5<\frac{q_{0}}{T}<28$, the absolute values of $\Delta_{B}^{\|}$ and $\Delta_{B}^{\perp}$ have two different behaviors. In both cases $\Delta_{B}^{\perp}>\Delta_{B}^{\|}$.}\label{fig10}
\end{figure}

In figures \ref{fig8} and \ref{fig9}, the anisotropy factor $\nu_{B}$ is plotted as a function of $q_{0}/T$ at $T=200$ MeV and $T=400$ MeV for weak  and moderate magnetic fields $eB=0.02,0.03$ GeV$^{2}$ and $eB=0.2,0.3$ GeV$^{2}$, respectively. Light (dark) green circles (squares) correspond to $eB=0.02,0.2$ GeV$^{2}$ ($eB=0.03,0.3$ GeV$^{2}$). In all these cases, the factor $\nu_{B}$ has a maximum value for  $\frac{q_{0}}{T}$ at the beginning of the relevant intervals near $[\frac{q_{0}}{T}]_{th}$, and decreases relatively fast towards $\nu_{B}\sim 1$. A comparison of the results for a fixed value of $T$ and each fixed $q_{0}/T$ shows that $\nu_{B}$ increases with increasing $eB$. Let us notice that the fact that for the whole interval of $q_{0}/T$, the anisotropy factor $\nu_{B}\geq 1$ indicates that in this interval $\Delta_{B}^{\perp}>\Delta_{B}^{\|}$ as well as $|\Delta_{B}^{\perp}|>|\Delta_{B}^{\|}|$. This seems to be valid for all chosen $T$ and $eB$. However, after more critical scrutiny, it turns out that for strong magnetic field $eB=0.3$ GeV$^{2}$, at $T=200$ MeV, in the interval $24.5<\frac{q_{0}}{T}<40$, the absolute values of $\Delta_{B}^{\|}$ and $\Delta_{B}^{\perp}$ have two different behaviors: Whereas in the regime $24.5<\frac{q_{0}}{T}<32$, we have
$|\Delta_{B}^{\perp}|<|\Delta_{B}^{\|}|$, for $32<\frac{q_{0}}{T}$, we obtain $|\Delta_{B}^{\perp}|>|\Delta_{B}^{\|}|$ (see figure \ref{fig10} for a demonstration of this behavior). In both cases, we have  $\Delta_{B}^{\perp}>\Delta_{B}^{\|}$. Having in mind that $\Delta_{B}^{\|}$ is always negative, the regime $\frac{q_{0}}{T}<32$ seems therefore to be invalid, because in this case $\Delta_{B}=\Delta_{B}^{\|}+\Delta_{B}^{\perp}$ becomes negative.
The specific behavior of $\nu_{B}$ in the
physically relevant regime of $\frac{q_0}{T}$, e.g. $\frac{q_{0}}{T}\geq 32$ in figure \ref{fig10}, is mainly related to the fact that,
for our specific choice of free parameters and in our one-loop approximation, $\Delta_{B}^{\|}$ is always
negative, while $\Delta_{B}^{\perp}$ is always positive. Hence, replacing $\Delta_{B}^{\|}$ in (\ref{O9}) by $-|\Delta_{B}^{\|}|$, the numerator
of $\nu_{B}$ can potentially be larger than its denominator. This is exactly what happens in the vicinity of $\frac{q_{0}}{T}\simeq 32$
in figure \ref{fig10}. Whereas at $\frac{q_{0}}{T}\simeq 32$, the numerator $\Delta_{B}^{\perp}-\Delta_{B}^{\|}=\Delta_{B}^{\perp}+|\Delta_{B}^{\|}|$ of $\nu_{B}$ is up to one order of magnitude larger than its denominator 
$\Delta_{B}^{\perp}+\Delta_{B}^{\|}=\Delta_{B}^{\perp}-|\Delta_{B}^{\|}|$ of $\nu_{B}$, for $\frac{q_{0}}{T}\leq 36$, the numerator and denominator of $\nu_{B}$ are in the same order of magnitude. This leads, for instance, to the specific behavior of $\nu_B$ demonstrated in figure \ref{fig10} for $\frac{q_0}{T}\leq 32$.
\par
\begin{figure}[tbp]
\centering
\includegraphics[width=10cm,height=7cm]{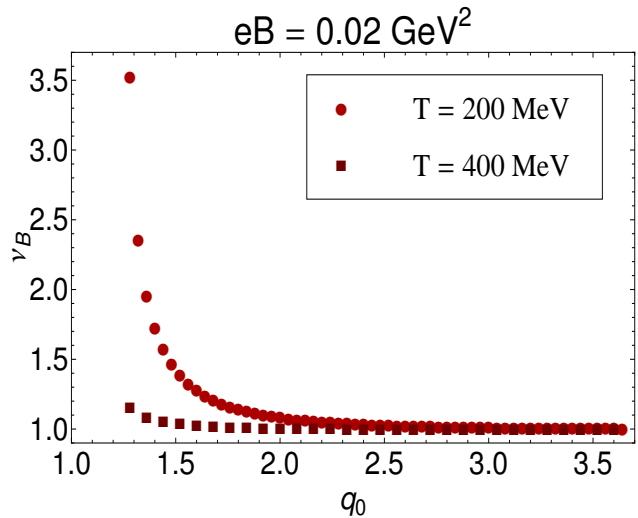}
\caption{(color online). The anisotropy factor $\nu_{B}$ from (\ref{O9}) is plotted at $eB=0.02$ GeV$^{2}$ and for $T=200$ MeV (light red circles) and $T=400$ MeV (dark red squares) as a function of $q_{0}$. It turns out that in the vicinity of threshold energy of virtual photons, $\nu_{B}$ decreases with increasing temperature.}\label{fig11}
\end{figure}

As concerns the effect of temperature on $\nu_{B}$, in figure \ref{fig11}, the anisotropy factor $\nu_{B}$ is plotted as a function of $q_{0}$ for fixed $eB=0.02$ GeV$^{2}$ and two different temperatures, $T=200$ MeV (light red circles) and $T=400$ MeV (dark red squares). It turns out that whereas for each fixed $\frac{q_{0}}{T}\sim [\frac{q_{0}}{T}]_{th}$, the anisotropy factor decreases with increasing temperature, for large enough $q_{0}$, it remains constant $\nu_{B}\sim 1$. In other words, the results presented in figure \ref{fig11}  show that, with decreasing $T$, the contribution of low energetic dileptons to $\nu_{B}$ increases, while the contribution of high energetic dileptons to $\nu_{B}$ remains constant.

\section{Summary and conclusions}\label{sec6}
\setcounter{equation}{0}
\par\noindent
Virtual photons (dileptons) are among the important electromagnetic probes which are used to reveal information on the evolution of the fireball of hot and dense QCD matter produced after (ultra-) relativistic HICs. It is, in particular, known that in early stages of noncentral collisions, very strong and time-dependent magnetic fields are produced  \cite{skokov2009, kharzeev2007}. Because of the finite electrical conductivity of the medium, these background magnetic fields are assumed to be constant, and aligned in a fixed spatial direction \cite{tuchin2013, rajagopal2014}. The latter feature is the origin of various anisotropies, among others, those appearing in the production rates of real photons in noncentral HICs; According to the arguments presented in \cite{kharzeev2012, fukushima2015}, these kinds of anisotropies are directly reflected in the transverse momentum dependence of flow coefficients, in particular, in the elliptic flow $v_{2}$ of real photons \cite{kharzeev2012} and heavy quarks \cite{fukushima2015}.
\par
In the present paper, we have, in particular, focused on the effect of constant and spatially fixed magnetic fields on dilepton production rates in a hot QGP. Following the systematic method presented in \cite{weldon1990}, where the differential multiplicity of dileptons $\Delta_{0}$ was computed in the absence of magnetic fields, we have derived the general structure of this quantity in a hot and magnetized QGP, denoted by $\Delta_{B}$ [see (\ref{A13}) for $\Delta_{0}$ and (\ref{A34}) for $\Delta_{B}$]. We have assumed that hot quarks and leptons which are involved in the process are magnetized. Similar to $\Delta_{0}$, $\Delta_{B}$ consists of a trace over the product of a leptonic part, ${\cal{L}}_{\mu\nu}$, and the imaginary part of the vacuum polarization tensor $\Pi^{\mu\nu}$. Both tensors are expressed in terms of a number of bases, presented in (\ref{A24}) and (\ref{D10}). They are separated into four groups, depending on whether their $\mu$ and $\nu$ indices are parallel or perpendicular to the direction of the background $B$ field. The fact that only two combinations of $\mu$ and $\nu$ indices, namely $(\mu,\nu)=(\|,\|)$ and $(\mu,\nu)=(\perp,\perp)$, survive the above mentioned trace operation leads to a separation of $\Delta_{B}$ into two parts, $\Delta_{B}^{\|}$ and $\Delta_{B}^{\perp}$ [see (\ref{D36})].
Combining these two contributions, a novel anisotropy factor $\nu_{B}$ is introduced in (\ref{O9}). The dependence of $\nu_{B}$ on the energy of (virtual) photons is studied in section \ref{sec5}. Here, it is shown that for each fixed $T$ and $B$, near a certain photon threshold energy $[q_{0}/T]_{th}$, we have $\Delta_{B}^{\perp}\gg \Delta_{B}^{\|}$. It is further shown that for fixed $T$ ($eB$), in the vicinity of these threshold energies, $\nu_{B}$ increases (decreases) with increasing $eB$ ($T$), while for $q_{0}\gg T$, it remains almost constant. Having in mind that dilepton production rates are directly related to flow coefficients $v_{n}$ of QGP, this novel anisotropy between $\Delta_{B}^{\perp}$ and $\Delta_{B}^{\|}$ may be interpreted as one of the microscopic sources of the observable macroscopic anisotropies, reflecting themselves in these coefficients.
\par
In section \ref{sec5}, we have also computed the ratio $\Delta_{B}/\Delta_{0}$ as a function of rescaled photon energy $q_{0}/T$ for various fixed $T$ and/or $eB$. We were mainly interested in the possible interplay between these two parameters on this ratio. We have, in particular, determined the (rescaled) threshold energies, $[q_{0}/T]_{th}$, for weak and moderate magnetic fields, $eB=0.02,0.03$ GeV$^{2}$ and $eB=0.2,0.3$ GeV$^{2}$ at $T=200,400$ MeV, and shown that at a fixed $T$, $[q_{0}/T]_{th}$ increases with increasing $eB$.\footnote{Threshold energies $[q_{0}]_{th}$ are different from the ``minimum'' energy which is necessary for producing dileptons. The latter is determined by LLL and is constant in $T$ and $eB$ (see the detailed discussion in section \ref{sec5a}).} As concerns the $T$ dependence of $[q_{0}]_{th}$ for fixed $eB$, however, we have observed that only for moderate magnetic fields $[q_{0}]_{th}$ decreases with increasing $T$, while for weak magnetic fields, it remains almost constant once $T$ increases.
\par
The final result for the one-loop approximated $\Delta_{B}$, presented in (\ref{D36})-(\ref{D41}), includes a summation over four Landau levels $n,k,p$ and $\ell$ from zero to a certain upper limit. In section \ref{sec5}, the dependence of our final numerical results on the choice of this upper limit is studied. Our detailed numerical analysis has, in particular, shown that for small values of $q_{0}/T$, once the magnetic field is chosen to be large enough, only the lowest Landau level contributes to $\Delta_{B}$. We quantified this statement in the example of $eB=0.02,0.03$ GeV$^{2}$ and $eB=0.2,0.3$ GeV$^{2}$, which are chosen so that they are approximately equal to the estimated values of $eB$, which is believed to be produced in RHIC and LHC.
\par
Our analytical results for $\Delta_{B}$ has another noteworthy feature, which clarifies the role played by positively and negatively charged quarks in producing dileptons. According to our final results from (\ref{D38}), positively charged quarks, e.g. up quarks, do not contribute to $\Delta_{B}^{\perp}$. Moreover, as it turns out, the total contribution of positively charged quarks to $\Delta_{B}$ is always smaller than that of negatively charged quarks. We have, in particular, shown that the interplay between the total contribution of positively and negatively charged quarks to $\Delta_{B}$ leads, for large enough magnetic fields, to a maximum value of $\Delta_{B}/\Delta_{0}$ in photon energies that are greater than certain photon threshold energies $[q_{0}/T]_{th}$. We have performed the above analysis for dielectrons as well as for dimuons, and shown that the difference between $\Delta_{B}/\Delta_{0}$ corresponding to dielectrons and dimuons maximizes in the vicinity of the corresponding thresholds $[q_{0}/T]_{th}$, and almost vanishes for $q_{0}\gg T$. In both cases, for $q_{0}$ near certain thresholds $[q_{0}/T]_{th}$, the production rate $\Delta_{B}$ in the presence of magnetic fields is larger than $\Delta_{0}$ for zero $eB$, while in the limit $q_{0}\gg T$, we have $\Delta_{B}\leq \Delta_{0}$.
\par
Let us remind at this stage that the analytical results presented in section \ref{sec4} for $\Delta_{B}$ are derived with the assumption that the background magnetic field and temperature are constant (in space and time), and that dileptons created in final stages of the process are hot and magnetized. The latter assumption is made because it is not clear how the lifetime of photons is affected by external magnetic fields. If it is shortened, then photons will not have enough time to escape the hot and magnetized medium before they are converted into dileptons, and, our computation will be then relevant. As concerns the first assumption, there are evidences that magnetic fields produced in HICs are time-dependent \cite{skokov2009, kharzeev2007}.
Thus, in a more realistic scenario, assuming that the lifetime of photons is not influenced by the background $B$ fields, in the stage in which dileptons are produced, the background magnetic field will be, most probably, rather weak and cannot affect the dynamics of dileptons.
Only in this case, the assumption concerning the effect of $B$ fields on created dileptons can be relaxed. Starting from this new assumption, the leptonic part ${\cal{L}}^{\mu\nu}$, appearing in (\ref{A34}), is to be replaced with $L_{\mu\nu}$ from (\ref{A3}). It would be interesting to explore the impact of this new assumption on the final results for the $q_{0}$-dependence of the ratio $\Delta_{B}/\Delta_{0}$ and the anisotropy factor $\nu_{B}$ for fixed $T$ or $B$.
\par
The analyses presented in this paper can be extended in many ways.
In \cite{taghinavaz2015}, e.g., we have systematically explored the complete quasi-particle spectrum of a magnetized plasma at finite temperature. We have, in particular, shown that for fixed $eB\ll T^{2}$ and in specific regimes of fermion energies, new poles arise in the one-loop corrected fermion propagator, in addition to the expected normal modes. These collective excitations, referred to as hot magnetized plasminos, are expected to play a crucial role in the production rates of dileptons. The effect of hot plasminos on DPR are studied in \cite{pisarski1990}, where it is shown that unexpected sharp peaks appear in the spectrum of DPR. These structures are believed to provide a unique signature of deconfined collective quarks in a QCD plasma \cite{thoma1999}. It would be interesting to combine \cite{taghinavaz2015} with \cite{pisarski1990}, and to study the effect of hot magnetized plasminos on the spectrum of dileptons at finite temperature and in the presence of uniform magnetic fields. This will be postponed to future publications.

\acknowledgments
The authors are grateful for useful discussions with M. Shokri and S.M.A. Tabatabaee.

\begin{appendix}
\section{Additional formulae from sections \ref{sec3} and \ref{sec4}}\label{app}
\subsection{Additional formulae from section \ref{sec3}}\label{app1}
\setcounter{equation}{0}
\par\noindent
In section \ref{sec3}, we have argued that the leptonic part of photon-to-dilepton process is given by
\begin{eqnarray*}
{\cal{L}}^{\mu\nu}_{}=\frac{1}{2}\sum\limits_{n,k=0}^{\infty}\sum\limits_{i,j=1}^{4}C_{ij}^{nk}
\Psi^{\mu\nu}_{ij,nk},
\end{eqnarray*}
[see (\ref{A21})]. Here, the coefficients $C_{ij}^{nk}$ read
\begin{eqnarray}\label{A22}
\begin{array}{rclcrclcrclcrcl}
C_{11}^{nk}&\equiv&A_{+}^{(1)}B_{+}^{(1)}+A_{-}^{(1)}B_{-}^{(1)},&~~&
C_{12}^{nk}&\equiv&A_{+}^{(1)}B_{-}^{(1)}+A_{-}^{(1)}B_{+}^{(1)},&&&&&&&&\\
C_{21}^{nk}&\equiv&A_{+}^{(1)}B_{+}^{(2)},&&\hspace{-0.2cm}
C_{22}^{nk}&\equiv&A_{-}^{(1)}B_{-}^{(2)},&&\hspace{-0.5cm}
C_{23}^{nk}&\equiv&A_{+}^{(1)}B_{-}^{(2)},&\qquad&
C_{24}^{nk}&\equiv&A_{-}^{(1)}B_{+}^{(2)},\nonumber\\
C_{31}^{nk}&\equiv&A_{+}^{(2)}B_{+}^{(1)},&&\hspace{-0.2cm}
C_{32}^{nk}&\equiv&A_{-}^{(2)}B_{-}^{(1)},&&\hspace{-0.5cm}
C_{33}^{nk}&\equiv&A_{+}^{(2)}B_{-}^{(1)},&\qquad&
C_{34}^{nk}&\equiv&A_{-}^{(2)}B_{+}^{(1)},\nonumber\\
C_{41}^{nk}&\equiv&A_{+}^{(2)}B_{+}^{(2)},&&\hspace{-0.2cm}
C_{42}^{nk}&\equiv&A_{-}^{(2)}B_{-}^{(2)},&&\hspace{-0.5cm}
C_{43}^{nk}&\equiv&A_{+}^{(2)}B_{-}^{(2)},&\qquad&
C_{44}^{nk}&\equiv&A_{-}^{(2)}B_{+}^{(2)},
\end{array}
\nonumber\\
\end{eqnarray}
with
\begin{eqnarray}\label{A23}
\begin{array}{rclcrcl}
A_{+}^{(1)}&\equiv& f_{n}(\xi_{x}^{p_{2}}) f_{k}(\bar{\xi}_{x}^{p_{1}}),&\qquad&
A_{-}^{(1)}&\equiv&\Pi_{n}\Pi_{k} f_{n-1}(\xi_{x}^{p_{2}}) f_{k-1}(\bar{\xi}_{x}^{p_{1}}),\\
A_{+}^{(2)}&\equiv&\Pi_{n} f_{n-1}(\xi_{x}^{p_{2}}) f_{k}(\bar{\xi}_{x}^{p_{1}}),&\qquad&
A_{-}^{(2)}&\equiv&\Pi_{k} f_{n}(\xi_{x}^{p_{2}}) f_{k-1}(\bar{\xi}_{x}^{p_{1}}),\\
B_{+}^{(1)}&\equiv& f_{k}(\bar{\xi}_{y}^{p_{1}})f_{n}(\xi_{y}^{p_{2}}) ,&\qquad&
B_{-}^{(1)}&\equiv&\Pi_{n}\Pi_{k}  f_{k-1}(\bar{\xi}_{y}^{p_{1}})f_{n-1}(\xi_{y}^{p_{2}}),\\
B_{+}^{(2)}&\equiv&\Pi_{k}f_{k-1}(\bar{\xi}_{y}^{p_{1}}) f_{n}(\xi_{y}^{p_{2}}) ,&\qquad&
B_{-}^{(2)}&\equiv&\Pi_{n}f_{k}(\bar{\xi}_{y}^{p_{1}}) f_{n-1}(\xi_{y}^{p_{2}}).
\end{array}
\end{eqnarray}
The tensor part of ${\cal{L}}^{\mu\nu}$ is given in (\ref{A24}).
\subsection{Additional formulae from section \ref{sec4}}\label{app2a}
\subsubsection{Additional formulae from section \ref{sec4A}}
\par\noindent
In section \ref{sec4A}, we have argued that the photonic part of $\Delta_{B}$ can be first expressed as
\begin{eqnarray*}
{\cal{T}}_{\mu\nu}^{(q)}=\sum_{i,j=1}^{4}A_{ij}^{(q)}~\Xi_{\mu\nu}^{(q)ij}.
\end{eqnarray*}
Here, the coefficients ${\cal{A}}_{ij}^{(q)}$ are given by
\begin{eqnarray}\label{D7}
\begin{array}{rclcrcl}
{\cal{A}}_{11}^{(q)}&\equiv&2\left({\cal{A}}_{+}^{1(q)}{\cal{B}}_{+}^{1(q)}+{\cal{A}}_{-}^{1(q)}{\cal{B}}_{-}^{1(q)}\right),&\qquad&
{\cal{A}}_{12}^{(q)}&\equiv&2\left({\cal{A}}_{+}^{1(q)}{\cal{B}}_{-}^{1(q)}+{\cal{A}}_{-}^{1(q)}{\cal{B}}_{+}^{1(q)}\right),\\
{\cal{A}}_{21}^{(q)}&\equiv&2{\cal{A}}_{+}^{1(q)}{\cal{B}}_{+}^{2(q)},&\qquad&
{\cal{A}}_{22}^{(q)}&\equiv&2{\cal{A}}_{-}^{1(q)}{\cal{B}}_{-}^{2(q)},\\
{\cal{A}}_{23}^{(q)}&\equiv&2{\cal{A}}_{+}^{1(q)}{\cal{B}}_{-}^{2(q)},&\qquad&
{\cal{A}}_{24}^{(q)}&\equiv&2{\cal{A}}_{-}^{1(q)}{\cal{B}}_{+}^{2(q)},\\
{\cal{A}}_{31}^{(q)}&\equiv&2{\cal{A}}_{+}^{2(q)}{\cal{B}}_{+}^{1(q)},&\qquad&
{\cal{A}}_{32}^{(q)}&\equiv&2{\cal{A}}_{-}^{2(q)}{\cal{B}}_{-}^{1(q)},
\\
{\cal{A}}_{33}^{(q)}&\equiv&2{\cal{A}}_{+}^{2(q)}{\cal{B}}_{-}^{1(q)},&\qquad&
{\cal{A}}_{34}^{(q)}&\equiv&2{\cal{A}}_{-}^{2(q)}{\cal{B}}_{+}^{1(q)},
\\
{\cal{A}}_{41}^{(q)}&\equiv&2{\cal{A}}_{+}^{2(q)}{\cal{B}}_{+}^{2(q)},&\qquad&
{\cal{A}}_{42}^{(q)}&\equiv&2{\cal{A}}_{-}^{2(q)}{\cal{B}}_{-}^{2(q)},
\\
{\cal{A}}_{43}^{(q)}&\equiv&2{\cal{A}}_{+}^{2(q)}{\cal{B}}_{-}^{2(q)},&\qquad&
{\cal{A}}_{44}^{(q)}&\equiv&2{\cal{A}}_{-}^{2(q)}{\cal{B}}_{+}^{2(q)}.
\end{array}
\end{eqnarray}
In the above expressions, $A_{\pm}^{1/2(q)}$ and  $B_{\pm}^{1/2(q)}$ are explicitly given by
\begin{eqnarray}\label{D8}
\begin{array}{rclcrcl}
{\cal{A}}^{1(+)}_{+}&=&f_{p}(\xi_{x}^{p})f_{\ell}(\xi_{x}^{\ell}),&\qquad&
{\cal{A}}^{1(+)}_{-}&=&\Pi_{p}\Pi_{\ell}f_{p-1}(\xi_{x}^{p})f_{\ell-1}(\xi_{x}^{\ell}),\\
{\cal{A}}^{2(+)}_{+}&=&\Pi_{p}f_{p-1}(\xi_{x}^{p})f_{\ell}(\xi_{x}^{\ell}),&\qquad&
{\cal{A}}^{2(+)}_{-}&=&\Pi_{\ell}f_{p}(\xi_{x}^{p})f_{\ell-1}(\xi_{x}^{\ell}),\\
{\cal{B}}^{1(+)}_{+}&=&f_{\ell}(\xi_{y}^{\ell})f_{p}(\xi_{y}^{p}),&\qquad&
{\cal{B}}^{1(+)}_{-}&=&\Pi_{p}\Pi_{\ell}f_{\ell-1}(\xi_{y}^{\ell})f_{p-1}(\xi_{y}^{p}),\\
{\cal{B}}^{2(+)}_{+}&=&\Pi_{\ell}f_{\ell-1}(\xi_{y}^{\ell})f_{p}(\xi_{y}^{p}),&\qquad&
{\cal{B}}^{2(+)}_{-}&=&\Pi_{p}f_{\ell}(\xi_{y}^{\ell})f_{p-1}(\xi_{y}^{p}),\\
\end{array}
\end{eqnarray}
for positive charged particles, and
\begin{eqnarray}\label{D9}
\begin{array}{rclcrcl}
{\cal{A}}^{1(-)}_{+}&=&f_{p-1}(\bar{\xi}_{x}^{p})f_{\ell-1}(\bar{\xi}_{x}^{\ell}),
&\qquad&
{\cal{A}}^{1(-)}_{-}&=&\Pi_{p}\Pi_{\ell}f_{p}(\bar{\xi}_{x}^{p})f_{\ell}(\bar{\xi}_{x}^{\ell}),\\
{\cal{A}}^{2(-)}_{+}&=&\Pi_{p}f_{p}(\bar{\xi}_{x}^{p})f_{\ell-1}(\bar{\xi}_{x}^{\ell}),&\qquad&
{\cal{A}}^{2(-)}_{-}&=&\Pi_{\ell}f_{p-1}(\bar{\xi}_{x}^{p})f_{\ell}(\bar{\xi}_{x}^{\ell}),\\
{\cal{B}}^{1(-)}_{+}&=&f_{\ell-1}(\bar{\xi}_{y}^{\ell})f_{p-1}(\bar{\xi}_{y}^{p}),&\qquad&
{\cal{B}}^{1(-)}_{-}&=&\Pi_{p}\Pi_{\ell}f_{\ell}(\bar{\xi}_{y}^{\ell})f_{p}(\bar{\xi}_{y}^{p}),\\
{\cal{B}}^{2(-)}_{+}&=&\Pi_{\ell}f_{\ell}(\bar{\xi}_{y}^{\ell})f_{p-1}(\bar{\xi}_{y}^{p}),&\qquad&
{\cal{B}}^{2(-)}_{-}&=&\Pi_{p}f_{\ell-1}(\bar{\xi}_{y}^{\ell})f_{p}(\bar{\xi}_{y}^{p}),\\
\end{array}
\end{eqnarray}
for negative charged particles. Here, as previously defined, $\xi_{x}^{p}\equiv \xi_{x,p}^{+}$ and $\bar{\xi}_{x}^{p}\equiv \xi_{x,p}^{-}$ and $\xi_{x,p}^{\pm}$ is given in (\ref{S5}). The tensor part of ${\cal{T}}_{\mu\nu}^{(q)}$ is presented in (\ref{D10}).
\subsubsection{Final result for the imaginary part of $[\Pi_{\mu\nu}]_{ij}^{nk}$}\label{app2b}
\par\noindent
In this part, we present the final result for the imaginary part of $[\Pi_{\mu\nu}]_{ij}^{nk}$ from \ref{sec4A}. For $(\mu,\nu)=(\|,\|)$ components, we arrive at
\begin{eqnarray}\label{D24}
\mbox{Im}[\Pi_{\mu\nu}]_{11}^{nk}&=&-\frac{N_{c}\alpha (e^{\beta q_{0}}-1)}{4(\qpar)^{2}}\sum_{q_{f}=\{u,d\}}q_{f}^{2}\sum_{p,\ell=0}^{\infty}\frac{\eta_{nkp\ell}^{(q)}{\cal{K}}_{11}^{(q)}}{\xi_{q}}
\left(
\begin{array}{cccc}
X_{00}^{11}&0&0&X_{03}^{11}\\
0&0&0&0\\
0&0&0&0\\
X_{30}^{11}&0&0&X_{33}^{11}
\end{array}
\right), \nonumber\\
\mbox{Im}[\Pi_{\mu\nu}]_{12}^{nk}&=&-\frac{N_{c}\alpha (e^{\beta q_{0}}-1)}{2}\sum_{q_{f}=\{u,d\}}q_{f}^{2}\sum_{p,\ell=0}^{\infty}\frac{\eta_{nkp\ell}^{(q)}\tilde{p}_{q}^{(2)}\tilde{\ell}_{q}^{(2)}
{\cal{K}}_{12}^{(q)}}{\xi_{q}}{\cal{N}}^{(q)}
g_{\mu_{\|}\nu_{\|}},
\end{eqnarray}
with the elements of the $X$-matrix, $X_{\mu\nu}^{11}\equiv {\cal{A}}_{\mu\nu}^{11}{\cal{N}}_{+}^{(q)}+{\cal{B}}_{\mu\nu}^{11}{\cal{N}}_{-}^{(q)}$,
and
\begin{eqnarray*}
{\cal{A}}_{00}^{11}&=&\xi_{q}[-\qpar(q_{0}^{2}+q_{3}^{2}+2q_{0}q_{3} {\cal{R}}_{q})+(q_{0}+q_{3}{\cal{R}}_{q})^{2}\xi_{q}]-4\ell (\qpar)^{2}|q_{f}eB|,\nonumber\\
{\cal{A}}_{03}^{11}&=&\xi_{q}[-\qpar(2q_{0}q_{3}+(q_{0}^{2}+q_{3}^{2}){\cal{R}}_{q})+(q_{0}+q_{3}{\cal{R}}_{q})(q_{3}+q_{0}{\cal{R}}_{q})\xi_{q}],\nonumber\\
{\cal{A}}_{30}^{11}&=&\xi_{q}[-\qpar(2q_{0}q_{3}+(q_{0}^{2}+q_{3}^{2}){\cal{R}}_{q})+(q_{0}+q_{3}{\cal{R}}_{q})(q_{3}+q_{0}{\cal{R}}_{q})\xi_{q}],\nonumber\\
{\cal{A}}_{33}^{11}&=&\xi_{q}[-\qpar(q_{0}^{2}+q_{3}^{2}+2q_{0}q_{3} {\cal{R}}_{q})+(q_{3}+q_{0}{\cal{R}}_{q})^{2}\xi_{q}]+4\ell (\qpar)^{2}|q_{f}eB|,\nonumber\\
\end{eqnarray*}
\begin{eqnarray}\label{D25}
{\cal{B}}_{00}^{11}&=&\xi_{q}[-\qpar(q_{0}^{2}+q_{3}^{2}-2q_{0}q_{3} {\cal{R}}_{q})+(q_{0}-q_{3}{\cal{R}}_{q})^{2}\xi_{q}]-4\ell (\qpar)^{2}|q_{f}eB|,\nonumber\\
{\cal{B}}_{03}^{11}&=&\xi_{q}[q_{0}^{2}(\qpar-\xi_{q}){\cal{R}}_{q}+q_{3}^{2}(\qpar-\xi_{q}){\cal{R}}_{q}+q_{0}q_{3}(-2\qpar+\xi_{q}+{\cal{R}}_{q}^{2}\xi_{q})],\nonumber\\
{\cal{B}}_{30}^{11}&=&\xi_{q}[q_{0}^{2}(\qpar-\xi_{q}){\cal{R}}_{q}+q_{3}^{2}(\qpar-\xi_{q}){\cal{R}}_{q}+q_{0}q_{3}(-2\qpar+\xi_{q}+{\cal{R}}_{q}^{2}\xi_{q})],\nonumber\\
{\cal{B}}_{33}^{11}&=&\xi_{q}[-\qpar(q_{0}^{2}+q_{3}^{2}-2q_{0}q_{3} {\cal{R}}_{q})+(q_{3}-q_{0}{\cal{R}}_{q})^{2}\xi_{q}]+4\ell (\qpar)^{2}|q_{f}eB|.
\end{eqnarray}
Here,
\begin{eqnarray}\label{D26}
{\cal{N}}^{(q)}_{+}&=&\frac{N_{f}(E_{\ell}^{+})N_{f}(E_{p}^{+})}{{\cal{R}}_{q}}-\frac{\qpar N_{f}(E_{\ell'}^{+})N_{f}(E_{p'}^{+})}{|2q_{0}q_{3}+(q_{0}^{2}+q_{3}^{2}){\cal{R}}_{q}|},\nonumber\\
{\cal{N}}^{(q)}_{-}&=&\frac{N_{f}(E_{\ell}^{-})N_{f}(E_{p}^{-})}{{\cal{R}}_{q}}-\frac{\qpar N_{f}(E_{\ell'}^{-})N_{f}(E_{p'}^{-})}{|2q_{0}q_{3}-(q_{0}^{2}+q_{3}^{2}){\cal{R}}_{q}|},
\end{eqnarray}
and  ${\cal{N}}^{(q)}\equiv {\cal{N}}^{(q)}_{+}+{\cal{N}}^{(q)}_{-}$.
Here, $E_{\ell/\ell'}^{\pm}$ and $E_{p/p'}^{\pm}$ are given in (\ref{D21}).
Similarly, for $(\mu,\nu)=(\perp,\perp)$ components, we arrive at
\begin{eqnarray}\label{D27}
\mbox{Im}[\Pi_{\mu\nu}]_{41}^{nk}
&=&-\frac{N_{c}\alpha (e^{\beta q_{0}}-1)}{2}\sum_{q_{f}=\{u,d\}}q_{f}^{2}\sum_{p,\ell=0}^{\infty}\frac{\eta_{nkp\ell}^{(q)}\tilde{p}_{q}^{(2)}\tilde{\ell}_{q}^{(2)}
{\cal{K}}_{41}^{(q)}}{\xi_{q}}{\cal{N}}^{(q)}\nonumber\\ &&\times\big[2g^{2\mu_{\perp}}g^{2\nu_{\perp}}+g^{\mu_{\perp}\nu_{\perp}}-is_{q}(g^{1\mu_{\perp}}g^{2\nu_{\perp}}+g^{1\nu_{\perp}}g^{2\mu_{\perp}})\big],
\nonumber\\
\mbox{Im}[\Pi_{\mu\nu}]_{42}^{nk}&=&
-\frac{N_{c}\alpha (e^{\beta q_{0}}-1)}{2}\sum_{q_{f}=\{u,d\}}q_{f}^{2}\sum_{p,\ell=0}^{\infty}\frac{\eta_{nkp\ell}^{(q)}\tilde{p}_{q}^{(2)}\tilde{\ell}_{q}^{(2)}
{\cal{K}}_{42}^{(q)}}{\xi_{q}}{\cal{N}}^{(q)}\nonumber\\
&&\times\big[2g^{2\mu_{\perp}}g^{2\nu_{\perp}}+g^{\mu_{\perp}\nu_{\perp}}+is_{q}(g^{1\mu_{\perp}}g^{2\nu_{\perp}}+g^{1\nu_{\perp}}g^{2\mu_{\perp}})\big],\nonumber\\
\mbox{Im}[\Pi_{\mu\nu}]_{43}^{nk}&=&
-\frac{N_{c}\alpha (e^{\beta q_{0}}-1)}{4}\sum_{q_{f}=\{u,d\}}q_{f}^{2}\sum_{p,\ell=0}^{\infty}\frac{\eta_{nkp\ell}^{(q)}(\xi_{q}-4\ell|q_{f}eB|)
{\cal{K}}_{43}^{(q)}}{\xi_{q}}{\cal{N}}^{(q)}\nonumber\\
&&\times \big[g^{\mu_{\perp}\nu_{\perp}}+is_{q}(g^{1\mu_{\perp}}g^{2\nu_{\perp}}-g^{1\nu_{\perp}}g^{2\mu_{\perp}})\big],\nonumber\\
\mbox{Im}[\Pi_{\mu\nu}]_{44}^{nk}&=&
-\frac{N_{c}\alpha (e^{\beta q_{0}}-1)}{4}\sum_{q_{f}=\{u,d\}}q_{f}^{2}\sum_{p,\ell=0}^{\infty}\frac{\eta_{nkp\ell}^{(q)}(\xi_{q}-4\ell|q_{f}eB|)
{\cal{K}}_{44}^{(q)}}{\xi_{q}}{\cal{N}}^{(q)}\nonumber\\ &&\times\big[g^{\mu_{\perp}\nu_{\perp}}-is_{q}(g^{1\mu_{\perp}}g^{2\nu_{\perp}}-g^{1\nu_{\perp}}g^{2\mu_{\perp}})\big].
\end{eqnarray}
In the above expressions ${\cal{K}}_{ij}^{(q)}$ for positive and negative charged particles are presented in (\ref{D39}) and (\ref{D40}).
\subsubsection{Final result for the leptonic part of $\Delta_{B}$}\label{app2c}
\par\noindent
The final result for the leptonic part of $\Delta_{B}$ from section \ref{sec4B} reads
\begin{eqnarray}\label{D33}
{\cal{L}}^{\mu\nu}_{11,nk}=\frac{1}{(\qpar)^{2}{\cal{Q}}\eta}
\left(
\begin{array}{cccc}
Y^{00}_{11}&0&0&Y^{03}_{11}\\
0&0&0&0\\
0&0&0&0\\
Y^{30}_{11}&0&0&Y^{33}_{11}
\end{array}
\right),\qquad
{\cal{L}}^{\mu\nu}_{12,nk}=\frac{4\tilde{p}_{1}^{(2)}\tilde{p}_{2}^{(2)}}{Q\eta}g^{\mu_{\|}\nu_{\|}},
\end{eqnarray}
with the elements of the $Y$-matrix
\begin{eqnarray}\label{D34}
Y^{00}_{11}&=&2\eta[q_{0}^{2}(\qpar-\eta)+q_{3}^{2}(\qpar-Q^{2}\eta)]+8k(\qpar)^{2}eB,\nonumber\\
Y^{03}_{11}&=&2\eta q_{0}q_{3}(-2\qpar+\eta+Q^{2}\eta),\nonumber\\
Y^{30}_{11}&=&2\eta q_{0}q_{3}(-2\qpar+\eta+Q^{2}\eta),\nonumber\\
Y^{33}_{11}&=&2\eta[q_{3}^{2}(\qpar-\eta)+q_{0}^{2}(\qpar-Q^{2}\eta)]-8k(\qpar)^{2}eB.
\end{eqnarray}
Moreover, we have
\begin{eqnarray}\label{D35}
{\cal{L}}^{\mu\nu}_{41,nk}&=&+\frac{4\tilde{p}_{1}^{(2)}\tilde{p}_{2}^{(2)}}{{\cal{Q}}\eta}\big[2g^{2\mu_{\perp}}g^{2\nu_{\perp}}+g^{\mu_{\perp}\nu_{\perp}}-i(g^{1\mu_{\perp}}g^{2\nu_{\perp}}+g^{1\nu_{\perp}}g^{2\mu_{\perp}})\big],\nonumber\\
{\cal{L}}^{\mu\nu}_{42,nk}&=&+\frac{4\tilde{p}_{1}^{(2)}\tilde{p}_{2}^{(2)}}{{\cal{Q}}\eta}\big[2g^{2\mu_{\perp}}g^{2\nu_{\perp}}+g^{\mu_{\perp}
\nu_{\perp}}+i(g^{1\mu_{\perp}}g^{2\nu_{\perp}}+
g^{1\nu_{\perp}}g^{2\mu_{\perp}})\big],\nonumber\\
{\cal{L}}^{\mu\nu}_{43,nk}&=&-\frac{2(\eta-4keB)}{{\cal{Q}}\eta}\big[g^{\mu_{\perp}\nu_{\perp}}+i(g^{1\mu_{\perp}}g^{2\nu_{\perp}}-g^{1\nu_{\perp}}g^{2\mu_{\perp}})\big],\nonumber\\
{\cal{L}}^{\mu\nu}_{44,nk}&=&-\frac{2(\eta-4keB)}{{\cal{Q}}\eta}\big[g^{\mu_{\perp}\nu_{\perp}}-i(g^{1\mu_{\perp}}g^{2\nu_{\perp}}-g^{1\nu_{\perp}}g^{2\mu_{\perp}})\big].
\end{eqnarray}
\subsubsection{Final results for ${\cal{K}}_{ij}^{(q)}$ from (\ref{D14})}\label{app2d}
\par\noindent
In this part, we will focus on the coefficients ${\cal{K}}_{ij}^{(q)}$ from (\ref{D14}). They arise from the integration over $\ell_{2}, x_{1}$ and $y_{1}$ in (\ref{D3}). The integration over $x_{1}$ and $y_{1}$ is performed by making use of (\ref{D11}) and (\ref{D12}) for positively and negatively charged particles. Because of the special character of the bases $\Xi_{\mu\nu}^{(q)ij}$ from (\ref{D10}), the results for ${\cal{K}}_{ij}^{(q)}$ can be separated into four groups: $(\mu,\nu)=(\|,\|)$, $(\mu,\nu)=(\|,\perp)$, $(\mu,\nu)=(\perp,\|)$ and $(\mu,\nu)=(\perp,\perp)$. For each group the contributions of positive and negative charges are to be computed separately. The corresponding expressions to $(\mu,\nu)=(\|,\|)$ and $(\mu,\nu)=(\perp,\perp)$ are already presented in (\ref{D39}) and (\ref{D40}). In what follows, for the sake of completeness,  we will present the results for $(\mu,\nu)=(\|,\perp)$ and $(\mu,\nu)=(\perp,\|)$.
\par\vspace{0.5cm}\par\noindent
\textit{\underline{${\cal{K}}_{ij}^{(q)}$ for $(\mu,\nu)=(\|,\perp)$}}
\par\vspace{0.5cm}\par\noindent
Positive charges:
\begin{eqnarray}\label{appA1}
{\cal{K}}_{21}^{(+)}&=&-\frac{\Pi_{\ell}}{\ell_{B+}} \frac{\sqrt{2\ell}}{p!\ell!}\kappa_{+}^{p+\ell-1-m_{1}-m_{3}}e^{-\kappa_{+}}(q_{2}+iq_{1})[U_{M_{1}-m_{1}+1}^{-m_{1}}(\kappa_{+})][U_{M_{3}-m_{3}+1}^{-m_{3}}(\kappa_{+})],\nonumber\\
{\cal{K}}_{22}^{(+)}&=&
-\frac{\Pi_{p}\Pi_{\ell}}{\ell_{B+}} \frac{p\sqrt{2\ell}}{p!\ell!}
\kappa_{+}^{p+\ell-1-m_{1}-m_{4}}e^{-\kappa_{+}}(q_{2}-iq_{1})[U_{M_{1}-m_{1}+1}^{-m_{1}+1}(\kappa_{+})][U_{M_{4}-m_{4}+1}^{-m_{4}}(\kappa_{+})],
\nonumber\\
{\cal{K}}_{23}^{(+)}&=&+\frac{\Pi_{p}}{\ell_{B+}} \frac{\sqrt{2p}}{p!\ell!}\kappa_{+}^{p+\ell-1-m_{1}-m_{4}}e^{-\kappa_{+}}(q_{2}-iq_{1})[U_{M_{1}-m_{1}+1}^{-m_{1}}(\kappa_{+})][U_{M_{4}-m_{4}+1}^{-m_{4}}(\kappa_{+})],\nonumber\\
{\cal{K}}_{24}^{(+)}&=&+\frac{\Pi_{p}\Pi_{\ell}}{\ell_{B+}} \frac{\ell\sqrt{2p}}{p!\ell!}\kappa_{+}^{p+\ell-1-m_{1}-m_{3}}e^{-\kappa_{+}}(q_{2}+iq_{1})
[U_{M_{1}-m_{1}+1}^{-m_{1}+1}(\kappa_{+})][U_{M_{3}-m_{3}+1}^{-m_{3}}(\kappa_{+})].
\end{eqnarray}
Negative charges:
\begin{eqnarray}\label{appA2}
{\cal{K}}_{21}^{(-)}&=&+\frac{\Pi_{\ell}}{\ell_{B-}} \frac{p\sqrt{2\ell}}{p!\ell!}\kappa_{-}^{p+\ell-1-m_{1}-m_{4}}e^{-\kappa_{-}}(q_{2}+iq_{1})
[U_{M_{1}-m_{1}+1}^{-m_{1}+1}(\kappa_{-})][U_{M_{4}-m_{4}+1}^{-m_{4}}(\kappa_{-})],\nonumber\\
{\cal{K}}_{22}^{(-)}&=&+\frac{\Pi_{p}\Pi_{\ell}}{\ell_{B-}} \frac{\sqrt{2\ell}}{p!\ell!}\kappa_{-}^{p+\ell-1-m_{1}-m_{3}}e^{-\kappa_{-}}(q_{2}-iq_{1})[U_{M_{1}-m_{1}+1}^{-m_{1}}(\kappa_{-})][U_{M_{3}-m_{3}+1}^{-m_{3}}(\kappa_{-})],\nonumber\\
{\cal{K}}_{23}^{(-)}&=&-\frac{\Pi_{p}}{\ell_{B-}} \frac{\ell\sqrt{2p}}{p!\ell!}\kappa_{-}^{p+\ell-1-m_{1}-m_{3}}e^{-\kappa_{-}}(q_{2}-iq_{1})
[U_{M_{1}-m_{1}+1}^{-m_{1}+1}(\kappa_{-})][U_{M_{3}-m_{3}+1}^{-m_{3}}(\kappa_{-})],\nonumber\\
{\cal{K}}_{24}^{(-)}&=&-\frac{\Pi_{p}\Pi_{\ell}}{\ell_{B-}} \frac{\sqrt{2p}}{p!\ell!}\kappa_{-}^{p+\ell-1-m_{1}-m_{4}}e^{-\kappa_{-}}(q_{2}+iq_{1})[U_{M_{1}-m_{1}+1}^{-m_{1}}(\kappa_{-})][U_{M_{4}-m_{4}+1}^{-m_{4}}(\kappa_{-})].
\end{eqnarray}
\par\vspace{0.5cm}\par\noindent
\textit{\underline{${\cal{K}}_{ij}^{(q)}$ for $(\mu,\nu)=(\perp,\|)$}}
\par\vspace{0.5cm}\par\noindent
Positive charges:
\begin{eqnarray}\label{appA3}
{\cal{K}}_{31}^{(+)}&=&+\frac{\Pi_{p}}{\ell_{B+}} \frac{\sqrt{2p}}{p!\ell!}\kappa_{+}^{p+\ell-1-m_{1}-m_{4}}e^{-\kappa_{+}}(q_{2}+iq_{1})[U_{M_{1}-m_{1}+1}^{-m_{1}}(\kappa_{+})][U_{M_{4}-m_{4}+1}^{-m_{4}}(\kappa_{+})],\nonumber\\
{\cal{K}}_{32}^{(+)}&=&+\frac{\Pi_{p}\Pi_{\ell}}{\ell_{B+}} \frac{\ell\sqrt{2p}}{p!\ell!}\kappa_{+}^{p+\ell-1-m_{1}-m_{3}}e^{-\kappa_{+}}(q_{2}-iq_{1})
[U_{M_{1}-m_{1}+1}^{-m_{1}+1}(\kappa_{+})][U_{M_{3}-m_{3}+1}^{-m_{3}}(\kappa_{+})],\nonumber\\
{\cal{K}}_{33}^{(+)}&=&-\frac{\Pi_{p}\Pi_{\ell}}{\ell_{B+}} \frac{p\sqrt{2\ell}}{p!\ell!}\kappa_{+}^{p+\ell-1-m_{1}-m_{4}}e^{-\kappa_{+}}(q_{2}+iq_{1})
[U_{M_{1}-m_{1}+1}^{-m_{1}+1}(\kappa_{+})][U_{M_{4}-m_{4}+1}^{-m_{4}}(\kappa_{+})],\nonumber\\
{\cal{K}}_{34}^{(+)}&=&-\frac{\Pi_{\ell}}{\ell_{B+}} \frac{\sqrt{2\ell}}{p!\ell!}\kappa_{+}^{p+\ell-1-m_{1}-m_{3}}e^{-\kappa_{+}}(q_{2}-iq_{1})[U_{M_{1}-m_{1}+1}^{-m_{1}}(\kappa_{+})][U_{M_{3}-m_{3}+1}^{-m_{3}}(\kappa_{+})].
\end{eqnarray}
Negative charges:
\begin{eqnarray*}
{\cal{K}}_{31}^{(-)}&=&-\frac{\Pi_{p}}{\ell_{B-}} \frac{\ell\sqrt{2p}}{p!\ell!}\kappa_{-}^{p+\ell-1-m_{1}-m_{3}}e^{-\kappa_{-}}(q_{2}+iq_{1})
[U_{M_{1}-m_{1}+1}^{-m_{1}+1}(\kappa_{-})][U_{M_{3}-m_{3}+1}^{-m_{3}}(\kappa_{-})],\nonumber\\
{\cal{K}}_{32}^{(-)}&=&-\frac{\Pi_{p}\Pi_{\ell}}{\ell_{B-}} \frac{\sqrt{2p}}{p!\ell!}\kappa_{-}^{p+\ell-1-m_{1}-m_{4}}e^{-\kappa_{-}}(q_{2}-iq_{1})[U_{M_{1}-m_{1}+1}^{-m_{1}}(\kappa_{-})][U_{M_{4}-m_{4}+1}^{-m_{4}}(\kappa_{-})],\nonumber\\
\end{eqnarray*}
\begin{eqnarray}\label{appA4}
{\cal{K}}_{33}^{(-)}&=&+\frac{\Pi_{p}\Pi_{\ell}}{\ell_{B-}} \frac{\sqrt{2\ell}}{p!\ell!}\kappa_{-}^{p+\ell-1-m_{1}-m_{3}}e^{-\kappa_{-}}(q_{2}+iq_{1})[U_{M_{1}-m_{1}+1}^{-m_{1}}(\kappa_{-})][U_{M_{3}-m_{3}+1}^{-m_{3}}(\kappa_{-})],\nonumber\\
{\cal{K}}_{34}^{(-)}&=&+\frac{\Pi_{\ell}}{\ell_{B-}} \frac{p\sqrt{2\ell}}{p!\ell!}\kappa_{-}^{p+\ell-1-m_{1}-m_{4}}e^{-\kappa_{-}}(q_{2}-iq_{1})
[U_{M_{1}-m_{1}+1}^{-m_{1}+1}(\kappa_{-})][U_{M_{4}-m_{4}+1}^{-m_{4}}(\kappa_{-})].
\end{eqnarray}
\section{Dilepton production rate in strong magnetic field limit}\label{appB}
\setcounter{equation}{0}
\par\noindent
In this appendix, we present the analytical expression for dilepton production rate $\Delta_{B}$ in strong magnetic field limit. To do this, we use the results already presented in previous sections, and set all internal and external Landau levels, ($p,\ell$) and ($n,k$),  equal to zero.
According to (\ref{A34}), $\Delta_{B}$ is given by the product of a photonic and a leptonic part. The final results for the photonic part, $\mbox{Im}[\Pi_{\mu\nu}]$, and the leptonic part, ${\cal{L}}^{\mu\nu}$, for generic Landau levels ($n,k$) and $(p,\ell)$, are presented in (\ref{D24})-(\ref{D27}) and (\ref{D33})-(\ref{D35}), respectively. As it turns out, in the LLL approximation, i.e. for $p=\ell=0$, the only nonvanishing contribution of $\mbox{Im}[\Pi_{\mu\nu}]$ arises from $\mbox{Im}[\Pi_{\mu\nu}]_{11}^{nk}$ with $n=k=0$ [see (\ref{D24})]. This implies the following general expression for $\Delta_{B}$ in the LLL approximation,
\begin{eqnarray}\label{appB1}
\Delta_{B}^{\mbox{\tiny{LLL}}}=\frac{2\alpha eB}{(q^{2})^{2}(e^{\beta q_{0}}-1)}\mbox{Im}[\Pi_{\mu\nu}(q)]_{11}^{00}{\cal{L}}_{11,00}^{\mu\nu},
\end{eqnarray}
where ${\cal{L}}_{11,nk}^{\mu\nu}$ with $n=k=0$ is given in (\ref{D33}). To arrive at the final expression of $\Delta_{B}^{\mbox{\tiny{LLL}}}$, let us first consider $\mbox{Im}[\Pi_{\mu\nu}]_{11}^{00}$ from (\ref{D24}). It is given in terms of
$\xi_{q}, {\cal{R}}_{q}, \eta_{nkp\ell}^{(q)}$ and ${\cal{K}}_{11}^{(q)}$, which are defined in section \ref{sec4}.
For $p=\ell=0$ as well as $n=k=0$, they are given by
\begin{eqnarray}\label{appB2}
\begin{array}{rclcrcl}
\xi_{q}&\stackrel{LLL}{\longrightarrow}&\qpar,&\qquad&
{\cal{R}}_{q}&\stackrel{LLL}{\longrightarrow}&R\equiv \sqrt{1-\frac{4m_{q}^{2}}{\qpar}},\\
\eta_{nkp\ell}^{(q)}&\stackrel{LLL}{\longrightarrow}&1,&\qquad&
{\cal{K}}_{11}^{(q)}&\stackrel{LLL}{\longrightarrow}&\frac{2e^{-\kappa_{+}}}{\ell_{B_{+}}^{2}}.\\
\end{array}
\end{eqnarray}
Plugging these expressions into (\ref{D24}), we immediately arrive at the following expression for the photonic part of $\Delta_{B}^{\mbox{\tiny{LLL}}}$ in the LLL approximation:
\begin{eqnarray}\label{appB3}
\mbox{Im}[\Pi_{\mu\nu}(q)]^{00}_{11}&=&\frac{8N_{c}\alpha (e^{\beta q_{0}}-1)m_{q}^{2}}{9(\qpar)^{2}}\frac{e^{-\kappa_{+}}}
{\ell_{B_{+}}^{2}}\left(\tilde{\cal{N}}_{+}+\tilde{\cal{N}}_{-}\right)(q_{\mu_{\|}}q_{\nu_{\|}}-g_{\mu_{\|}\nu_{\|}}\qpar).
\end{eqnarray}
Here, $\widetilde{\cal{N}}_{+}$ and $\widetilde{\cal{N}}_{-}$ are given by [see (\ref{D26}) and set $p=\ell=0$]
\begin{eqnarray}\label{appB4}
\widetilde{\cal{N}}_{+}&=&\frac{N_{f}(\tilde{E}_{\ell}^{+})N_{f}(\tilde{E}_{p}^{+})}{R}-\frac{\qpar N_{f}(\tilde{E}_{\ell'}^{+})N_{f}(\tilde{E}_{p'}^{+})}{\bigg|2q_{0}q_{3}+(q_{0}^{2}+q_{3}^{2}){R}\bigg|},\nonumber\\
\widetilde{\cal{N}}_{-}&=&\frac{N_{f}(\tilde{E}_{\ell}^{-})N_{f}(\tilde{E}_{p}^{-})}{R}-\frac{\qpar N_{f}(\tilde{E}_{\ell'}^{-})N_{f}(\tilde{E}_{p'}^{-})}{\bigg|2q_{0}q_{3}-(q_{0}^{2}+q_{3}^{2}){R}\bigg|},
\end{eqnarray}
with
$\tilde{E}_{\ell}^{\pm}=\tilde{E}_{p}^{\mp}=-\tilde{E}_{\ell'}^{\pm}=
\frac{q_{0}\pm q_{3}{R}}{2}$ as well as
$\tilde{E}_{p'}^{\pm}=\frac{3q_{0}\pm q_{3}{R}}{2}$ [see (\ref{D21}) and set $p=\ell=0$]. The appearance of a factor $(q_{\mu_{\|}}q_{\nu_{\|}}-g_{\mu_{\|}\nu_{\|}}\qpar)$ on the r.h.s. of (\ref{appB2}) is a guarantee for the gauge invariance of our result in LLL. Same factor appears also in \cite{fukushima2011} and very recently in \cite{mamo2012, munshi2016}.
\par
As concerns the leptonic part in the LLL, let us consider  ${\cal{L}}_{11,nk}^{\mu\nu}$ with $n=k=0$ from (\ref{D33}). It is given in terms of $\eta$ and ${\cal{Q}}$, which are defined in section \ref{sec4}. For $n=k=0$, they are given by
\begin{eqnarray}\label{appB5}
\begin{array}{rclcrcl}
\eta&\stackrel{LLL}{\longrightarrow}&\qpar,&\qquad&
{\cal{Q}}&\stackrel{LLL}{\longrightarrow}&Q\equiv \sqrt{1-\frac{4m_{\ell}^{2}}{\qpar}}.
\end{array}
\end{eqnarray}
Plugging these expressions into (\ref{D34}), ${\cal{L}}_{11,00}^{\mu\nu}$ is given by
\begin{eqnarray}\label{appB6}
{\cal{L}}_{11,00}^{\mu\nu}=\frac{8m_{\ell}^{2}}{(\qpar)^{2}Q}(q^{\mu_{\|}}q^{\nu_{\|}}-g^{\mu_{\|}\nu_{\|}}\qpar).
\end{eqnarray}
Plugging at this stage (\ref{appB3}) and (\ref{appB6}) into (\ref{appB1}), the final analytical result for $\Delta_{B}^{\mbox{\tiny{LLL}}}$ reads
\begin{eqnarray}\label{appB7}
\Delta_{B}^{\mbox{\tiny{LLL}}}=\frac{128N_{c}eB\alpha^{2}m_{\ell}^{2}m_{q}^{2}}{9(\qpar)^{2}(q^{2})^{2}Q}\frac{e^{-\kappa_{+}}}{\ell_{B_{+}}^{2}}(\widetilde{\cal{N}}_{+}+\widetilde{\cal{N}}_{-}).
\end{eqnarray}
Let us notice that in the LLL the factors $R=\left(1-\frac{4m_{q}^{2}}{\qpar}\right)^{-1}$ in $\widetilde{\cal{N}}_{\pm}$ and $Q=\left(1-\frac{4m_{\ell}^{2}}{\qpar}\right)^{-1}$ in the denominator of (\ref{appB7}) fix the threshold value for dilepton production in the strong field limit to $\qpar>4m_{q}^{2}$ and $\qpar>4m_{\ell}^{2}$. As expected, in contrast to our results in section \ref{sec5}, where the contributions of all levels $n=k=0,\cdots, 10$ and $p=\ell=0,\cdots, 10$ to $\Delta_{B}$ from (\ref{D36})-(\ref{D41}) were considered, the threshold value of dilepton production in the LLL does not depend on the ratio $T^{2}/eB$, which, according to our descriptions in section \ref{sec5}, fixes the upper limit of the summation over $n,k,p,\ell$ through $\lfloor \frac{T^{2}}{eB}\rfloor$.
\end{appendix}

\end{document}